\documentstyle[11pt]{article}

\input amssym 
\catcode`\@=11\relax
\newwrite\@unused
\def\typeout#1{{\let\protect\string\immediate\write\@unused{#1}}}

\def\psglobal#1{
\immediate\special{ps: plotfile #1 }}
\def\psfiginit{\typeout{psfiginit}
\immediate\psglobal{figtex.pro}%
\special{ps:: /TeXMagnification {\the\mag} def}
}

\def\@nnil{\@nil}
\def\@empty{}
\def\@psdonoop#1\@@#2#3{}
\def\@psdo#1:=#2\do#3{\edef\@psdotmp{#2}\ifx\@psdotmp\@empty \else
    \expandafter\@psdoloop#2,\@nil,\@nil\@@#1{#3}\fi}
\def\@psdoloop#1,#2,#3\@@#4#5{\def#4{#1}\ifx #4\@nnil \else
       #5\def#4{#2}\ifx #4\@nnil \else#5\@ipsdoloop #3\@@#4{#5}\fi\fi}
\def\@ipsdoloop#1,#2\@@#3#4{\def#3{#1}\ifx #3\@nnil
       \let\@nextwhile=\@psdonoop \else
      #4\relax\let\@nextwhile=\@ipsdoloop\fi\@nextwhile#2\@@#3{#4}}
\def\@tpsdo#1:=#2\do#3{\xdef\@psdotmp{#2}\ifx\@psdotmp\@empty \else
    \@tpsdoloop#2\@nil\@nil\@@#1{#3}\fi}
\def\@tpsdoloop#1#2\@@#3#4{\def#3{#1}\ifx #3\@nnil
       \let\@nextwhile=\@psdonoop \else
      #4\relax\let\@nextwhile=\@tpsdoloop\fi\@nextwhile#2\@@#3{#4}}
\def\psdraft{
	\def\@psdraft{0}
	\def\@psdraftspecial{100}
}
\def\psdraftspecial{
	\def\@psdraft{0}
	\def\@psdraftspecial{0}
}
\def\psfull{
	\def\@psdraft{100}
}
\psfull

\newif\if@prologfile
\newif\if@postlogfile
\newif\if@bbllx
\newif\if@bblly
\newif\if@bburx
\newif\if@bbury
\newif\if@height
\newif\if@width
\newif\if@rheight
\newif\if@rwidth
\newif\if@clip
\newif\if@right
\newif\if@left
\newif\if@toplines
\newif\if@box
\newif\if@caption
\newif\if@surround
\newif\if@captionwidth
\newif\if@captionwrite
\newif\if@captionopen
\def\@p@@sclip#1{\@cliptrue}
\def\@p@@sfile#1{
		\def\@p@sfile{#1}
}
\def\@p@@sfigure#1{
		\def\@p@sfile{#1}
}
\def\@p@sfake{\hbox to 0pt{\hss Whatever\hss}}
\def\@p@@sbbllx#1{
		\@bbllxtrue
		\@d@mscratch=#1
		\edef\@p@sbbllx{\number\@d@mscratch}
}
\def\@p@@sbblly#1{
		\@bbllytrue
		\@d@mscratch=#1
		\edef\@p@sbblly{\number\@d@mscratch}
}
\def\@p@@sbburx#1{
		\@bburxtrue
		\@d@mscratch=#1
		\edef\@p@sbburx{\number\@d@mscratch}
}
\def\@p@@sbbury#1{
		\@bburytrue
		\@d@mscratch=#1
		\edef\@p@sbbury{\number\@d@mscratch}
}
\def\@p@@sheight#1{
		\@heighttrue
		\@d@mscratch=#1
   		\edef\@p@sheight{\number\@d@mscratch}
}
\def\@p@@swidth#1{
		\@widthtrue
		\@d@mscratch=#1
		\edef\@p@swidth{\number\@d@mscratch}
}
\def\@p@@srheight#1{
		\@rheighttrue
		\@d@mscratch=#1
		\edef\@p@srheight{\number\@d@mscratch}
}
\def\@p@@srwidth#1{
		\@rwidthtrue
		\@d@mscratch=#1
		\edef\@p@srwidth{\number\@d@mscratch}
}
\def\@p@@sright#1{\@righttrue \@surroundtrue}
\def\@p@@sleft#1{\@lefttrue \@surroundtrue}
\def\@p@@sextraheight#1{\@d@mextraheight=#1}
\def\@p@@sbox#1{\@boxtrue}
\def\@p@@scaption#1{\@captiontrue}
\def\@p@@stoplines#1{
		\@toplinestrue
		\@c@ttoplines=#1
}
\def\@p@@scaptionwidth#1{
		\@captionwidthtrue
	  	\@d@mcaptionwidth=#1
}
\def\@p@@scaptionwrite#1{
		\global\@captionwritetrue
		\global\@w@rname=\expandafter{\jobname_captions.tex}
		\typeout{Captions are written to \the\@w@rname}
}

\def\@p@@sprolog#1{\@prologfiletrue\def\@prologfileval{#1}}
\def\@p@@spostlog#1{\@postlogfiletrue\def\@postlogfileval{#1}}
\def\@cs@name#1{\csname #1\endcsname}
\def\@setparms#1=#2,{\@cs@name{@p@@s#1}{#2}}
\def\ps@init@parms{
		\@bbllxfalse \@bbllyfalse
		\@bburxfalse \@bburyfalse
		\@heightfalse \@widthfalse
		\@rheightfalse \@rwidthfalse
		\def\@p@sbbllx{}\def\@p@sbblly{}
		\def\@p@sbburx{}\def\@p@sbbury{}
		\def\@p@sheight{}\def\@p@swidth{}
		\def\@p@srheight{}\def\@p@srwidth{}
		\def\@p@sfile{}
		\def\@p@scost{10}
		\def\@sc{}
		\@prologfilefalse
		\@postlogfilefalse
		\@clipfalse
		\@rightfalse \@leftfalse
		\@boxfalse \@captionfalse
		\@toplinesfalse \@surroundfalse
		\@d@mextraheight=0pt
 		\@c@ttoplines=0
		\@pshape={} \def\@p@srheight@total{}
		\@captionwidthfalse \@d@mcaptionwidth=0pt
}
\def\parse@ps@parms#1{
	 	\@psdo\@psfiga:=#1\do
		   {\expandafter\@setparms\@psfiga,}}
\newif\ifno@bb
\newif\ifnot@eof
\newread\ps@stream
\newtoks\@linetok
\def\bb@missing{
	\typeout{psfig: searching \@p@sfile \space  for bounding box}
	\openin\ps@stream=\@p@sfile
	\no@bbtrue
	\not@eoftrue
	\catcode`\%=12
	\loop
		\read\ps@stream to \line@in
		\global\@linetok=\expandafter{\line@in}
		\ifeof\ps@stream \not@eoffalse \fi
		\@bbtest{\@linetok}
		\if@bbmatch\not@eoffalse\expandafter\bb@cull\the\@linetok\fi
	\ifnot@eof \repeat
	\catcode`\%=14
}	
\catcode`\%=12
\newif\if@bbmatch
\def\@bbtest#1{\expandafter\@a@\the#1
\long\def\@a@#1
     \ifx\@bbtest#2\@bbmatchfalse\else\@bbmatchtrue\fi}
\long\def\bb@cull#1 #2 #3 #4 #5 {
	\@d@mscratch=#2 bp\edef\@p@sbbllx{\number\@d@mscratch}
	\@d@mscratch=#3 bp\edef\@p@sbblly{\number\@d@mscratch}
	\@d@mscratch=#4 bp\edef\@p@sbburx{\number\@d@mscratch}
	\@d@mscratch=#5 bp\edef\@p@sbbury{\number\@d@mscratch}
	\no@bbfalse
}
\catcode`\%=14
\def\compute@bb{
		\no@bbfalse
		\if@bbllx \else \no@bbtrue \fi
		\if@bblly \else \no@bbtrue \fi
		\if@bburx \else \no@bbtrue \fi
		\if@bbury \else \no@bbtrue \fi
		\ifno@bb \bb@missing \fi
		\ifno@bb \typeout{FATAL ERROR: no bb supplied or found}
			\no-bb-error
		\fi
		\count203=\@p@sbburx
		\count204=\@p@sbbury
		\advance\count203 by -\@p@sbbllx
		\advance\count204 by -\@p@sbblly
		\edef\@bbw{\number\count203}
		\edef\@bbh{\number\count204}
}
\def\in@hundreds#1#2#3{\count240=#2 \count241=#3
		     \count100=\count240	
		     \divide\count100 by \count241
		     \count101=\count100
		     \multiply\count101 by \count241
		     \advance\count240 by -\count101
		     \multiply\count240 by 10
		     \count101=\count240	
		     \divide\count101 by \count241
		     \count102=\count101
		     \multiply\count102 by \count241
		     \advance\count240 by -\count102
		     \multiply\count240 by 10
		     \count102=\count240	
		     \divide\count102 by \count241
		     \count200=#1\count205=0
		     \count201=\count200
			\multiply\count201 by \count100
		     	\advance\count205 by \count201
		     \count201=\count200
			\divide\count201 by 10
		     	\multiply\count201 by \count101
			\advance\count205 by \count201
		     \count201=\count200
			\divide\count201 by 100
			\multiply\count201 by \count102
			\advance\count205 by \count201
		     \edef\@result{\number\count205}
}
\def\compute@wfromh{
		\in@hundreds{\@p@sheight}{\@bbw}{\@bbh}
		\edef\@p@swidth{\@result}
}
\def\compute@hfromw{
		\in@hundreds{\@p@swidth}{\@bbh}{\@bbw}
		\edef\@p@sheight{\@result}
}
\def\compute@handw{
		\if@height
			\if@width
			\else
				\compute@wfromh
			\fi
		\else
			\if@width
				\compute@hfromw
			\else
				\edef\@p@sheight{\@bbh}
				\edef\@p@swidth{\@bbw}
			\fi
		\fi
}
\def\compute@resv{
		\if@rheight \else \edef\@p@srheight{\@p@sheight} \fi
		\if@rwidth \else \edef\@p@srwidth{\@p@swidth} \fi
		\edef\@p@srheight@total{\@p@srheight}
}
\newtoks\@pshape
\def\@c@ttoplines{\count120}
\def\@c@theightcount{\count121}
\def\@c@tshapecount{\count122}
\newdimen\@d@mwidthshape
\newdimen\@d@mextraheight
\newdimen\@d@mscratch
\def\compute@parshape{
	\if@right
		\if@left
	   		\typeout{error: Can't have both left and right set}
			\@leftfalse
		\fi
	\fi
	\@d@mscratch=\@p@swidth truesp
	\divide \@d@mscratch by 19
	\multiply \@d@mscratch by 20
	\edef\@p@swidthdimen{\the\@d@mscratch}
	\@c@tshapecount=\@c@ttoplines
 	\@d@mscratch=\baselineskip
	\multiply \@d@mscratch by \@c@ttoplines
	\advance \@d@mscratch by .4\baselineskip
    	\edef\@p@stopdistance{\the\@d@mscratch }
	\@d@mscratch=\@p@sheight truesp
	\divide \@d@mscratch by 2
	\edef\@p@shalfboxheight{\the\@d@mscratch}
	\if@toplines
		\loop \@pshape=\expandafter{\the\@pshape 0pt \hsize}
		\advance\@c@ttoplines by -1
		\ifnum\@c@ttoplines>0 \repeat
	\fi
   	\@c@theightcount=\@p@srheight@total
	\advance \@c@theightcount by \@d@mextraheight
	\divide  \@c@theightcount by \baselineskip
	\advance \@c@theightcount by 1
    	\advance \@c@tshapecount by \@c@theightcount
	\advance \@c@theightcount by -1
	\@d@mwidthshape=\hsize
     	\advance \@d@mwidthshape by -\@p@swidthdimen
	\if@left
		\def\@moveshape{0pt}
		\ifnum\@c@theightcount>0
		      	\loop
			\@pshape=%
			\expandafter{\the\@pshape %
					\@p@swidthdimen \@d@mwidthshape}
			\advance \@c@theightcount by -1
			\ifnum\@c@theightcount>0 \repeat
		\else
			\advance \@c@tshapecount by 1
		\fi
	\fi
	\if@right
		\@d@mscratch=\hsize
		\advance \@d@mscratch by -\@p@swidth truesp
		\edef\@moveshape{\@d@mscratch}
		\ifnum\@c@theightcount>0
			\loop
			\@pshape=\expandafter{\the\@pshape 0pt \@d@mwidthshape}
			\advance \@c@theightcount by -1
			\ifnum\@c@theightcount>0 \repeat
		\else
			\advance \@c@tshapecount by 1
		\fi
	\fi
	\ifnum \@p@srheight=0
		\@pshape={}
		\@c@tshapecount = 0
	\else
	 	\@pshape=\expandafter{\the\@pshape 0pt \hsize}
	\fi
}
\def\@p@ssetsurroundboxes{
	\global\parshape=\count122 \the\@pshape		
 	\moveright\@moveshape
	\vbox to 0pt\bgroup\hskip0pt\vskip\@p@stopdistance
}
\newtoks\@captiontok
\newbox\@b@xcaption
\newdimen\@d@mcaptionwidth
\newdimen\@d@mcaptionheight
\newwrite\@w@rcaption
\newtoks\@w@rname
\def\setcaption#1{\@captiontok={#1}}
\def\@set@caption{
	\setbox\@b@xcaption\vbox{\hsize\@d@mcaptionwidth
	\tolerance=9000 \baselineskip=11.4pt
	\noindent\relax\the\@captiontok}
	\if@captionwrite
		\if@captionopen
		\else
			\global\@captionopentrue
			\immediate\openout\@w@rcaption=\the\@w@rname
		\fi
		\immediate\write\@w@rcaption{\the\@captiontok}
		\immediate\write\@w@rcaption{}
	\fi
}
\def\compute@caption{
	\if@captionwidth
	\else
		\@d@mcaptionwidth = \@p@swidth truesp
		\divide \@d@mcaptionwidth by 20
		\multiply \@d@mcaptionwidth by 17
	\fi
	\@set@caption
	\@d@mcaptionheight=\ht\@b@xcaption
	\if@rheight
	\else
		\count100=\@p@srheight
	   	\advance \count100 by \@d@mcaptionheight
	   	\advance \count100 by \bigskipamount
	   	\advance \count100 by \medskipamount
	   	\edef\@p@srheight@total{\number\count100}
	\fi
}
\newif\if@alreadyjtem \@alreadyjtemfalse
\def\newpar{\hfil\vadjust{\vskip\parskip}%
	{\count100=\parskip
	\count101=\baselineskip
	\divide\count101 by 10  \multiply\count101 by 3
	\advance \count100 by \count101
	\divide\count100 by \baselineskip
	\advance\count100 by \prevgraf
	\global\prevgraf=\count100}%
	\break\if@alreadyjtem\else\indent\fi%
}
\let\sav@par=\par
\def\jtem#1{%
    	\if@alreadyjtem\else\bgroup\fi
	\def\par{\sav@par\egroup\sav@par}
	\if@alreadyjtem\else\leftskip\parindent\fi
	\@alreadyjtemtrue
	\noindent\hskip0pt
	\llap{#1\ }\ignorespaces
}
\def\compute@sizes{%
	\compute@bb
	\compute@handw
  	\compute@resv
	\if@caption
		\compute@caption
	\fi
	\if@surround
		\compute@parshape
	\fi
}
\def\@p@sdospecials{
	\ifnum\@p@scost<\@psdraft
	       	\typeout{psfig: including \@p@sfile \space }
	\fi
	\special{ps::[begin] 	\@p@swidth \space \@p@sheight \space
			\@p@sbbllx \space \@p@sbblly \space
			\@p@sbburx \space \@p@sbbury \space
			startTexFig \space }
	\ifnum\@p@scost<\@psdraft
		\if@clip
			\typeout{(clip)}
			\special{ps:: \@p@sbbllx \space \@p@sbblly \space
				\@p@sbburx \space \@p@sbbury \space
			    	doclip \space }
		\fi
	\fi
	\if@box
		\typeout{(box)}
  		\special{ps:: \@p@sbbllx \space \@p@sbblly \space
			\@p@sbburx \space \@p@sbbury \space
		    	dobox \space }
	\fi
	\ifnum\@p@scost<\@psdraft
		\if@prologfile
	    		\special{ps: plotfile \@prologfileval \space }
		\fi
		\special{ps: plotfile \@p@sfile \space }
    		\if@postlogfile
			\special{ps: plotfile \@postlogfileval \space }
		\fi
	\fi
	\special{ps::[end] endTexFig \space }
}
\newif\if@putinvbox

\def\psfig#1{{%
	\ifhmode%
		\vbox\bgroup
		\@putinvboxtrue
	\else
		\@putinvboxfalse
	\fi
       	\ps@init@parms
	\parse@ps@parms{#1}
       	\compute@sizes
	\if@surround
		\psfig@for@surround
	\else
		\psfig@for@regular
	\fi
	\if@putinvbox
       		\egroup
	\fi
}}
\def\psfig@for@surround{%
	\@p@ssetsurroundboxes
	\ifnum\@p@scost<\@psdraft
		\@p@sdospecials
		\vbox to \@p@srheight true sp{\vss}
       	\else
		\if@box
			\@p@sdospecials
		\fi
		\vbox to \@p@srheight true sp{
			\vskip\@p@shalfboxheight
			\hbox to \@p@srwidth true sp{
				\hss
				\ifnum\@p@scost<\@psdraftspecial
					\@p@sfile
				\else
					\@p@sfake
				\fi
      				\hss
			}
		\vss
		}
	\fi
	\if@caption
		\medskip
		\hbox to \@p@srwidth true sp{
			\hss
			\box\@b@xcaption
			\hss
		}
 		\medskip
	\fi
	\vss\egroup
	\vskip-\parskip
}

\def\psfig@for@regular{%
	\if@putinvbox
	\else
		\vskip\parskip
	\fi
	\ifnum\@p@scost<\@psdraft
		\@p@sdospecials
		\vbox to \@p@srheight true sp{%
			\hbox to \@p@srwidth true sp{
			\hfil
			}
		\vfil
		}
       	\else
		\if@box
			\@p@sdospecials
		\fi
	    	\vbox to \@p@srheight true sp{
			\vss
			\hbox to \@p@srwidth true sp{
				\hss
				\ifnum\@p@scost<\@psdraftspecial
					\@p@sfile
				\else
					\@p@sfake
				\fi
				\hss
			}
		    	\vss
		}
	\fi
	\if@caption
		\medskip
		\hbox to \@p@srwidth true sp{
			\hss
			\box\@b@xcaption
			\hss
		}
		\bigskip
	\fi
	\if@putinvbox
	\else
		\vskip-\parskip
	\fi
}
\catcode`\@=12\relax
   
\psfiginit

\font\scriptsizebbfont=msbm7 scaled \magstep 1
\font\subscriptsizebbfont=msbm7 scaled \magstep 1
\font\footnotesizebbfont=msbm9 scaled \magstep 0
\font\smallbbfont=msbm7 scaled \magstep 2
\font\bbfont=msbm9 scaled \magstep1  
 1
 2
 3
 4

\def\scriptsizeBbb#1{\hbox{\scriptsizebbfont #1}}
\def\subscriptsizeBbb#1{\hbox{\subscriptsizebbfont #1}}
\def\footnotesizeBbb#1{\hbox{\footnotesizebbfont #1}}
\def\smallBbb#1{\hbox{\smallbbfont #1}}
\def\Bbb#1{\hbox{\bbfont #1}}

\begin{document}

\enlargethispage{23cm}

\begin{titlepage}

$ $

\vspace{-2.5cm}  

\noindent\hspace{-1cm}
\parbox{6cm}{\footnotesize Revised: November 1998 \newline
              Micron MPC P166 by Marco Monti}\
   \hspace{7.5cm}\
   \parbox{5cm}{ {\tt hep-th/9801175}\newline
                     {ut-ma/980004}}

\vspace{1.5cm}     

\centerline{\large\bf
 On extremal transitions of Calabi-Yau threefolds}
\vspace{1ex}
\centerline{\large\bf
 and the singularity of the associated $7$-space from rolling}

\vspace{1.2cm}     


\begin{center}
\parbox[t]{5cm}{
 \centerline{\large Volker Braun\footnotemark}
 \vspace{1.1em}
 \centerline{\it Department of Physics}
 \centerline{\it University of Texas at Austin}
 \centerline{\it Austin, Texas 78712} }
 \footnotetext{E-mail: vrbraun@zippy.ph.utexas.edu}  \
\hspace{2cm} \
\parbox[t]{5cm}{
 \centerline{\large Chien-Hao Liu\footnotemark}
 \vspace{1.1em}
 \centerline{\it Department of Mathematics}
 \centerline{\it University of Texas at Austin}
 \centerline{\it Austin, Texas 78712}  }
 \footnotetext{E-mail: chienliu@math.utexas.edu}
\end{center}

\vspace{1.7em}   

\centerline{({\sl In memory of Professor Frederick J.\ Almgren Jr.})}
          
\vspace{1.7em}   

\begin{quotation}
\centerline{\bf Abstract}
\vspace{.2cm}    

\baselineskip 12pt  
{\small
M-theory compactification leads one to consider $7$-manifolds
obtained by rolling Calabi-Yau threefolds in the web of Calabi-Yau
moduli spaces. The resulting $7$-space in general has singularities
governed by the extremal transition undergone. After providing some
background in Sec.\ 1, the simplest case of conifold transitions is
studied in Sec.\ 2. In Sec.\ 3 we employ topological methods,
Smale's classification theorem of smooth simply-connected spin
closed $5$-manifolds, and a computer code in the Appendix to
understand the $5$-manifolds that appear as the link of the
singularity of a singuler Calabi-Yau threefolds from a Type II
primitive contraction of a smooth one.
From this we obtain many locally admissible extremal transition
pairs of Calabi-Yau threefolds, listed in Sec.\ 4.
Their global realization will require further study.
As a mathematical byproduct in the pursuit of the subject, we obtain
a formula to compute the topology of the boundary of the tubular
neighborhood of a Gorenstein rational singular del Pezzo surface
embedded in a smooth Calabi-Yau threefold as a divisor.
} 
\end{quotation}

\medskip   

\baselineskip 12pt


\medskip   

\noindent {\small
MSC number 1991: 81T30, 57N15, 14J25, 14B05, 14J30.
} 

\medskip   

\baselineskip 11pt  

{\footnotesize
\noindent{\bf Acknowledgements.}
C.-H.L.\ would like to thank
 Orlando Alvarez and William Thurston
for influential educations. We both would like to thank 
 Philip Candelas
for educations and group meetings,
 Jacques Distler, Daniel Freed, and Xenia de la Ossa 
for comments and discussions;
 Robert Gompf
for his course on manifold theory and many
helps/discussions/guidance-of-literatures for the work and
 Se\'{a}n Keel
for the lessons/literatures on del Pezzo surfaces; 
 Hung-Wen Chang, Thomas Curtright, Martin Halpern,
 Rafael Nepomechie, Lorenzo Sadun, Neil Turok
for helps and advices;
 Virek Narayanan, Pirjo Pasanen, and Govindan Rajesh
for drawing our attention to physics literatures;
 Sharad Agnihotri, Alex Avran, Carlos Cadavid,
 Kevin Folt\'{i}nek, Vadim Kaplunovsky, Margaret Symington,
and participants of the Geometry and String Theory Seminar
for the inspiring lectures/discussions/debates;
 Nita Goldrick, Patrick Goetz, and Marco Monti
for the assistances.
Finally C.-H.L.\ thanks Ling-Miao Chou for the moral support.

} 

\noindent
\underline{\hspace{20em}}

$^1${\footnotesize E-mail: vrbraun@zippy.ph.utexas.edu}

$^2${\footnotesize E-mail: chienliu@math.utexas.edu}

\end{titlepage}

\newpage

\begin{titlepage}

$ $

\vspace{2cm}

\begin{quote}
 {\it ``Anything that's worth doing is worth doing badly.''}$^{\ast}$

 \hspace{5cm} ------ Frederick J.\ Almgren Jr.\ (1933-1997)
\end{quote}

\vspace{6cm}

\noindent
$^{\ast}$D.\ Mackenzie:
{\it Fred Almgren (1933-1997), Lover of mathematics, family, and
    life's} \newline
{\it adventures},
{\sl Notices Amer.\ Math.\ Soc.}\ {\bf 44} (1997), pp.\ 1102 - 1106.

\bigskip

\baselineskip 11pt  
\noindent {\footnotesize
{\bf Notes from C.H.L.}\ {\it [In memory of a teacher]}.
I got some acquaintance with Prof.\ Almgren from three occasions
at Princeton: firstly as a TA for a session of calculus course that
he taught also; secondly in his joint course/seminar with
Prof.\ Frank Morgan; and thirdly in an annual Mayday relay between
the math departments of Princeton and Rutgers, in which he, his
wife Prof.\ Jean Taylor, and their daughter all participated.
After the relay we had a picnic at his home. Since he is one of the
few professors at Princeton, whom I had more contact during my
study there, I cannot help but
feeling bad when knowing of his passing away the spring 1997.
However, it was not until I read the above quote from {\sl Notices}
half a year later, as recalled by his son, that I realized that no
wonder he has such a warmth that projects also to people around him.
What a power both to work and to life that can be hidden in this
motto! And what an encouragement this motto provides to a bewildered
student who tries to understand the amazingly beautiful and yet
bafflingly difficult Final Theory of Nature - Superstrings! Thus I
like to dedicate this humble piece of work to him for an
unforgettable memory and gratitudes.
}

\end{titlepage}

\newpage
$ $

\vspace{-6em}  

\centerline{\sc
 Calabi-Yau threefolds and $7$-spaces}

\vspace{4em}

\baselineskip 14pt  

\begin{flushleft}
{\Large\bf 0. Introduction and outline.}
\end{flushleft}

\begin{flushleft}
{\bf Introduction.}
\end{flushleft}
M-theory anticipates the space-time to be $11$-dimensional
compactified on a $7$-dimensional space. If one requires this
$7$-space to be closed and satisfy the first order constraint
from the string $\beta$-function, i.e.\ Ricci flatness,
then recently Joyce has constructed a class of closed
$7$-manifolds that admit torsion-free $G_2$ structures and, hence,
Ricci flat metrics ([Jo1], [Jo2], [D-T]). On the other hand,
from the spirit of the work by Ho\v{r}ava and Witten
([H-W1], [H-W2]), more likely this $7$-space is compact yet with
boundary that are Calabi-Yau complex threefolds. While this
$7$-space itself serves geometrically as an interpolating space
of different Calabi-Yau threefolds, the M-theory built thereon
serves physically as an interpolating theory for the string
theories on the boundary Calabi-Yau threefolds. If one is willing
to accept this point of view, then certainly a key question is
that; {\it What could this $7$-space be?}

First notice that the oriented cobordism ring is trivial in
dimension $6$ ([M-S]), which means that any two oriented
closed $6$-manifolds, in particular Calabi-Yau threefolds, bound
an oriented $7$-manifold. From this point view and following
Joyce's spirit, a possible kind of $7$-space that could be
relevant are Ricci-flat compact $7$-manifolds with Calabi-Yau
boundaries.

On the other hand, in view of the phenomenon of enhanced gauge
symmetry for strings moving in a singular Calabi-Yau threefolds,
one may allow this $7$-space to have some reasonable kind of
singularities, i.e.\ non-manifold points.
Taking all these into account and balancing them, one leads to
a very special class of $7$-spaces that are closely related
to Calabi-Yau threefolds:
{\it $7$-space from rolling Calabi-Yau threefolds in the web of
 Calabi-Yau moduli spaces}.
Among many details one may like to understand for such $7$-spaces,
we focus in this paper on one single issue:
{\it its possible isolated singularities},
which arise when the degenarate Calabi-Yau
threefold that appears in the extremal transition (i.e.\
rolling through a shared boundary point of one Calabi-Yau moduli
space and entering another) has an isolated singularity.

After providing some mathematical background in Sec.\ 1 for
physicists, we discuss in Sec.\ 2 the resulting singularity in
the $7$-space from rollong when passing through a conifold.
We then move on in Sec.\ 3 and 4 to consider the more complicated
extremal transitions that involve pinching a del Pezzo surface
embedded in a Calabi-Yau threefold to a point. After showing that
the boundary $5$-manifold of a tubular neighborhood of a certain
class of del Pezzo surfaces embedded in a smooth Calabi-Yau
threefold is simply-connected and working out its second integral
homology, using Smale's classification theorem of smooth
simply-connected spin closed $5$-manifolds, we can lay down many
topologically locally admissible extremal transitions, each of
which contributes to a possible isolated singularity in the
$7$-space from rolling. Exactly which of them will appear globally
and, when it does, what is the relation of the $6$-dimensional
link of the singularity in the $7$-space to the phenomenon of
enhanced gauge symmetry will require further study in the future.

\bigskip

From a pure mathematical aspect, a main result that we
obtain in this paper, which serves as a tool in the pursuit of the
above issue, is the following:

\bigskip

\noindent
{\bf Proposition 3.2.8 [homology of boundary].} {\it
 Let $Z$ be a Gorenstein rational singular del Pezzo surface
 with $\mbox{\rm Pic}\,(Z)={\Bbb Z}$ that is embedded in a
 smooth Calabi-Yau threefold $X$ as a divisor.
 Denote the minimal resolution
 of $Z$ by $Z_r$, which is a smooth del Pezzo surface obtained
 by ${\Bbb C}P^2$ blown up at $r$ many points, and the vanishing
 cycle of the resolution by $E$ with irreducible components
 $E_1,\cdots,E_r$. Then $\partial\nu_X(Z)$ is a simply-connected
 spin $5$-manifold with the second integral homology $H_2$
 isomorphic to
 ${\Bbb Z}^r/\hspace{-.2ex}
  \mbox{\raisebox{-.4ex}{$\{{\Bbb Z}^r\,\cdot\,
    (A^t\,U^{-1}\,A)\}$}}$,
 where $U$ is the Seifert matrix associated to the singularity
 with respect to $(E_1,\cdots, E_r)$ and $A$ is the matrix whose
 $i$-th row is the coefficients of $E_i$ with respect to any basis
 for $K_{Z_r}^{\perp}$ in $H_2(Z_r;{\Bbb Z})$.
 (In the above expression, elements in ${\Bbb Z}^r$ are integral
  row vectors and the $i$-th column of $U$ is the coefficients of
  $U(E_i)$ with respect to $(E_1,\cdots,E_r)$.)
} 

\bigskip

\noindent
{\bf Convention.}
Since both real and complex manifolds are involved in this article,
to avoid confusion, a {\it real} $n$-dimensional manifold will be
called an {\it $n$-manifold} while a {\it complex} $n$-dimensional
manifold an {\it $n$-fold}.

\bigskip      

\begin{flushleft}
{\bf Outline.}
\end{flushleft}
{\small
\baselineskip 11pt  

\begin{quote}
  1. Essential mathematical backgrounds.
  \vspace{-1ex}
  \begin{quote}
   \hspace{-1.3em}
   1.1 Geometric operations, Smale's work on $5$-manifolds,
       and singularity.

   \hspace{-1.3em}
   1.2 Essence of Calabi-Yau threefolds.
  \end{quote}

  2. Lessons from conifold transitions.
  \vspace{-1ex}
  \begin{quote}
   \hspace{-1.3em}
   2.1 Conifold transition I.

   \hspace{-1.3em}
   2.2 Conifold transition II: flops.
  \end{quote}

  3. The type II primitive K\"{a}kler deformations.
  \vspace{-1ex}
  \begin{quote}
   \hspace{-1.3em}
   3.1 The case of a smooth del Pezzo surface.

   \hspace{-1.3em}
   3.2 The case of a rational singular del Pezzo surface.
  \end{quote}

  4. Topologically admissible singularities from
     pinching a del Pezzo surface.

  Appendix. A computer code.
\end{quote}
} 

\bigskip

\newpage
\baselineskip 14pt  

\section{Essential mathematical backgrounds.}

In Sec.\ 1.1 we collect some mathematical background involved
in this article for the convenience of physicists.
In Sec.\ 1.2 the essence of Calabi-Yau threefolds needed for the
work is provided.
Some expository articles or textbooks are also referred.

\bigskip

\noindent
{\it Caution.}
The term {\it link} has two different meanings in this paper:
(1) a collection of $S^n$ embedded disjointly in $S^{2n+1}$; or
(2) the base for a cone-type neighborhood of a cone point in
    a topological space, e.g.\ the link of any point in a
    closed $n$-manifold is an $(n-1)$-sphere.
Both are standard and should be distinguishable from the context.

\bigskip

\subsection{Geometric operations, Smale's work on 5-manifold,
 and singularity.}

\noindent $\bullet$
{\bf General fundamentals.}
Readers are referred to [B-T], [G-P], [Hirs], [Mu], [Sp], and [Sw]
for algebraic and differential topology;
and to [B-P-VV], [G-H], [Sh1], and [Sh2] for
complex algebraic geometry.

\bigskip

\noindent $\bullet$
{\bf Handle decompositions of manifolds.}
([G-S], [Ki], [Po], and [Sm1];
 also [Mi] for the related Morse theory.) Let $D^r$ be the
$r$-dimensional ball. Then a {\it handle decomposition} of a
compact $n$-manifold $M^n$ is a sequence
$$
 D^n\,=\,M_0\;\subset\;\cdots\;\subset\;
  M_{i-1}\;\subset\;M_i\;\subset\;\cdots\;\subset M_k\,=\,M^n\,,
$$
where $M_i$ is obtained from $M_{i-1}$ by adding a
$k_i$-handle, i.e.\
$M_i=M_{i-1}\cup_{f_i} (D^{k_i}\times D^{n-k_i})$ where the
attaching map 
$f_i:\partial D^{k_i}\times D^{n-k_i} \rightarrow \partial M_{i-1}$
is an embedding. In terms of Morse theory, this corresponds to
a non-degenerate critical point with index $k_i$ of a Morse function
on $M^n$. Its effect to the $(n-1)$-manifold $\partial M_{i-1}$ is
that an embedded $(k_i-1)$-sphere - i.e.\ $f_i(\partial D^{k_i})$ -
gets shrunk to a point and then gets expanded in transverse
directions to an embedded $(n-k_i-1)$-sphere -
i.e.\ $\partial D^{n-k_i}$ - in $\partial M_i$.
({\sc Figure} 1-1).
\begin{figure}[htbp]
\setcaption{{\sc Figure 1-1.}
\baselineskip 14pt
 In this illustration, a $1$-handle (darkly shaded) is attached to
 $M_i$ on its boundary $\partial M_i$ (lightly shaded).
}
\centerline{\psfig{figure=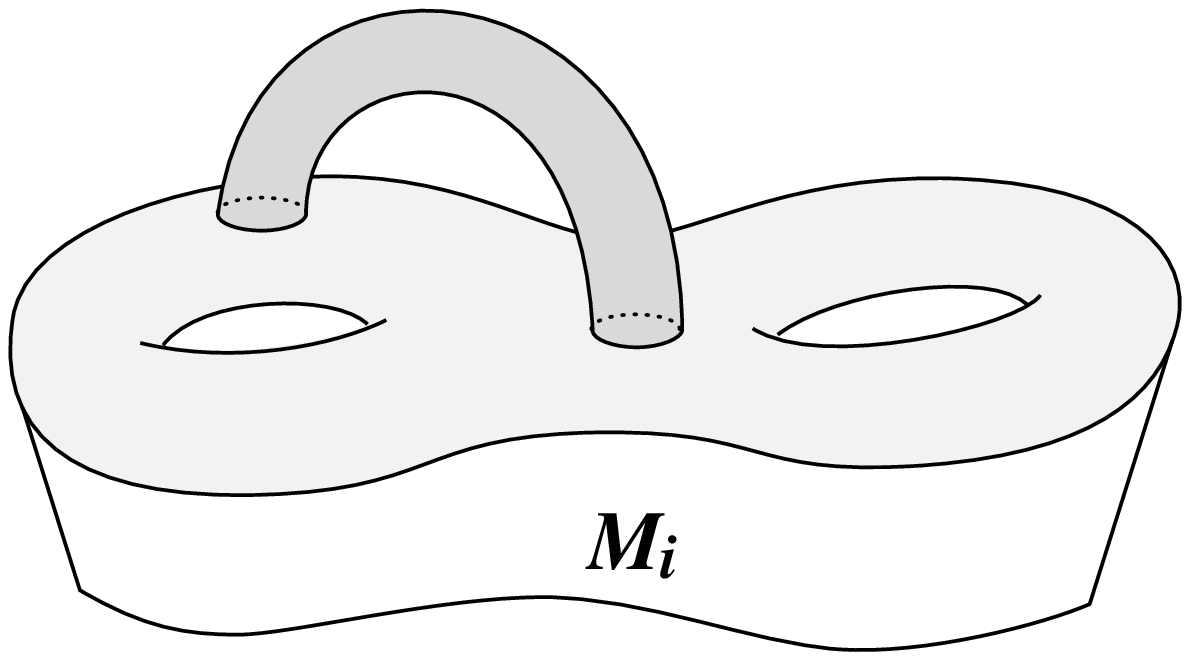,width=13cm,caption=}}
\end{figure}

\bigskip                

\noindent $\bullet$
{\bf Fiber sum and plumbing.} ([Go], [G-S].)
Let $\pi_i:{\cal E}_i\rightarrow M^n_i$, $i=1,2$, be fibrations
over $n$-manifolds with the same generic fibers $F$ and let
$p_i\in M_i$ such that the fibers $\pi_i^{-1}(p_i)$ are generic.
Take tubular neighborhoods $\nu_{{\cal E}_i}(\pi_i^{-1}(p_i))$ of
these fibers and a fiber-preserving homeomorphism $\rho$ between
the boundary of ${\cal E}_i-\nu_{{\cal E}_i}(\pi_i^{-1}(p_i))$.
One can then glue the latter spaces together by $\rho$ and
construct a new space ${\cal E}_1\sharp_f{\cal E}_2$, which now
admits a fibration over the connected sum $M^n_1\sharp M^n_2$ with
the same generic fiber $F$. This space
${\cal E}_1\sharp_f{\cal E}_2$ is called a {\it fiber sum} of
$\pi_1:{\cal E}_1\rightarrow M^n_1$ and
$\pi_2:{\cal E}_2\rightarrow M^n_2$. ({\sc Figure} 1-2(a).)

Let $\pi:{\cal E}\rightarrow M^n$ be a $D^n$-bundle over an
$n$-manifold that may have several components. Let
$(D_{11}, D_{12}), \cdots, (D_{k1},D_{k2})$ be a disjoint
collection of pairs of disjoint $n$-disks whose closure lie in
the interior of $M^n$. Since ${\cal E}$ restricted to these
$n$-disks are the trivial bundle $D_{js}\times D^n$, one can
identify $\pi^{-1}(D_{j1})$ with $\pi^{-1}(D_{j2})$ for each
$j$ using a map that preserves the product structures but
interchanges the factors. The resulting new manifold-with-boundary
$W^{2n}$ is called a {\it plumbing} of
$\pi:{\cal E}\rightarrow M^n$ ({\sc Figure} 1-2(b)). For example,
when an $n$-manifold $M$ is immersed in a $2n$-manifold $N$ with
only transverse crossings, its tubular neighborhood $\nu_{N}(M)$ is
then a plumbing of the $D^n$-bundle associated to the normal bundle
over $M$ from the immersion.

Associated to $W^{2n}$ is a diagram $\Gamma$, the
{\it plumbing diagram}, whose vertices $v_i$ correspond to the
connected components $M_{i}$ of $M^n$ and edges
$e_j$ correspond to the occurrences of plumbing associated to
the pairs $(D_{j1},D_{j2})$. There is a natural continuous pinching
map $\phi$ from $W^{2n}$ onto $\Gamma$ defined as follows: Let
$\partial D_{js}\times [0,\epsilon]$ be a collar of
$\partial D_{js}$ in $M^n-\cup_{j,s}D_{js}$. Define $\phi$ first on
$M^n$ by sending $M_{i}-(\mbox{all the $n$-disks with the collar})$
to $v_i$, $D_{j1}\cup D_{j2}$ to the mid-point of $e_j$, and then
extending the $\phi$ so far defined to the whole $M^n$ by
the projection
$\partial D_{js}\times [0,\epsilon]\rightarrow [0,\epsilon]$ with
$[0,\epsilon]$ identified with the half-edge of $e_j$ at the
vertex that corresponds to the component of $M^n$ that contains
$D_{js}$. By pre-composition with the bundle projection $\pi$,
one then obtains a map, still denoted by $\phi$, from $W^{2n}$
onto $\Gamma$. ({\sc Figure} 1-2(b).)

\bigskip

\begin{figure}[htbp]
\setcaption{{\sc Figure 1-2.}
\baselineskip 14pt
 In (a), a fiber sum of the fibrations ${\cal E}_i\rightarrow M_i$
 is shown. In (b), a plumbing of the two disk-bundles are shown.
 The associated plumbing diagram $\Gamma$ and the natural map
 $\phi$ are also indicated. }
\centerline{\psfig{figure=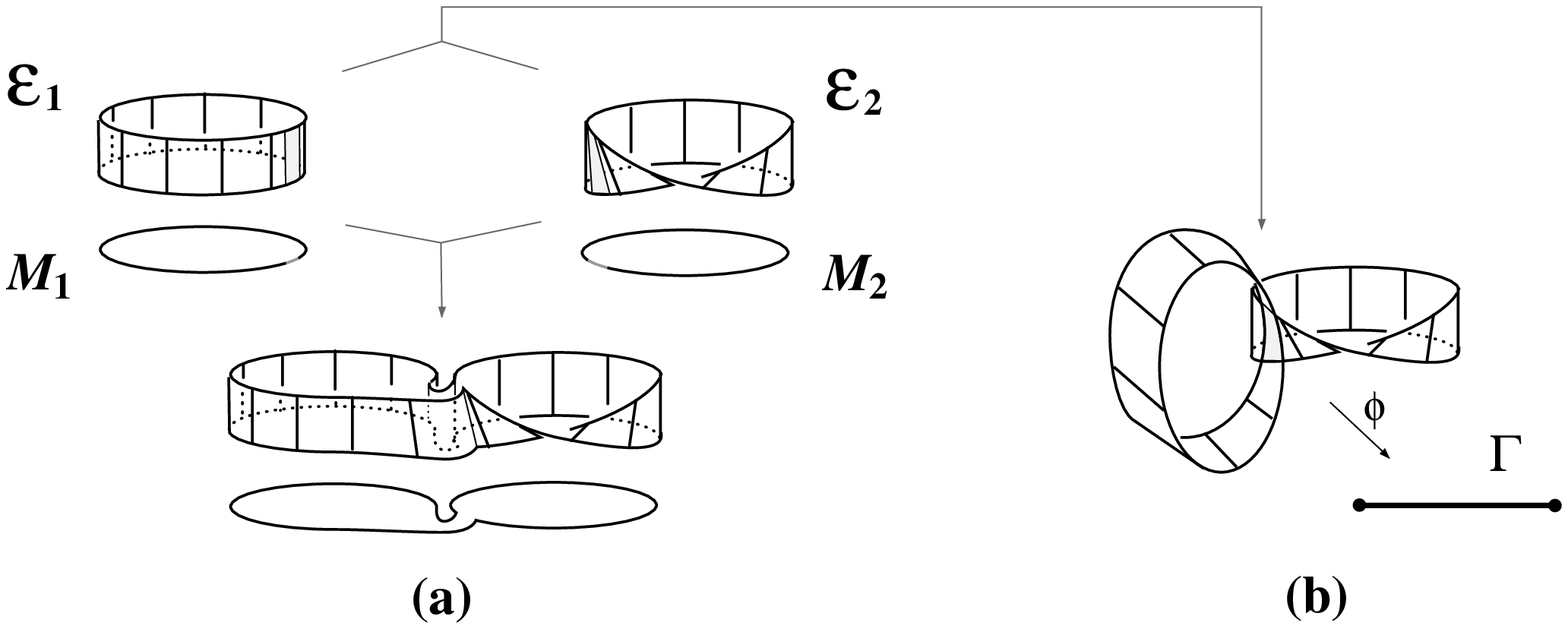,width=13cm,caption=}}
\end{figure}

\bigskip

\noindent $\bullet$
{\bf Smale's work on 5-manifolds.} ([Sm2])

\bigskip

\noindent
{\bf Fact 1.1.1 [Smale].} {\it Smooth simply-connected spin closed
 $5$-manifolds $M^5$ are classified by their second integral
 homology group $H_2(M^5;{\Bbb Z})$ and are realizable as the
 boundary of $6$-dimensional handlebodies with only one $0$-handle
 and some $3$-handles. The detail follows:
 \begin{quote}
 \hspace{-1.7em}\parbox[t]{1.5em}{(1)} \parbox[t]{13.5cm}{
 Let
 $H_2(M^5;{\Bbb Z})\;
  =\; H_2^{\mbox{\scriptsize ({\it free})}}(M^5;{\Bbb Z})\,
   \oplus\,H_2^{\mbox{\scriptsize ({\it tor})}}(M^5;{\Bbb Z})$
 be a direct sum decomposition of $H_2(M^5;{\Bbb Z})$ into
 a free and torsion part. Then the torsion part is the direct 
 sum of two identical groups. In notation,
 $$
  H_2^{\mbox{\scriptsize ({\it tor})}}(M^5;{\Bbb Z})\;
 =\; \mbox{$\frac{1}{2}$}\,
       H_2^{\mbox{\scriptsize ({\it tor})}}(M^5;{\Bbb Z})\,
   \oplus\, \mbox{$\frac{1}{2}$}\,
       H_2^{\mbox{\scriptsize ({\it tor})}}(M^5;{\Bbb Z})\,.
 $$
 } 

 \hspace{-1.7em}\parbox[t]{1.5em}{(2)} \parbox[t]{13.5cm}{
 The correspondence
 $$
  M^5 \; \longmapsto \;
   H_2^{\mbox{\it\scriptsize (free)}}(M^5;{\Bbb Z})\,
    \oplus\,\mbox{$\frac{1}{2}$}\,
       H_2^{\mbox{\it\scriptsize (tor)}}(M^5;{\Bbb Z})
 $$
 is bijective between the set of diffeomorphism classes
 of smooth simply-connected spin $5$-manifolds and the set of
 isomorphism classes of finitely generated abelian groups.
 } 

 $ $ 

 \hspace{-1.7em}\parbox[t]{1.5em}{(3)} \parbox[t]{13.5cm}{
 Let $H^6$ be the $6$-dimensional handlebody obtained by one
 $0$-handle attached with $k$ $3$-handles. By handle sliding, one
 may assume that all the attaching $S^2$ are in the boundary $S^5$
 of the $0$-handle. Let $Q$ be the associated anti-symmetric
 linking matrix of these $S^2$'s in $S^5$. Recall that $Q$ can be
 transformed uniquely into the following block diagonal form
 ([Bou])
 $$
 \mbox{\it Diag}\,\left\{\,
   \left[\begin{array}{cc} 0 & \alpha_1 \\ -\alpha_1 & 0
         \end{array}\right],\,\cdots,\,
   \left[\begin{array}{cc} 0 & \alpha_r \\ -\alpha_r & 0
         \end{array}\right],\,
   {\rm O}_{k-2r} \,\right\}
 $$
 under the transformation $Q\rightarrow SQS^t$ for some $S$ in
 $\mbox{\it GL}\,(k,{\Bbb Z})$, where $\alpha_i$ are positive
 integers that satisfy $\alpha_1|\,\alpha_2|\,\cdots\,|\,\alpha_r$
 and ${\rm O}_{k-2r}$ is the $(k-2r)\times(k-2r)$ zero-matrix.
 Then
 $$
 \partial H^6\;
  \cong\; M^5_{\alpha_1}\,\sharp\,\cdots\,
           \sharp\,M^5_{\alpha_r}\,
           \sharp\,(k-2r)\,S^2\times S^3\,,
 $$
 where $M^5_{\alpha}$ is the boundary of the $6$-dimensional
 $3$-handlebody obtained by one $0$-handle attached with two
 $3$-handles whose attaching $2$-spheres form a link with
 linking matrix
 {\scriptsize $\left(
   \begin{array}{rc} 0 & \alpha \\[-.2ex] -\alpha & 0
   \end{array} \right)$}. The $5$-manifold $\partial H^6$ is
 simply-connected and spin; and conversely every closed
 simply-connected spin $5$-manifold arises this way.
 } 
\end{quote}
}

\bigskip

\noindent {\it Remark 1.1.2.}
Note that $M^5_0=S^2\times S^3$, $M^5_1=S^5$, and, for $\alpha>1$,
$H_2(M^5_{\alpha};{\Bbb Z})={\Bbb Z}_{\alpha}\oplus{\Bbb Z}_{\alpha}$.

\bigskip

\noindent $\bullet$
{\bf Rational surface singularities.}
([A-GZ-V], [B-P-VV], [Di], [H-K-K], [Mi2], [Sl1], and [Sl2].)
These are exactly the quotient singularities
$({\Bbb C}^2\hspace{-.4ex}/
   \hspace{-.2ex}\mbox{\raisebox{-.4ex}{$G$}},0)$
for finite groups $G\subset \mbox{\it SU}\,(2)$ and hence their
links are homeomorphic to
$S^3\hspace{-.4ex}/\hspace{-.2ex}\mbox{\raisebox{-.4ex}{$G$}}$.
Their resolution graphs coincide with the Dynkin diagrams of the
simple complex Lie algebras of type $A_k$, $D_k$, $E_6$, $E_7$,
and $E_8$ respectively.  For this reason, they are also called the 
{\it $A$-$D$-$E$ singularities}. Such singularity is special in
that its minimal resolution is the same as its smooth deformation
([H-K-K]), which is the Milnor fiber $F$ appears in the Milnor
fibration associated to the singularity. The topology of $F$ is a
plumbing of a collection of the tangent $D^2$-bundle of $S^2$ with
the plumbing diagram the above mentioned Dynkin diagram. The
boundary $\partial F$ of $F$ is the same as the link 
$S^3\hspace{-.4ex}/\hspace{-.2ex}\mbox{\raisebox{-.4ex}{$G$}}$
of the singularity. Recall that the total space of the Milnor
fibration in the spherical form is a mapping torus
$F\times [0, 2\pi]/
  \hspace{-.2ex}\mbox{\raisebox{-.4ex}{$(
  F\times\{0\}\stackrel{\tau}{\sim} F\times\{2\pi\})$}}$,
by which the monodromy homeomorphism $\tau$ is defined. It is
noteworthy that, up to homotopy, the restriction of $\tau$ to the
boundary $\partial F$ is the identity map for these singularities.
Readers are referred particularly to [Di] for more details.

\bigskip

\subsection{Essence of Calabi-Yau threefolds.}

\noindent $\bullet$
{\bf Calabi-Yau threefolds.}
([A-G-M], [A-L], [B-C-dlO], [C-dlO], [C-G-H], [C-H-S-W], [Gr],
[H\"{u}], [Ti], [Vo], [Wil1], and [Wil2])
A {\it Calabi-Yau threefold} $X$ is a compact K\"{a}hler threefold
with $c_1=0$ (or equivalently trivial canonical line bundle $K_X$).
They have moduli that form a space which itself is K\"{a}hler.
Two different points in the Calabi-Yau moduli space ${\cal M}_X$
associated to the smooth manifold $X$ can be connected by
deformations of the complex and K\"{a}hler structures on $X$. The
boundary points of ${\cal M}_X$ correspond to singular Calabi-Yau
threefolds $\overline{X}$ that can be obtained from $X$ by such
deformations.
Readers are referred particularly to [Gr] and [H\"{u}] for more
details.

\bigskip

\noindent $\bullet$
{\bf Extremal transitions.}
([A-G-E], [C-G-G-K], [Gr], [G-M-S], [G-M-V], [H\"{u}],
 [M-V1], [M-V2], and [Wil2].)
Two Calabi-Yau moduli spaces
${\cal M}_X$ and ${\cal M}_{X^{\prime}}$ can be connected with
each other if they happen to share some common boundary points,
which means that $X$ and $X^{\prime}$ have a same degeneration
$\overline{X}$. Indeed it has been conjectured that all the
Calabi-Yau moduli spaces associated to different smooth topologies
are related this way and form a connected {\it web}
${\cal M}_{\mbox{\scriptsize\rm CY}}$ of Calabi-Yau moduli spaces.
Thus one may roll a smooth Calabi-Yau threefold $X$, degenerating
it to a singular $\overline{X}$ and then rolling on to another
smooth $X^{\prime}$ with possibly different topology - all
happening in ${\cal M}_{\mbox{\scriptsize\rm CY}}$. From mirror
symmetry, the corresponding conformal field theory on these
Calabi-Yau spaces that occur in the rolling may still be
well-defined and smoothly deformed accordingly. Such transmutation
is called an {\it extremal transition} of Calabi-Yau threefolds.

There are six generic types of degenerations of Calabi-Yau
threefolds whose combinations lead to various extremal transitions
discussed in the literature. To explain them, let
$(X,Z)\rightarrow (\overline{X},\overline{Z})$ be a degeneration
of a smooth $X$ to $\overline{X}$ by collapsing a subspace $Z$
of $X$ to $\overline{Z}$. In all these six cases, $Z$ is a
fibration over $\overline{Z}$ and the collapsing comes from
pinching all the fibers:
\begin{itemize}
 \item
 From the deformation of the complex structures on $X$:
 \begin{quote}
  \hspace{-1.9em}\parbox[t]{2em}{(${\rm C}_0$)}\
  \parbox[t]{13cm}{
  $Z$ is an embedded $3$-sphere $S^3$ in $X$ and
  $\overline{Z}$ is a point in $\overline{X}$.}
 \end{quote}

 \item
 From the deformation of the K\"{a}hler structures on $X$ to
 one that sits in the codimension-one wall of the K\"{a}hler
 cone ([Wil2]):
 \begin{quote}
  \hspace{-1.9em}\parbox[t]{2em}{(${\rm C}_1$)}\
  \parbox[t]{13cm}{
  ({\it Type I primitive contraction} or {\it small contraction}):
  $Z$ is a finite disjoint union of a collection of embedded
  ${\Bbb C}{\rm P}^1$s in $X$ and
  $\overline{Z}$ is a finite set of points in $\overline{X}$. }

  $ $

  \hspace{-1.9em}\parbox[t]{2em}{(${\rm C}_2$)}\
  \parbox[t]{13cm}{
  ({\it Type II primitive contraction}):
  $Z$ is a generalized del Pezzo surface in $X$ and
  $\overline{Z}$ is a point in $\overline{X}$.  }

  $ $

  \hspace{-1.9em}\parbox[t]{2em}{(${\rm C}_3$)}\
  \parbox[t]{13cm}{
  ({\it Type III primitive contraction}):
  $Z$ is a conic bundle in $X$ over a complex curve $C$ and 
  $\overline{Z}$ is the curve $C$ embedded in $\overline{X}$.  }

  $ $ 

  \hspace{-1.9em}\parbox[t]{2em}{(${\rm C}_4$)}\
  \parbox[t]{13cm}{
  $X$ admits a K3 or ${\Bbb T}^4$ fibration over a complex curve
  $C$; $Z=X$ and $\overline{Z}=C$.        }

  $ $ 

  \hspace{-1.9em}\parbox[t]{2em}{(${\rm C}_5$)}\
  \parbox[t]{13cm}{
  $X$ admits an elliptic fibration over a complex surface $S$;
  $Z=X$ and $\overline{Z}=S$.   }
 \end{quote}
\end{itemize}

\bigskip

\noindent $\bullet$
{\bf 7-spaces from rolling Calabi-Yau threefolds.} Continuing
the above discussion, let
$X_t\rightarrow\overline{X}=X_0$, $t\in [-\epsilon,0]$ be
a rolling of Calabi-Yau threefolds that degenerate
$X=X_{-\epsilon}$ to $\overline{X}$ by shrinking the fibers of
a fibration $\pi:Z\rightarrow\overline{Z}$ of a subspace $Z$ in
$X$ over a subspace $\overline{Z}$ in $\overline{X}$.
Naturally associated to this shrinking is a cone bundle
$\Lambda_{\overline{Z}}$ over $\overline{Z}$, whose fiber over
$\overline{z}\in\overline{Z}$ is the cone over the restriction of
a tubular neighborhood $\nu_X(Z)$ of $Z\subset X$ on
$\pi^{-1}(\overline{z})$. The resulting $7$-space $Y^7_-$ is
homeomorphic to $X\times[-\epsilon,0]$ with
$\Lambda_{\overline{Z}}$ attached canonically to $X\times\{0\}$
along the common subspace $\nu_X(Z)$ of both $X$ and
$\Lambda_{\overline{Z}}$; in notation,
$Y^7_-=X\times[-\epsilon,0]\cup_{\nu_X(Z)}\Lambda_{\overline{Z}}$
({\sc Figure 1-3}).
\begin{figure}[htbp]
\setcaption{{\sc Figure 1-3}.
\baselineskip 14pt
   The $7$-space $Y^7_-$ that arises from degenerating Calabi-Yau
   threefolds is topologically realizable as
   $X\times[-\epsilon,0]\cup_{\nu_X(Z)}\Lambda_{\overline{Z}}$.
   (In general, the fibration $\pi:Z\rightarrow\overline{Z}$ may
   have singular fibers.) }
\centerline{\psfig{figure=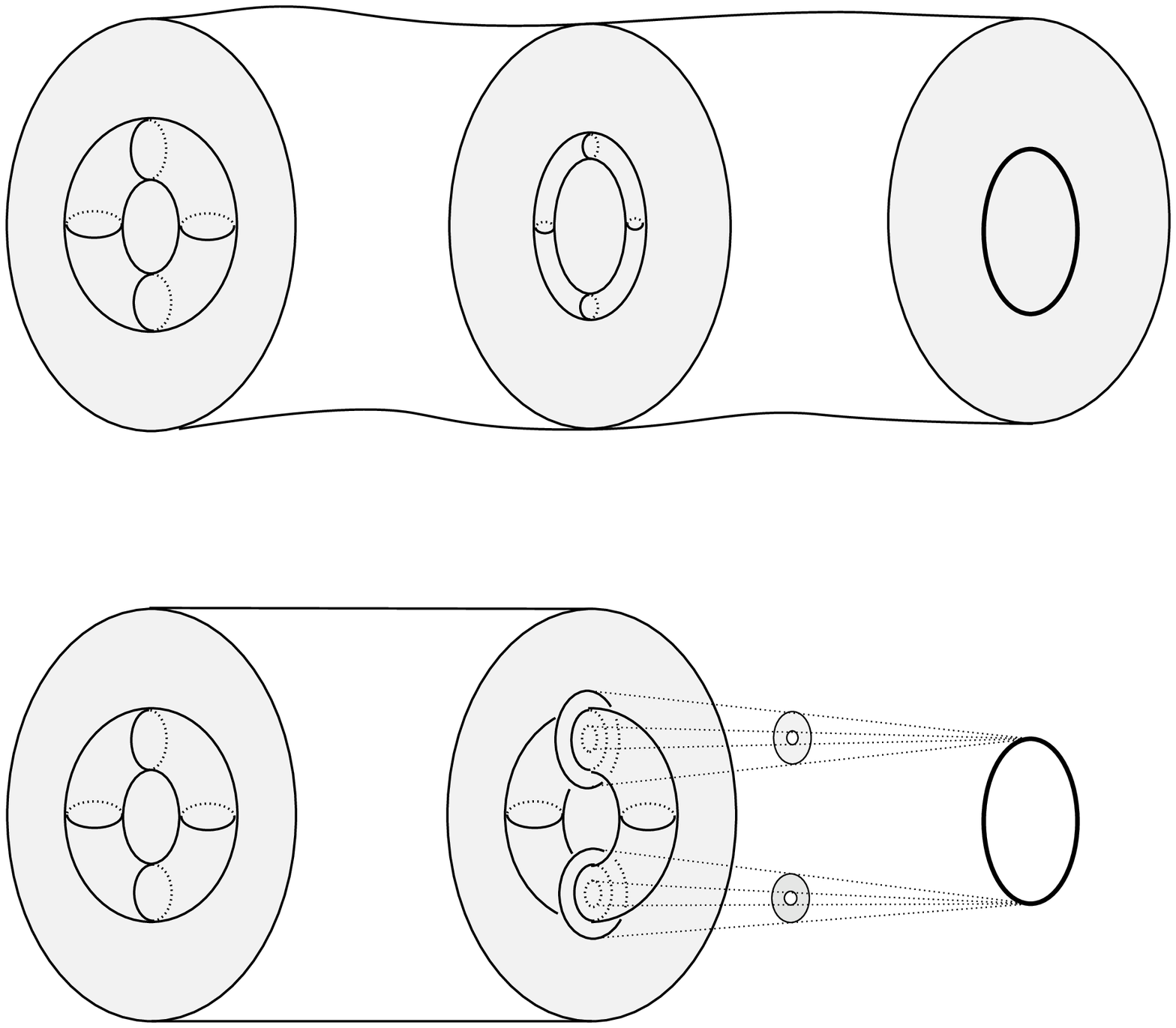,width=13cm,caption=}}
\end{figure}
When an extremal transition is obtained by a rolling
$(X,Z)\rightarrow(\overline{X},\overline{Z})
  \leftarrow(X^{\prime},Z^{\prime})$,
parametrized by $t\in[-\epsilon,\epsilon]$, where both the
degenerations
$(X,Z)$, $(X^{\prime},Z^{\prime})
 \rightarrow(\overline{X},\overline{Z})$ are of the six types,
then the resulting $7$-space, $Y^7_-$ from
$(X,Z)\rightarrow(\overline{X},\overline{Z})$
and $Y^7_+$ from
$(X^{\prime},Z^{\prime})\rightarrow(\overline{X},\overline{Z})$,
can be sewed along the shared $\overline{X}$ and form a $7$-space
$Y^7$. This is our {\it $7$-space from rolling}.
From the discussion, it is clear that

\bigskip

\noindent
{\bf Lemma 1.2.1 [topological matching condition].} {\it
 A necessary condition for an extremal transition 
 $(X,Z)\rightarrow(X^{\prime},Z^{\prime})$ to be possible is that 
 the two $5$-manifolds
 $\partial\nu_X(Z)$ and $\partial\nu_{X^{\prime}}(Z^{\prime})$
 are homeomorphic.
} 

\bigskip

When $\overline{Z}$ is a point $p_0$, in general it induces then
an isolated singularity, still denoted by $p_0$, in the $7$-space
$Y^7$, whose link is a $6$-manifold
$\nu_X(Z)\cup_{\partial}\nu_{X^{\prime}}(Z^{\prime})$, obtained
by pasting $\nu_X(Z)$ and $\nu_{X^{\prime}}(Z^{\prime})$ along
their homeomorphic boundary.

\bigskip

\section{Lessons from conifold transitions.}

A special case of examples of extremal transitions of Calabi-Yau
threefolds that has been discussed quite extensively in physics
literature is the conifold transition
(cf.\ [C-G-G-K], [Gr], [G-M-S], [H\"{u}], and [Str]).
Hence let us start with this example and try to extract some
lessons from it.

\bigskip

\subsection{Conifold transition I.}

A complex $3$-dimensional {\it conifold} is a threefold except at
some ordinary double points ({\it nodes}). Each of these nodes is
modelled on the germ of the quadric hypersurface $xy-uv=0$ in
${\Bbb C}^3$ at the origin. This is an $A_1$-singularity and can be
realized as
a real cone over $S^2\times S^3$. Such degeneration arises from
either process (${\rm C}_0$) or (${\rm C}_1$). Let
$\overline{Z}=\{p_0\}$ and $Y$ be the $7$-space associated to the
transition $({\rm C}_0{\rm C}_1^{-1})$ that deforms first the
complex structure of $X$ to shrink down an $S^3$ to get a conifold
$\overline{X}$ and then resolves the singularity by a small
resolution that blows up an $S^2$ transverse to the blown-down
$S^3$ to obtain a new non-singular $X^{\prime}$. Then $p_0$ is at
worst a cone-point in $Y$ with the link coming from the pasting of
$\nu_X(S^3)$ in $X$ and $\nu_{X^{\prime}}(S^2)$ in $X^{\prime}$
along their boundary. From [At], the local topology around the
node $p_0$ in $\overline{X}$ indicates that the former is
homeomorphic to $S^3\times D^3$, the latter homeomorphic to
$S^2\times D^4$, and the pasting homeomorphism $h$ from
$\partial\nu_X(S^3)$ to $\partial\nu_{X^{\prime}}(S^2)$ is given
by
$$
\begin{array}{cccccl}
 h & : & S^3\times S^2 & \longrightarrow & S^2\times S^3 &\\[1ex]
   &   & (a,b) & \longmapsto & (b,a)  &.
\end{array}
$$
This shows that indeed the link of $p_0$ is a $6$-sphere $S^6$
since
$$
 S^6\; = \;\partial D^7\; = \;\partial(D^4\times D^3)\;
       =\; S^3\times D^3\,\cup_{S^3\times S^2}\,D^4\times S^2\,.
$$
Furthermore, comparing with the setting of
handle-decomposition/Morse-theory of manifolds, one notices that   
the blown-down $S^3$ may be regarded as the attaching/descending
sphere while the blown-up $S^2$ as the belt/ascending sphere of a
$4$-handle $D^4\times D^3$ in $7$ dimensions (cf.\ [G-S], [Ki]).
In other words, $Y$ is simply $X\times[-\epsilon,0]$ attached
with a $4$-handle to the boundary $X\times\{0\}$.

Similarly the reverse process $({\rm C}_1{\rm C}_0^{-1})$ from
$X^{\prime}$ to $X$ corresponds to attaching a $3$-handle
$D^3\times D^4$ to the boundary $X^{\prime}\times\{0\}$ of
$X^{\prime}\times[0,\epsilon^{\prime}]$. In summary

\bigskip

\noindent
{\bf Lemma 2.1.1 [conifold I].} {\it 
 The conifold transition of type either $({\rm C}_0{\rm C}_1^{-1})$
 or $({\rm C}_1{\rm C}_0^{-1})$ for Calabi-Yau threefolds does not
 lead to non-manifold points on the resulting $7$-space $Y$.
} 

\bigskip

\noindent
{\it Remark 2.1.2.} As we shall see that this is likely the only
 case of extremal transitions that does lead to $7$-manifolds
 without any singularities. In view of the rich variations of the
 topologies of Calab-Yau threefolds, this already provides us
 with a rich class of $7$-manifolds that could play roles in
 M-theory.

\bigskip

\subsection{Conifold transition II: flops.}

On the other hand, the neighborhood of a node on a complex
$3$-conifold can also be realized as a complex cone over
${\Bbb C}{\rm P}^1\times{\Bbb C}{\rm P}^1$. One may deform the
K\"{a}hler structure of $X$ in such a way that first blows down a
${\Bbb C}{\rm P}^1$ to get a conifold $\overline{X}$ and then
blows up another ${\Bbb C}{\rm P}^1$ transverse to the previous
${\Bbb C}{\rm P}^1$. Such procedure is called a {\it flop} and
corresponds to a nontrivial process $({\rm C}_1{\rm C}_1^{-1})$.
Let $\overline{Z}=\{p_0\}$ and $Y$ the resulting $7$-space.
Then again $p_0$ is at worst a cone-point in $Y$. Its link
$\mbox{\rm Lk}\,(p_0)$ in $Y$ now comes from the pasting of
$\nu_X(S^2)$ and $\nu_{X^{\prime}}(S^2)$ along their boundary
$S^2\times S^3$. To see what this $6$-manifold is, we need to
recall some details of this flop from [At].

Let ${\Bbb L}_i$ be the tautological line bundle over
${\Bbb C}{\rm P}^i$ and $V$ be the quadric surface $\{xy-uv=0\}$.
The variety $V$ contains two transverse families of complex lines.
Each of the families is parametrized by ${\Bbb C}{\rm P}^1$ and
each gives a ruling for $V$. Let us denote the first family by
${\Bbb C}{\rm P}^1_{\alpha}$ and call the lines it parameterizes
by {\it $\alpha$-lines}. Similarly for ${\Bbb C}{\rm P}^1_{\beta}$
and {\it $\beta$-lines} ([G-H], [W-W]). This gives an isomorphism
between $V$ and
${\Bbb C}{\rm P}^1_{\alpha}\times{\Bbb C}{\rm P}^1_{\beta}$.
In terms of these, the line bundle ${\Bbb L}_3|_V$ is represented
by $(-1,-1)$ in the Picard group
$\mbox{\rm Pic}(V)=H_2(V;{\Bbb Z})={\Bbb Z}\oplus{\Bbb Z}$ with
the canonical basis given by the pair of homology classes
$([{\Bbb C}{\rm P}^1_{\alpha}]\,,\,[{\Bbb C}{\rm P}^1_{\beta}])$.
If one pinches each of the $\beta$-lines in the zero-section of
${\Bbb L}_3|_V$ to a point, then the result is a smooth complex
threefold $E_{\alpha}$ isomorphic to the total space of
${\Bbb L}_1\oplus{\Bbb L}_1$ over ${\Bbb C}{\rm P}^1_{\alpha}$.
On the other hand, if one pinches each of the $\alpha$-lines in
the zero-section of ${\Bbb L}_3|_V$ to a point, then the result is
a smooth complex threefold $E_{\beta}$ isomorphic to the total 
space of ${\Bbb L}_1\oplus{\Bbb L}_1$ over
${\Bbb C}{\rm P}^1_{\beta}$. Pinching further the zero-section of
$E_{\alpha}$ or the zero-section of $E_{\beta}$ is the same as
pinching the zero-section of ${\Bbb L}_3|_V$ altogether to a point.
The resulting space is exactly the complex cone $OV$ in
${\Bbb C}^3$ over $V$, which describes the neighborhood of the
node $p_0$ in $\overline{X}$. Tranforming from $E_{\alpha}$ to
$OV$ and then to $E_{\beta}$ is precisely the type-II conifold
transition described above.

From this picture, one has
$$
 \mbox{\rm Lk}\,(p_0)\; \cong\;
  \left.\rule[-.2ex]{0ex}{2ex}
  \partial \left(
   \nu_{{\subscriptsizeBbb L}_3|_V}(\mbox{\rm zero-section})\right)
  \times [0,1]\right/\hspace{-.2ex}\mbox{\raisebox{-.2ex}{$\sim$}}\,,
$$
where the relation $\sim$ is defined by first regarding
$\partial \left(
 \nu_{{\subscriptsizeBbb L}_3|_V}(\mbox{\rm zero-section})\right)$,
denoted by $M^5$ in the following discussion, naturally as an
$S^1$-bundle over $V$ and then performing the following pinchings:
\begin{quote}
 \hspace{-1.8em}(1)
 Pinch each of the $S^1$-fibers of $M^5\times\{0\}$ to a point,
 which transforms $M^5\times\{0\}$ to
 ${\Bbb C}{\rm P}^1_{\alpha}\times{\Bbb C}{\rm P}^1_{\beta}$; and
 then pinch ${\Bbb C}{\rm P}^1_{\beta}$. These together transform
 $M^5\times\{0\}$ to ${\Bbb C}{\rm P}^1_{\alpha}$.

 \hspace{-1.8em}(2)
 Pinch each of the $S^1$-fibers of $M^5\times\{1\}$ to a point,
 which transforms $M^5\times\{1\}$ to
 ${\Bbb C}{\rm P}^1_{\alpha}\times{\Bbb C}{\rm P}^1_{\beta}$; and
 then pinch ${\Bbb C}{\rm P}^1_{\alpha}$. These together transform
 $M^5\times\{1\}$ to ${\Bbb C}{\rm P}^1_{\beta}$.
\end{quote}

With these geometric pictures in mind, we can now show that
                
\bigskip

\noindent
{\bf Lemma 2.2.1 [conifold II].} {\it
 The conifold transition for Calabi-Yau threefolds via a flop leads
 to a non-manifold point $p_0$ on the resulting $7$-space $Y$.
 The link of $p_0$ in $Y$ is homeomorphic to ${\Bbb C}{\rm P}^3$.
} 

\bigskip

\noindent {\it Proof.}
Let $(z_1, z_2, z_3, z_4)$ be the coordinates for
${\Bbb C}^4={\Bbb C}^2\oplus{\Bbb C}^2$, $S^3$
(resp.\ ${S^3}^{\prime}$) be the unit sphere in the first
(resp.\ second) ${\Bbb C}^2$ factor of ${\Bbb C}^4$, and $S^3_t$
be the sphere of radius $t$ at the origin in ${\Bbb C}^2$. Consider
the Hopf fibration and the compatible foliation of $S^7$ with
generic leaf $S^3\times S^3$ from the realization of $S^7$ as the
join $S^3\ast{S^3}^{\prime}$ ([Ro], [Sp]):
$$
\begin{array}{cccccl}
 S^1 & \rightarrow & S^7=S^3\ast {S^3}^{\prime}
     & \hookleftarrow & S^3_{\sqrt{1-t^2}}\times S^3_t & \\[1ex]
 & & \downarrow &  & \downarrow & \\[1ex]
 & & {\Bbb C}{\rm P}^3 & \hookleftarrow & M^5_t &,
\end{array}
$$
where $M^5_t$ is by definition the image of
$S^3_{\sqrt{1-t^2}}\times S^3_t$ in ${\Bbb C}{\rm P}^3$ under
the Hopf map. Then the ${\Bbb T}^2$-action on $S^7$, given by
$$
 \tau_{\theta,\theta^{\prime}}\;:\;
   (z_1,\, z_2,\, z_3,\, z_4)\; \longmapsto\;
     (e^{i\theta}z_1,\, e^{i\theta}z_2,\,
           e^{i\theta^{\prime}}z_3,\, e^{i\theta^{\prime}}z_4)\,,
$$
leaves the Hopf fibration and leaves of the
$S^3\times S^3$-foliation invariant. The restriction of the action
to the diagonal subgroup $\Delta$ gives indeed the Hopf map. Since
this action is free, each $S^3_{\sqrt{1-t^2}}\times S^3_t$ is thus
realized as a ${\Bbb T}$-bundle over the quotient
${\Bbb C}{\rm P}^1\times{\Bbb C}{\rm P}^1$ with fiber the orbit.
Consequently, its image $M^5_t$ in ${\Bbb C}{\rm P}^3$ is the
total space of an $S^1$-bundle over
${\Bbb C}{\rm P}^1\times{\Bbb C}{\rm P}^1$. The symmetry of the
construction implies that this $S^1$-bundle must correspond to
a line bundle $(k,k)$, $k\in{\Bbb Z}$, in
$\mbox{\rm Pic}({\Bbb C}{\rm P}^1\times{\Bbb C}{\rm P}^1)$.

Let us consider the leaf, say,
$S^3_{\sqrt{1/2}}\times S^3_{\sqrt{1/2}}$.
Since the image in ${\Bbb C}{\rm P}^3$ of $S^3\cup{S^3}^{\prime}$
is a disjoint union
${\Bbb C}{\rm P}^1\cup{{\Bbb C}{\rm P}^1}^{\prime}$, by
construction the complement of $M^5_t$ in 
${\Bbb C}{\rm P}^3$ is the union
$\nu_{{\subscriptsizeBbb C}{\rm P}^3}({\Bbb C}{\rm P}^1)\cup
 \nu_{{\subscriptsizeBbb C}{\rm P}^3}({{\Bbb C}{\rm P}^1}^{\prime})$
with $M^5_t$ as the shared boundary. This implies that $M^5_t$ is
homeomorphic to $S^2\times S^3$ since
$\nu_{{\subscriptsizeBbb C}{\rm P}^3}({\Bbb C}{\rm P}^1)$ is
bundle-isomorphic to ${\Bbb L}_1\oplus{\Bbb L}_1$ by the adjunction
formula and the latter is trivial as a real vector bundle due its
vanishing second Stiefel-Whitney class. Now let $(k,k)_{S^1}$ be
the total space of the $S^1$-bundle associated to the complex line
bundle $(k,k)$ in
$\mbox{\rm Pic}({\Bbb C}{\rm P}^1\times{\Bbb C}{\rm P}^1)$. Then
in the smooth category ([Di])
\begin{itemize}
\item
$(0,0)_{S^1}\;\cong\; S^2\times S^2\times S^1$;

\item
$(-1,-1)_{S^1}\;\cong\;(1,1)_{S^1}\;\cong\; S^2\times S^3$; and

\item
For $k>1$, $(-k,-k)_{S^1}\;\cong\;(k,k)_{S^1}\;\cong\;
 (1,1)_{S^1}/\hspace{-.2ex}\mbox{\raisebox{-.4ex}{${\Bbb Z}_k$}}$,
whose fundamental group is ${\Bbb Z}_k$.
\end{itemize}
Consequently, $M^5_t$ is bundle-isomorphic to $(-1,-1)_{S^1}$.
In particuler, $M^5_{\sqrt{1/2}}$ is homeomorphic to
$\partial \left(
 \nu_{{\subscriptsizeBbb L}_3|_V}(\mbox{\rm zero-section})\right)$.

Starting with $M^5_{\sqrt{1/2}}$, if we let $t\searrow 0$,
then the whole space shrinks to ${\Bbb C}{\rm P}^1$ in exactly the
same way as $M^5\times\{0\}$ is pinched to
${\Bbb C}{\rm P}^1_{\alpha}$. Similarly, when $t\nearrow 1$,
$M^5_{\sqrt{1/2}}$ to ${{\Bbb C}{\rm P}^1}^{\prime}$ is the
same as $M^5\times\{1\}$ to ${\Bbb C}{\rm P}^1_{\beta}$.
Consequently, $\mbox{\rm Lk}(p_0)$ as described is indeed a
${\Bbb C}{\rm P}^3$. This concludes the proof.

\noindent
\hspace{14cm} $\Box$  

\bigskip

\noindent
{\it Remark 2.2.2.}
 Notice also that the trivial $({\rm C}_1{\rm C}_1^{-1})$ process
 that blows down an $S^2$ and blow the same one up again leads to
 a cone-point in $Y$ with link $S^2\times S^4$. Such fake
 transitions can be removed by perturbing the corresponding path
 in the Calabi-Yau moduli space a little bit. 

\bigskip

\noindent
{\it Remark 2.2.3 [Conifold transition, real and complex join].}
 Given two complex manifolds $M_1$ and $M_2$, and a complex
 line bundle $L$ over $M_1\times M_2$. One can do the one-point
 compactification for each fiber to get a
 $\widehat{\Bbb C}$-bundle $\hat{L}$ over $M_1\times M_2$, where
 $\widehat{\Bbb C}={\Bbb C}\cup\{\infty\}$. $\hat{L}$ has two
 distinguished sections: the zero-section $\sigma_0$ and the
 infinity-section $\sigma_{\infty}$. Both sections are naturally
 isomorphic to $M_1\times M_2$. If one pinches $\sigma_0$ to $M_1$
 and $\sigma_{\infty}$ to $M_2$, then one get a new space
 $M_1\ast_{\subscriptsizeBbb C}M_2$, a {\it complex join} of
 $M_1$ and $M_2$. This generalizes the construction for the
 real join $\ast$. In terms of these, one obtains a unified
 picture for $\mbox{\rm Lk}(p_0)$ in conifold transitions:
 (1) For Type-I, an $S^2$ and an $S^3$ are transmuted to each
 other; and $\mbox{\rm Lk}(p_0)$ is the join $S^2\ast S^3=S^6$.
 (2) For Type-II, a transverse pair of ${\Bbb C}{\rm P}^1$'s are
 transmuted to each other; and $\mbox{\rm Lk}(p_0)$ is a complex
 join ${\Bbb C}{\rm P}^1\ast_{\subscriptsizeBbb C}
                            {\Bbb C}{\rm P}^1={\Bbb C}{\rm P}^3$.

\bigskip
 
\section{The type II primitive K\"{a}hler deformations.}

Recall the topological matching condition in Sec.\ 1.2, which
in the current case states that,
if both $(X,Z)$ and $(X^{\prime},Z^{\prime})$ degenerate to a same
$(\overline{Z},p_0)$, then the two $5$-manifolds
$\partial\nu_X(Z)$ and $\partial\nu_X^{\prime}(Z^{\prime})$
have to be homeomorphic. One can use this to rule out many
impossible transitions and only focuses on those transitions
that are topologically admissibe in order to make deeper and
more effective study. For this reason and due to Wilson's
classification of the K\"{a}hler deformation associated to
the wall of codimension $1$ of the K\"{a}ler cone of a
Calabi-Yau threefold ([Wil2]), in this section we shall spend our
efforts in understanding the topology of $\partial\nu_X(Z)$
for $(X,Z)$, where $Z$ a del Pezzo surface embedded in $X$. 

\bigskip

\subsection{The case of smooth del Pezzo surfaces.}

A {\it del Pezzo surface} $Z$ is characterized by the feature
that its anticanonical bundle is ample ([De], [Wa3]). Such
surfaces, when smooth, can only be the exceptional one
$Z_e={\Bbb C}{\rm P}^1\times{\Bbb C}{\rm P}^1$ or the blow-up
$Z_r$ of ${\Bbb C}{\rm P}^2$ at $r$, $0\le r\le 8$, distinct
points in general positions (i.e.\ no three of them are
collinear, no six of them lie in a conic, and no cubic through
seven of them with a double point at the eighth). Such a
realization provides a standard basis
$(\varepsilon_0,\, \varepsilon_1\, \ldots,\, \varepsilon_r)$ for
$H_2(Z_r;{\Bbb Z})=\mbox{\rm Pic}\,(Z_r)$, where $\varepsilon_0$
is the canonical generator $[{\Bbb C}{\rm P}^1]$ of
$H_2({\Bbb C}{\rm P}^2;{\Bbb Z})$ lifted to $Z_r$ and
$\varepsilon_i$, which is also a ${\Bbb C}{\rm P}^1$ with
self-intersection number $(-1)$, are from the exceptional
divisors of the blow-up. With respect to
this basis, the intersection form $Q|_{Z_r}$ of $Z_r$ is given by
$$
 Q|_{Z_r}\; =\; \langle 1\rangle\oplus r\langle -1\rangle\,;
$$
and the canonical divisor of $Z_r$ is known to be 
$$
 K_{Z_r}\; =\; -3\, \varepsilon_0\, +\, \varepsilon_1\,
     +\, \cdots\, +\, \varepsilon_r\,.
$$
In the following discussion, we shall not distinguish these
classes with either the embedded ${\Bbb C}{\rm P}^1$ (or their
union) in $Z_r$ or the complex line bundle (and its real
${\Bbb R}^2$- or $D^2$-bundle) over $Z_r$ they correspond.

When $Z$ embeds in a Calabi-Yau threefold $X$, since   
$K_Z=K_X|_Z\otimes\nu_X(Z)$ and $K_X$ is trivial for Calabi-Yau
manifolds, the tubular neighborhood $\nu_X(Z)$ of $Z$ in $X$ has
to be isomorphic to $K_Z$, which restricts the possible
embeddings. For $Z=Z_r$, the structure of its canonical divisor
implies that $\nu_X(Z_r)$ is indeed a {\it fiber sum} of
${\Bbb L}_2^{\otimes 3}$ and $r$ many ${\Bbb L}_2^{-1}$'s,
where recall that ${\Bbb L}_2$ is the tautological line bundle
over ${\Bbb C}{\rm P}^2$. Notice that the total space of the
$S^1$-bundle associated to ${\Bbb L}_2^{-1}$ is $S^5$ and the
total space of the $S^1$-bundle associated to
${\Bbb L}_2^{\otimes 3}$ is the lens space
$S^5/\hspace{-.2ex}\mbox{\raisebox{-.4ex}{${\Bbb Z}_3$}}$, where
${\Bbb Z}^3$ is the cyclic subgroup of order three contained
in $U(1)$ that gives the Hopf fibration of $S^5$. Hence the
boundary $5$-manifold $\partial\nu_X(Z_r)$ is the fiber sum
$(S^5/\hspace{-.2ex}\mbox{\raisebox{-.4ex}{${\Bbb Z}_3$}})\,
 \sharp_f\, r S^5$
over ${\Bbb C}{\rm P}^2\,\sharp\,r\,\overline{{\Bbb C}{\rm P}^2}$.
With the fact that $S^5=S^3\ast S^1$, one has the following two
collections of $S^1$-bundles, which contain all the spaces
involved in doing the fiber sum:
$$
\begin{array}{ccccccc}
 S^1 & \rightarrow & S^5-\nu_{S^5}(S^1)  & \cong  & S^3\times D^2
   & \stackrel{\partial}{\hookleftarrow} & S^3\times S^1 \\[.2ex]
 & & \downarrow & & & & \downarrow \\
 & & {\Bbb C}{\rm P}^2-D^4  & =
   & \nu_{{\subscriptsizeBbb C}{\rm P}^2}({\Bbb C}{\rm P}^1)
   & \stackrel{\partial}{\hookleftarrow} & S^3
\end{array}
$$
and
$$
\begin{array}{cccccccccl}
 S^1 & \rightarrow & S^5-\nu_{S^5}(S^3) & =
   & \nu_{S^5}(S^1)  & = & D^4\times S^1
   & \stackrel{\partial}{\hookleftarrow} & S^3\times S^1 & \\[.2ex]
 & & \downarrow & & & & & & \downarrow & \\
 & & {\Bbb C}{\rm P}^2
     -\nu_{{\subscriptsizeBbb C}{\rm P}^2}({\Bbb C}{\rm P}^1)
   & & & = & D^4
   & \stackrel{\partial}{\hookleftarrow} & S^3 &.
\end{array}
$$
Comparing with the decomposition
$$
 S^5\;=\;S^3\times D^2\,\cup_{\,S^3\times S^1}\,D^4\times S^1\,,
$$
one recognizes that the fiber sum with $r$-many $S^5$'s is indeed
the surgery on $r$ distinct $S^1$-fibers of
$S^5/\hspace{-.2ex}\mbox{\raisebox{-.4ex}{${\Bbb Z}_3$}}$ with
respect to the local trivialization of their tubular neighborhood
from the bundle structure. Together with the homotopy sequence
$$
\begin{array}{cccccl}
 \pi_1(S^1) & \rightarrow & \pi_1(\partial\nu_X(Z_r)) & \rightarrow
  & \pi_1({\Bbb C}{\rm P}^2\,\sharp\,r\,\overline{{\Bbb C}{\rm P}}^2)
  & \\[.2ex]
 \|  & & & & \| & \\[.2ex]
 {\Bbb Z} & & & & 0 &,
\end{array}
$$
it also follows that, for $r\ne 0$, $\partial\nu_X(Z_r)$ is
simply-connected since the only possible generator for
$\pi_1(\partial\nu_X(Z_r))$ comes from an $S^1$-fiber of
$\partial\nu_X(Z_r)$, which can be homotoped into the fiber
over, say, $\nu_{Z_r}(\varepsilon_1)$ and then homotoped to a
point.

\bigskip

\noindent
{\bf Lemma 3.1.1.} {\it For $r\ne 0$, $\partial\nu_X(Z_r)$ is
 homeomorphic to the connected sum $\sharp\,r\, S^2\times S^3$.}

\bigskip

\noindent
{\it Proof.} Since $\pi_1(\partial\nu_X(Z_r))=0$,
$H_2(\partial\nu_X(Z_r);{\Bbb Z})
 =\pi_2(\partial\nu_X(Z_r))={\Bbb Z}^r$
by the homotopy sequence. On the other hand, since $c_1(X)=0$ and
the normal bundle of $\partial\nu_X(Z_r)$ in $X$ is trivial,
$\partial\nu_X(Z_r)$ is a spin $5$-manifold. The lemma then
follows from the classification of closed simply-connected spin
$5$-manifolds by Smale [Sm2] (cf.\ Sec.\ 1.1).

\noindent
\hspace{14cm} $\Box$  

\noindent
{\it Remark 3.1.2.} For $r=0$,
$\partial\nu_X({\Bbb C}{\rm P}^2)=
 S^5/\hspace{-.2ex}\mbox{\raisebox{-.4ex}{${\Bbb Z}_3$}}$
since $K_{{\subscriptsizeBbb C}{\rm P}^2}={\cal O}(-3)
        ={\Bbb L}|_{{\subscriptsizeBbb C}{\rm P}^2}^{\otimes 3}$.
For the exceptional del Pezzo,
$\partial\nu_X({\Bbb C}{\rm P}^1\times{\Bbb C}{\rm P}^1)=
 (S^2\times S^3)/\hspace{-.2ex}\mbox{\raisebox{-.4ex}{${\Bbb Z}_2$}}$
(cf.\ proof of Lemma 2.2.1).

\bigskip

\noindent
{\it Remark 3.1.3.}
Notice that, for $r\ne 0$, $H_2(\partial\nu_X(Z_r);{\Bbb Z})$ is
isomorphic to the kernel of the homomorphism from 
$\pi_2({\Bbb C}{\rm P}^2\,\sharp\,r\,\overline{{\Bbb C}{\rm P}}^2)$
to $\pi_1(S^1)$ in the homotopy sequence. Consequently,
$H_2(\partial\nu_X(Z_r);{\Bbb Z})\;
 =\; \{\:\xi\in H_2(Z_r;{\Bbb Z})\:|\:K_{Z_r}\cdot\,\xi=0\:\}\;
 =\; K_{Z_r}^{\perp}$ with respect to $Q_{Z_r}$.

\bigskip

\subsection{The case of a rational singular del Pezzo surface.}

We shall start with a review of some necessary facts on a certain
rational singular del Pezzo surfaces $Z$ and then turn to show
that for such $Z$, $\partial\nu_X(Z)$ is simply-connected when it
is embedded in a Calabi-Yau threefold $X$. Since it is also spin,
we then compute $H_2(\partial\nu_X(Z);{\Bbb Z})$, which, by
Smale's classification theorem [Sm2] in Fact 1.1.1, completely
determines the topology of $\partial\nu_X(Z)$. For the clarity
of notation, in many places, we shall denote $\partial\nu_X(Z)$
by $M^5(Z,X)$.

\bigskip

\begin{flushleft}
{\bf The setup and the simply-connectedness of
 $\partial\nu_X(Z)(=M^5(Z,X))$.}
\end{flushleft}
Let $Z$ be a Gorenstein rational singular del Pezzo surface with
$\mbox{\rm Pic}(Z)={\Bbb Z}$. For such $Z$, it is known
([Fu], [K-MK], [M-Z]) that $K_Z^2=9-r$, $r\ge 3$, and that
the minimal resolution of $Z$ is a smooth del Pezzo surface
of topology $Z_r$. The types and combinations of the singularities
that can appear together on $Z$ are very restricted; the details
are given in the following list\footnotemark:
\begin{quote}
 \hspace{-1.9em}
 \parbox[t]{3em}{$r=3$} \hspace{1em}
 \parbox[t]{12cm}{$A_1+A_2$}

 \hspace{-1.9em}
 \parbox[t]{3em}{$r=4$} \hspace{1em}
 \parbox[t]{12cm}{$A_4$}

 \hspace{-1.9em}
 \parbox[t]{3em}{$r=5$} \hspace{1em}
 \parbox[t]{12cm}{$2A_1+A_3$, $\;D_5$}

 \hspace{-1.9em}
 \parbox[t]{3em}{$r=6$} \hspace{1em}
 \parbox[t]{12cm}{$A_1+A_5$, $\;3A_2$, $\;E_6$}

 \hspace{-1.9em}
 \parbox[t]{3em}{$r=7$} \hspace{1em}
 \parbox[t]{12cm}{$A_1+2A_3$, $\;A_2+A_5$, $\;A_7$,
                  $\;3A_1+D_4$, $\;A_1+D_6$, $\;E_7$ }

 \hspace{-1.9em}
 \parbox[t]{3em}{$r=8$} \hspace{1em}
 \parbox[t]{12cm}{$2A_1+2A_3$, $\;A_1+A_2+A_5$, $\;A_1+A_7$,
                  $\;4A_2$, $\;2A_4$, $\;A_8$, $\;2A_1+D_6$,
                  $\;A_3+D_5$, $\;A_1+E_7$, $\;A_2+E_6$, $\;2D_4$,
                  $\;D_8$, $\;E_8$}
\end{quote}
\footnotetext{Readers may notice that the complete list given in
 [Fur] and [M-Z] differ slightly:
 Assuming that the $(r=8,A_2+E_7)$ in [Fur] is a typo for
 $(r=8,A_2+E_6)$, [Fur] has $(r=7,A_3+D_4)$ that does
 not appear in [M-Z] while [M-Z] has $(r=1, A_1)$ and $(3A_1+D_4)$
 that do not appear in [Fur]. On the other hand, as will be
 clear below, there is an abelian group naturally associated
 to a singular del Pezzo surface, independent of whether it is
 embeddable in a smooth Calabi-Yau threefold or not, whose order is
 an integer square (cf.\ Proposition 3.2.8 and Corollary 3.2.9).
 Following the discussion below, this order is $8$ for
 $(r=7,A_3+D_4)$ and is $16$ for $(r=7,3A_1+D_4)$. For this reason,
 we favor [M-Z] with $(r=1, A_1)$ deleted. If there is going to be
 any change of the list we adapt here, it may influence the lists
 or tables based on it but not the abstract discussions below.
 I would like to thank Prof.\ Se\'{a}n Keel for a discussion of
 this.
} 
where, for example, $2A_1+A_3$ means that $Z$ has three isolated
surface singularities: two of type $A_1$ and one of type $A_3$.
Let $E$ be the exceptional divisor in $Z_r$ associated to the
resolution and $E_i$ be the irreducible components of $E$.
Then each $E_i$ is a $(-2)$ rational curve in $Z_r$ and their 
intersection diagram is the Dynkin diagram as indicated by
the label $A_j$, $D_j$, or $E_j$. The tubular neighborhood
$\nu_{Z_r}(E)$ of $E$ in $Z_r$ is thus a plumbing of a collection
of the tangent $D^2$-bundle of $S^2$ with the same Dynkin diagram
as the plumbing diagram. For $Z$ Gorenstein, the intersection
number of these exceptional divisors in $Z_r$ with the canonical
divisor $K_{Z_r}$ are all zero. Since our discussion is in
topological category, by tubing if necessary, one can represent
the homology class of the divisor $K_{Z_r}$ by a surface in $Z_r$
that is truely disjoint from $E$. For the following discussions,
$K_{Z_r}$ will always be represented by such a surface when
needed. To save notations, we also use $K_{Z_r}$ for the canonical
line bundle of $Z_r$, which should be clear from the context.

When $Z$ embeds in a smooth Calabi-Yau threefolds $X$ as a
divisor, its tubular neighborhood $\nu_X(Z)$ is still
well-defined - though no longer a regular bundle over $Z$ -,
for example, by the union of all small enough balls with center
at $Z$ with respect to some Riemannian metric on $X$.
The complement $\nu_X(Z)-Z$ remains having the product structure
$\partial\nu_X(Z)\times(0,\epsilon)$. Though, due to the
singularity of $Z$, the canonical line bundle $K_Z$ of $Z$ is no
longer homeomorphic to the entire $\nu_X(Z)$, one may notice that
the proof of the adjunction formula in [G-H] applies without
difficulty to $Z-\nu_Z(\mbox{\rm Sing}(Z))$ in
$X-\nu_X(\mbox{\rm Sing}(Z))$ (cf.\ [Di]). Thus one can conclude
that $\nu_{X-\nu_X({\rm Sing}(Z))}(Z-\nu_Z(\mbox{\rm Sing}(Z)))$
remains the same as the restriction $K_Z|_{Z-\nu_Z({\rm Sing}(Z))}$.
Furthermore, since $Z$ is Gorenstein and hence the restriction of
$K_Z$ on $\nu_Z(\mbox{\rm Sing}(Z))$ is trivial, the restriction
$K_Z|_{Z-\nu_Z({\rm Sing}(Z))}$ is bundle isomorphic to 
$K_{Z_r}|_{Z_r-\nu_{Z_r}(E)}$. On the other hand, since $Z$ is a
divisor in the smooth $X$ and hence is locally principal, in a
small neighborhood ${\cal U}$ of
$q\in\mbox{\rm Sing}(Z)$ in $X$, $Z$ is the same as a germ
$\psi$ of the corresponding isolated analytic hypersurface
singularity in ${\Bbb C}^3$ at the origin, described by some
polynomial equation $\psi=0$.
Let $\Delta$ be a small enough disk around the origin in
${\Bbb C}$. Then, as a function on ${\cal U}$,
$$
 \psi\;:\; \psi^{-1}(\Delta)\subset{\cal U} \;
          \longrightarrow\; \Delta
$$
gives a fibration of $\nu_X(q)$ over $\Delta$, which is the
same as the Milnor fibration associated to the singularity $q$.
Furthermore, since $\psi=0$ describes $\nu_{Z}(q)$ in $X$, one
can identify this $\Delta$ as the fiber $D^2$ of the bundle 
$\nu_{X-\nu_X({\rm Sing}(Z))}(Z-\nu_Z(\mbox{\rm Sing}(Z)))$ over
$X-\nu_X({\rm Sing}(Z))$ restricted to
$\partial(X-\nu_X({\rm Sing}(Z)))$.
(Indeed these $\psi$ together give a trivialization of the bundle
restriced to a neighborhood of 
$\partial(X-\nu_X({\rm Sing}(Z)))=\partial\nu_X({\rm Sing}(Z))$.)
Consequently, $\nu_X(Z)$ is obtained from $K_{Z_r}$ by removing
its restriction to $\nu_{Z_r}(E)$ and then pasting in a collection
of the total space of the Milnor fibration associated to the 
isolated hypersurface singularities $\mbox{\rm Sing}(Z)$ in $X$
(cf.\ {\sc Figure 3-1}).
To understand the boundary of $\nu_X(Z)$, we
need to rephrase this picture in terms of associated circle
bundles.

Let $\Xi^5$ be the associated circle bundle of $K_{Z_r}$ over
$Z_r$. For the clarity of notation, let us denote
$\partial\nu_X(Z)$ by $M^5(Z,X)$, $\Xi^5|_{Z_r-\nu_{Z_r}(E)}$ by
$W^5_{\mbox{\scriptsize\rm reg}}(Z,X)$, and the complement of
$W^5_{\mbox{\scriptsize\rm reg}}(Z,X)$ in $M^5(Z,X)$ by
$W^5_{\mbox{\scriptsize\rm Sing}}(Z,X)$.
Then, following the previous discussions and notations, each
component of $W^5_{\mbox{\scriptsize\rm Sing}}(Z,X)$ corresponds
to a singularity of $Z$ and is homeomorphic to the total space of
the Milnor fibration in the slightly modified spherical form:
$$
 \varphi\;:\;S^5-\nu_{S^5}(\{\psi=0\})\;\longrightarrow\; S^1\,,
  \hspace{2em} \varphi\;=\;\psi/|\psi|\,.
$$
It is known (cf.\ Sec.\ 1.1) that the intersection of
$\{\psi=0\}$ with $S^5$ is homeomorphic to a quotient
$S^3\hspace{-.4ex}/\hspace{-.2ex}\mbox{\raisebox{-.4ex}{$G$}}$,
where $G$ is a finite subgroup in $\mbox{\it SU}(2)$ associated
to the singularity in $Z$. Hence the boundary of
$S^5-\nu_{S^5}(\{\psi=0\})$ is homeomorphic to
$S^3\hspace{-.4ex}/\hspace{-.2ex}\mbox{\raisebox{-.4ex}{$G$}}
 \times S^1$.

Now let
$[S^3\hspace{-.4ex}/\hspace{-.2ex}\mbox{\raisebox{-.4ex}{$G$}},
  S^1]$ be the set of homotopy classes of maps from
$S^3\hspace{-.4ex}/\hspace{-.2ex}\mbox{\raisebox{-.4ex}{$G$}}$ to 
$S^1$. Then observe that
$$
 [S^3\hspace{-.4ex}/\hspace{-.2ex}\mbox{\raisebox{-.4ex}{$G$}},
   S^1]
 =H^1(S^3\hspace{-.4ex}/\hspace{-.2ex}\mbox{\raisebox{-.4ex}{$G$}}
      ;{\Bbb Z})\;
 =\; \mbox{\it Hom}(H_1(
   S^3\hspace{-.4ex}/\hspace{-.2ex}\mbox{\raisebox{-.4ex}{$G$}};
     {\Bbb Z}),{\Bbb Z})
    \oplus \mbox{\rm Ext}(H_0(
      S^3\hspace{-.4ex}/\hspace{-.2ex}\mbox{\raisebox{-.4ex}{$G$}},
       {\Bbb Z}),{\Bbb Z})\;
 =\; 0
$$
since
$H_1(S^3\hspace{-.4ex}/\hspace{-.2ex}\mbox{\raisebox{-.4ex}{$G$}};
  {\Bbb Z})$
is torsion and
$H_0(S^3\hspace{-.4ex}/\hspace{-.2ex}\mbox{\raisebox{-.4ex}{$G$}};
  {\Bbb Z})$
is free. Thus all maps from
$S^3\hspace{-.4ex}/\hspace{-.2ex}\mbox{\raisebox{-.4ex}{$G$}}$ to 
$S^1$ are homotopic to each other. Consequently, no matter how we
trivialize $W^5_{\mbox{\scriptsize\rm reg}}(Z,X)$ over
$Z_r-\nu_{Z_r}(K_{Z_r})$, we can always homotope the
trivialization so that over $\partial(Z_r-\nu_{Z_r}(E))$
it coincides with the product structure
$S^3\hspace{-.4ex}/\hspace{-.2ex}\mbox{\raisebox{-.4ex}{$G$}}
  \times S^1$ from the Milnor fibration of each component of
$\partial W^5_{\mbox{\scriptsize\rm Sing}}(Z,X)$
({\sc Figure 3-1}).
\begin{figure}[htbp]
\setcaption{{\sc Figure 3-1.}
\baselineskip 14pt 
 The tubular neighborhood $\nu_X(Z)$ of a singular del Pezzo
 surface $Z$ embedded in a smooth Calabi-Yau threefold $X$ is
 indicated. The $W^5_{\mbox{\scriptsize\rm Sing}}(Z,X)$ part 
 of the boundary $\partial\nu_X(Z)$ is shaded. The way it is
 realized as the total space of the Milnor fibration is shown
 on the right. While $W^5_{\mbox{\scriptsize\rm reg}}(Z,X)$ is
 an $S^1$-bundle over the open complex surface
 $Z-\nu_Z(\mbox{\rm Sing}(Z))$, each connected component of 
 $W^5_{\mbox{\scriptsize\rm Sing}}(Z,X)$ is an open surface
 bundle over $S^1$ with fiber $F$ homeomorphic to some connected
 component of $\nu_{Z_r}(E)$.
 Up to homotopy, both give the same product structure
 $S^3\hspace{-.4ex}/\hspace{-.2ex}\mbox{\raisebox{-.4ex}{$G$}}
   \times S^1$ for the components of their shared boundary
 $4$-manifold. Living in the Milnor fiber $F$ is the exceptional
 divisor $E$ from the resolution of the singularity of $Z$.
 Its irreducible components $E_i$ are indicated by thick line
 segments due to the suppression of dimensions. Shown in the
 figure is a $D_6$-chain. The monodromy homeomorphism $\tau$ and
 the Milnor fibration map $\varphi$ are also indicated.
 If $\tau$ is replaced by the identity map, then the topology of
 $\partial\nu_X(Z)$ is changed and becomes $\Xi^5$, the
 $S^1$-bundle associated to the canonical line bundle $K_{Z_r}$
 of the minimal resolution $Z_r$ of $Z$.
} 
\centerline{\psfig{figure=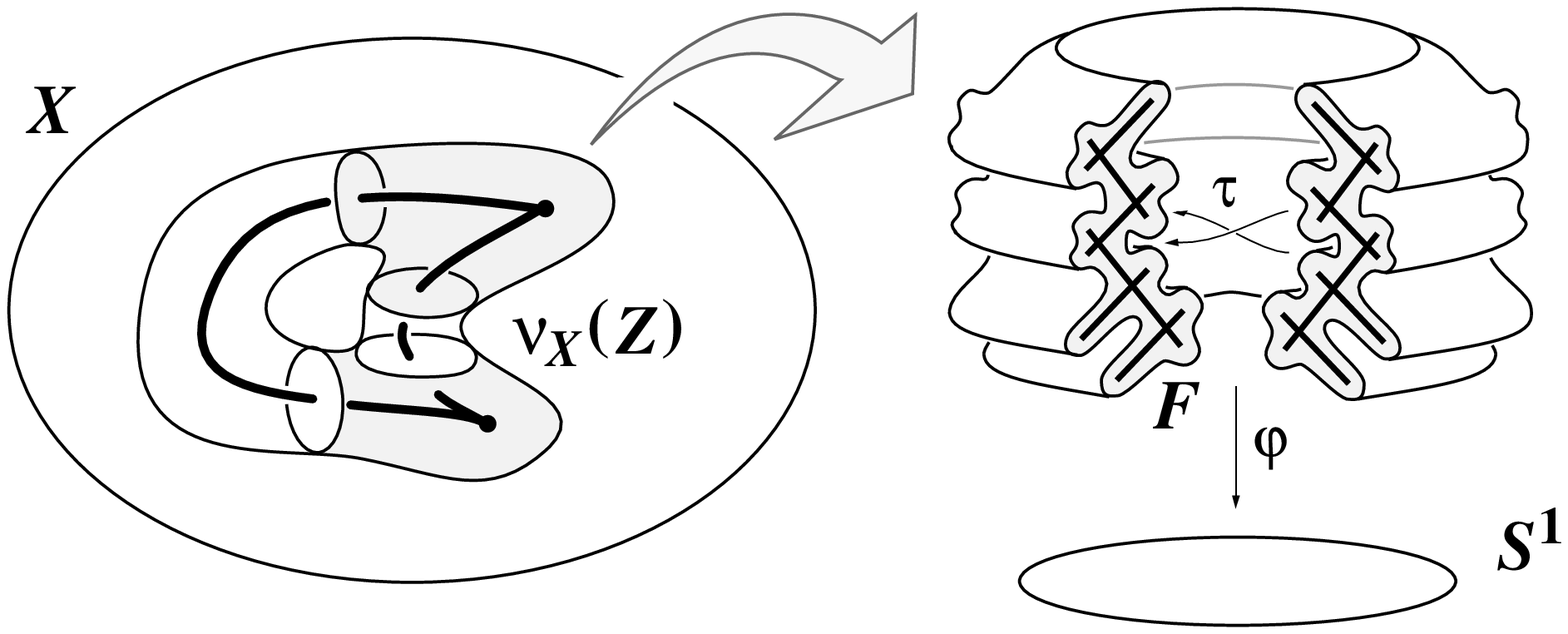,width=13cm,caption=}}
\end{figure}
Consequently, one has the following lemma:

\bigskip

\noindent
{\bf Lemma 3.2.1.} {\it
$M^5(Z,X)$ can be obtained from $\Xi^5$ by the
following cut-and-paste: First cut $\Xi^5$ along a lifting
$\widehat{\nu_{Z_r}(E)}$ of $\nu_{Z_r}(E)$ into $\Xi^5$. The
resulting manifold has a boundary which is the double
$D\widehat{\nu_{Z_r}(E)}$ of $\widehat{\nu_{Z_r}(E)}$. Next close
up this boundary by the orientation-reversing automorphism of
$D\widehat{\nu_{Z_r}(E)}$ induced by the monodromy homeomorphism
$\tau$ on $\nu_{Z_r}(E)$ to itself 
} ({\sc Figure 3-2}). 
\begin{figure}[htbp]
\setcaption{{\sc Figure 3-2.}
\baselineskip 14pt
 How $M^5(Z,X)$ is obtained from $\Xi^5$.
}   
\centerline{\psfig{figure=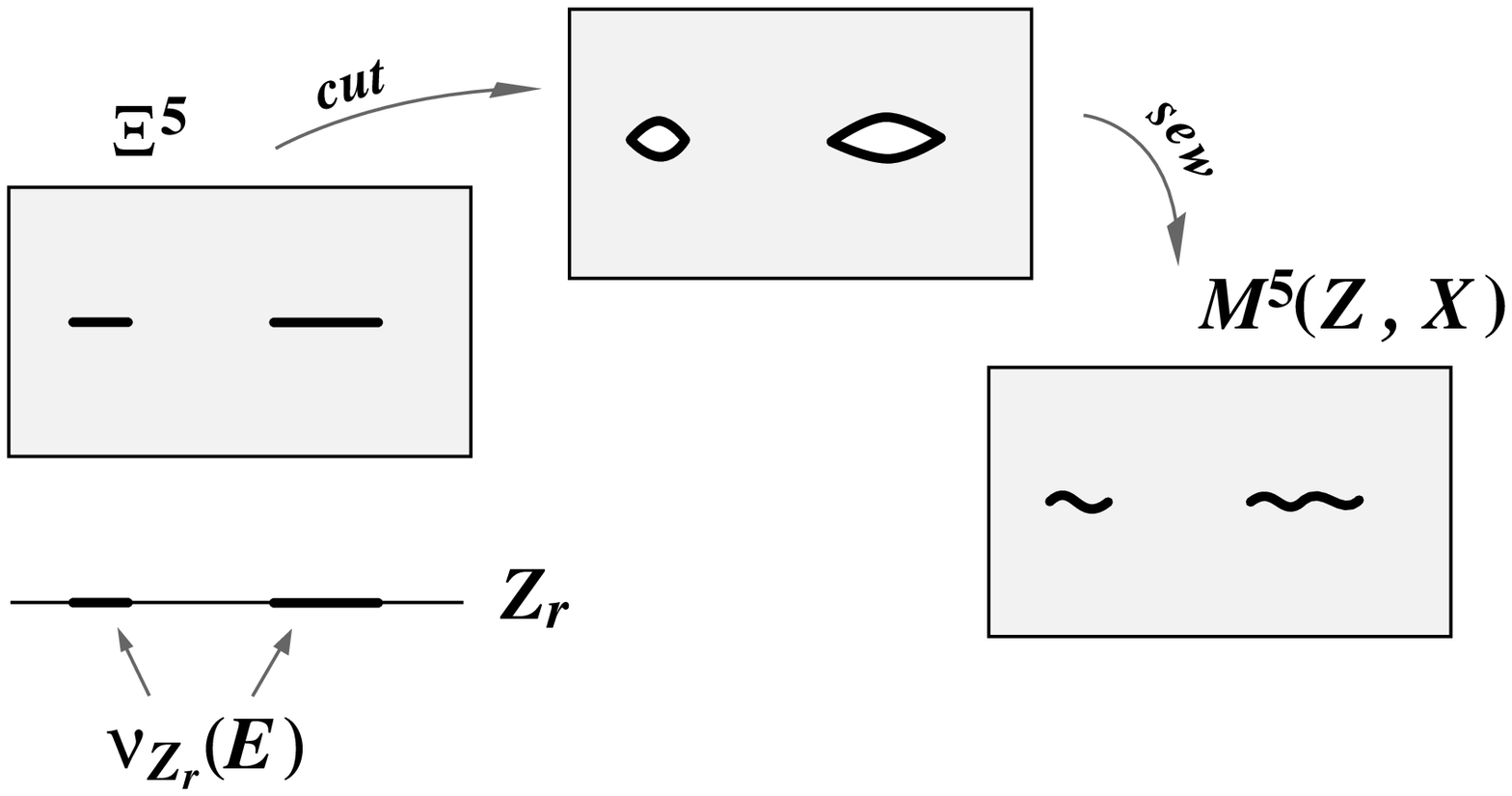,width=13cm,caption=}}
\end{figure}

\bigskip

\noindent
{\bf Corollary 3.2.2.} {\it $M^5(Z,X)$ is simply-connected.} 

\bigskip

\noindent
{\it Proof.}
Let $\widehat{E}$ be the lifting of $E$ into
$\widehat{\nu_{Z_r}(E)}$. Notice that $\widehat{E}$ is a
$2$-cycle in $M^5(Z,X)$ that consists of a collection of
embedded $S^2$ that intersect at worst transversely at some
isolated points. Thus, given a loop $\gamma$ in $M^5(Z,X)$, up to
homotopy we may assume that $\gamma$ is disjoint from
$\widehat{E}$ and hence from $\widehat{\nu_{Z_r}(E)}$. Lemma 3.2.1
implies that we may now regard $\gamma$ as a loop in $\Xi^5$.
Since $\Xi^5$ is simply-connected from the discussion in
Sec.\ 3.1, $\gamma$ bounds an embedded $2$-disk $D^2$ in $\Xi^5$.
By dimension count, we may assume that $D^2$ is disjoint from
$\widehat{E}$ and hence from $\widehat{\nu_{Z_r}(E)}$ by a
further homotopy if necessary. This shows that $\gamma$ in
$M^5(Z,X)$ indeed bounds a $2$-disk and hence concludes the
proof.

\noindent
\hspace{14cm} $\Box$  

\bigskip

Furthermore, since it bounds the spin $6$-manifold $\nu_X(Z)$,
$M^5(Z,X)$ is also spin. By Fact 1.1.1 in Sec.\ 1.1, its topology
is determined by the group structure of its second integral
homology $H_2(M^5(Z,X);{\Bbb Z})$, which we shall now turn to.

\bigskip

\begin{flushleft}
{\bf The integral homology $H_2(M^5(Z,X);{\Bbb Z})$.}
\end{flushleft}
The fact that $M^5(Z,X)$ is obtained from
$\Xi^5-\widehat{\nu_{Z_r}(E)}$ by closing up the boundary using
the monodromy homeomorphism $\tau$ suggests that one may relate
$H_2(M^5(Z,X);{\Bbb Z})$ to
$H_2(\Xi^5-\widehat{\nu_{Z_r}(E)};{\Bbb Z})$ and the monodromy
operator $T$. We shall now work out the detail of this relation.

First notice that, if one denotes the lifting of $E$ to
$\widehat{\nu_{Z_r}(E)}$ by $\widehat{E}$, then, from the fact
that
$$
\begin{array}{c}
 \nu_{\Xi^5}(\widehat{E})-\widehat{\nu_{Z_r}(E)}\;
   \cong\; \widehat{\nu_{Z_r}(E)}\,
                 \times\,((-\epsilon,0)\cup(0,\epsilon))\;
   \cong\; \nu_{M^5(Z,X)}(\widehat{E})
                  -\widehat{\nu_{Z_r}(E)},\\[1ex]
 \nu_{\Xi^5}(\widehat{E})-\widehat{E}\;
   \cong\; D\widehat{\nu_{Z_r}(E)}\,\times\,(0,\epsilon)\;
   \cong\; M^5(Z,X)-\widehat{E},
\end{array}
$$
and that the boundary $D\widehat{\nu_{Z_r}(E)}$ of
$\Xi^5(Z,X)-\widehat{\nu_{Z_r}(E)}$ has a collar, one has 
$$
\begin{array}{ccccc}
 \Xi^5-\widehat{\nu_{Z_r}(E)} 
   & \cong   & \Xi^5-\nu_{\Xi^5}(\widehat{E})
   & \cong   & \Xi^5-\widehat{E}     \\[.2ex]
 \| & & & & \\[.2ex]
 M^5(Z,X)-\widehat{\nu_{Z_r}(E)}
   & \cong   & M^5(Z,X)-\nu_{M^5(Z,X)}(\widehat{E})
   & \cong   & M^5(Z,X)-\widehat{E}\,.
\end{array}
$$
Thus by the Lefschetz duality one obtains
$$
 H_2(\Xi^5-\widehat{\nu_{Z_r}(E)};{\Bbb Z})\;
 =\; H_2(\Xi^5-\widehat{E};{\Bbb Z})\;
 =\; H^3(\Xi^5,\widehat{E};{\Bbb Z})\,. 
$$
Now recall the exact sequence associated to the pair
$(\Xi^5,\widehat{E})$:
$$ {\small
\begin{array}{cccccccccccccl}
 \cdots & \rightarrow
  & H^2(\Xi^5;{\smallBbb Z})  & \rightarrow
  & H^2(\widehat{E};{\smallBbb Z})
  & \stackrel{\delta^{\ast}}{\rightarrow}
  & H^3(\Xi^5,\widehat{E};{\smallBbb Z})  & \rightarrow
  & H^3(\Xi^5;{\smallBbb Z})  & \rightarrow
  & H^3(\widehat{E};{\smallBbb Z})
  & \stackrel{\delta^{\ast}}{\rightarrow}  & \cdots &   \\[.2ex]
 & & \hspace{3ex}\|\,\mbox{\sc p.d.}  & & & & \| &
  & \hspace{3ex}\|\,\mbox{\sc p.d.}  & & \| & & \\[.2ex]
 & & H_3(\Xi^5;{\smallBbb Z})  & & &
  & H_2(\Xi^5-\widehat{E};{\smallBbb Z})  &
  & H_2(\Xi^5;{\smallBbb Z})  & & 0  & & ,
\end{array} } 
$$
where {\sc p.d.} stands for the Poincar\'{e} duality,
one then has the following short exact sequence
$$
 0\; \rightarrow\;
 \mbox{\raisebox{.4ex}{$H^2(\widehat{E},{\Bbb Z})$}}
  \hspace{-.2ex}/\hspace{.1ex}
  \mbox{\raisebox{-.8ex}{$\mbox{\rm ker}\,\delta^{\ast}$}}\;
 \rightarrow\; H_2(\Xi^5-\widehat{E};{\Bbb Z})\; \rightarrow\;
 H_2(\Xi^5;{\Bbb Z})\; \rightarrow \; 0\,.
$$

\bigskip

\noindent
{\bf Lemma 3.2.3.} {\it 
 The above short exact sequence splits.
} 

\bigskip

\noindent {\it Proof.}
Recall that $E\cdot K_{Z_r}=0$,
$H_2(\Xi^5;{\Bbb Z})=K_{Z_r}^{\perp}$, and that $\Xi^5$ over
$Z_r-K_{Z_r}$ is trivial. Thus there exists a partial section
of $\Xi^5$, defined only over $Z_r-K_{Z_r}$ and whose image is
disjoint from $\widehat{E}$, that lifts $K_{Z_r}^{\perp}$
isomorphically onto $H_2(\Xi^5;{\Bbb Z})$. Since the image of
this section is contained in $\Xi^5-\widehat{E}$, the composition
of the section with the inclusion map of its image into
$\Xi^5-\widehat{E}$ induces then a homomorphism from
$K_{Z_r}^{\perp}$ into $H_2(\Xi^5-\widehat{E};{\Bbb Z})$. By
construction, the composition of this homomorphsm with that
induced by the projection map from $\Xi^5$ onto $Z_r$ is the
identity map on $K_{Z_r}^{\perp}$. This shows that the above
short exact sequence splits.

\noindent
\hspace{14cm} $\Box$  

\bigskip

Consequently,
$$
 H_2(\Xi^5-\widehat{E};{\Bbb Z})\;
 =\; H_2(\Xi^5;{\Bbb Z})\, \oplus\,
     \mbox{\raisebox{.4ex}{$H^2(\widehat{E};{\Bbb Z})$}}
      \hspace{-.2ex}/\hspace{.1ex}
      \mbox{\raisebox{-.8ex}{$\mbox{\rm ker}\,\delta^{\ast}$}}
$$
and we shall now study the second factor of this
decomposition.

The universal coefficient theorem implies that  
$$
 H^2(\widehat{E};{\Bbb Z})\;
 =\; \mbox{\it Hom}\,(H_2(\widehat{E};{\Bbb Z}),{\Bbb Z})
   \oplus \mbox{\it Ext}\,(H_1(\widehat{E};{\Bbb Z}),{\Bbb Z})\;
 =\; \mbox{\it Hom}\,(H_2(\widehat{E};{\Bbb Z}),{\Bbb Z})
$$
since $H_1(\widehat{E};{\Bbb Z})=0$. Thus
$$
 \mbox{\raisebox{.4ex}{$H^2(\widehat{E},{\Bbb Z})$}}
  \hspace{-.2ex}/\hspace{.1ex}
   \mbox{\raisebox{-.8ex}{$\mbox{\rm ker}\,\delta^{\ast}$}}\;
 =\; \mbox{\raisebox{.4ex}{
  $\mbox{\it Hom}\,(H_2(\widehat{E};{\Bbb Z}),{\Bbb Z})$}}
   \hspace{-.2ex}/\hspace{.1ex}
     \mbox{\raisebox{-.8ex}{$\mbox{\rm Im}\,
      (H^2(\Xi^5;{\Bbb Z})\,\rightarrow\,
                             H^2(\widehat{E};{\Bbb Z}))$}}\,.
$$
Since $H_2(\widehat{E};{\Bbb Z})$ is freely generated by the
lifting $\widehat{E}_i$ to $\widehat{E}$ of all the irreducible
components $E_i$ of $E$,
$H^2(\widehat{E};{\Bbb Z})
    =\mbox{\it Hom}\,(H_2(\widehat{E};{\Bbb Z}),{\Bbb Z})$
is freely generated by the dual basis $\widehat{E}_i^{\ast}$,
$i=1,\ldots,r$, that sends $\widehat{E}_i$ to $1$ and all other
$\widehat{E}_j$ to $0$. Under the Lefschetz duality, the image
$\delta^{\ast}\widehat{E}_i^{\ast}$ in
$H^3(\Xi^5,\widehat{E};{\Bbb Z})$ is turned into the class in
$H_2(\Xi^5-\widehat{E};{\Bbb Z})$ represented by the meridian
$2$-sphere $\widehat{S}_i$ of $\widehat{E}_i$ in
$\Xi^5-\widehat{E}\subset\Xi^5$. (Cf.\ {\sc Figure 3-3}.)
\begin{figure}[htbp]
\setcaption{{\sc Figure 3-3.}
\baselineskip 14pt
 A tubular neighborhood of $E$ embedded in some $N$ is shown.
 Let $E_i$ be an irreducible component of $E$. Then a meridian of
 $E_i$ in $N$ is the boundary sphere of a transverse disk to
 $E_i$. In our problem, $N$ can be $\Xi^5$, $M^5(Z,X)$, or $Z_r$.
} 
\centerline{\psfig{figure=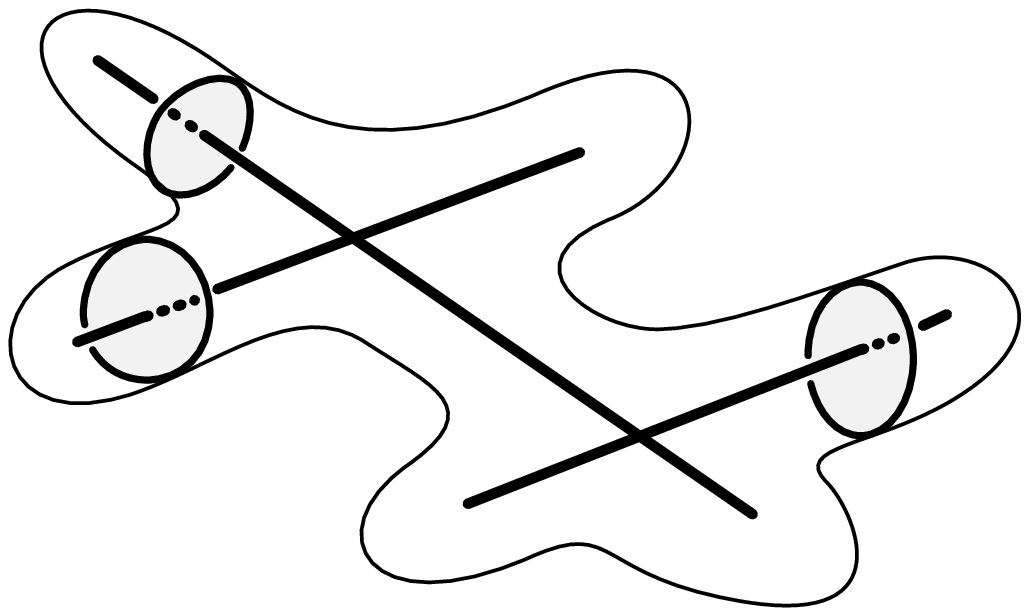,width=13cm,caption=}}
\end{figure}

The universal coefficient theorem implies also that  
$$
 H^2(\Xi^5;{\Bbb Z})\;
 =\; \mbox{\it Hom}\,(H_2(\Xi^5;{\Bbb Z}),{\Bbb Z})
   \oplus \mbox{\it Ext}\,(H_1(\Xi^5;{\Bbb Z}),{\Bbb Z})\;
 =\; \mbox{\it Hom}\,(H_2(\Xi^5;{\Bbb Z}),{\Bbb Z})
$$
since $H_1(\Xi^5;{\Bbb Z})=0$. Taken into account the Poincar\'{e}
duality, which identifies $H^2(\Xi^5;{\Bbb Z})$ with
$H_3(\Xi^5;{\Bbb Z})$ by considering the intersection number of
a $3$-cycle with all the $2$-cycles, one can work out a set of
generators for $H^2(\Xi^5;{\Bbb Z})$ represented by a set of
$3$-cycles and hence the image of $H^2(\Xi^5;{\Bbb Z})$
in $H^2(\widehat{E};{\Bbb Z})$ as follows:
\begin{itemize}
 \item
 Let $e_1,\cdots,e_r$ be a basis for $K_{Z_r}^{\perp}$ and
 $\xi_1,\cdots,\xi_r$ be a dual collection of $2$-cycles in $Z_r$
 represented by maps, still denoted by $\xi_i$, from $S^2$ into
 $Z_r$, such that $\xi_i\,\cdot\,e_j=\delta_{ij}$. By the
 unimodularness of the intersection form $Q_{Z_r}$ of $Z_r$, such
 a dual collection always exists but not unique; the non-uniqueness
 comes exactly by an arbitrary multiple of $K_{Z_r}$. The natural
 bundle homomorphism $\widetilde{\xi}_i$ from the total space of
 the pullback $S^1$-bundle $\xi_i^{\ast}\Xi$ over $S^2$ into
 $\Xi^5$ gives then a collection of $3$-cycles
 $\widetilde{\xi}_1,\cdots,\widetilde{\xi}_r$ that form a
 basis for
 $H_3(\Xi^5;{\Bbb Z})\stackrel{\mbox{\sc p.d.}}{=}H^2(\Xi^5;{\Bbb Z})$.
 By construction,
 $\widetilde{\xi}_i\cdot\widehat{E}_j=\xi_i\cdot E_j=\delta_{ij}$.

 \item
 Since $E_i\cdot K_{Z_r}=0$, $E_i$ can be expressed uniquely as
 $\sum_{j=1}^r\,a_{ij}\,e_j$ for some integers $a_{ij}$. In terms
 of this, using the intersection property of $\widetilde{\xi_i}$
 with $\widehat{E}_j$ above, one has
 $$
  \widetilde{\xi}_i\,\cdot\,\widehat{E_j}\;=\;a_{ji}
  \hspace{2em} \mbox{and hence} \hspace{2em}
  \widetilde{\xi}_i\;
   =\;\sum_{j=1}^r\,a_{ji}\,\widehat{E}_j^{\ast}\,.
 $$
 The image 
 $\mbox{\rm Im}\,(H^2(\Xi^5;{\Bbb Z})\rightarrow
                                   H^2(\widehat{E};{\Bbb Z}))$
 is then spanned by $\widetilde{\xi}_1,\cdots,\widetilde{\xi}_r$
 over ${\Bbb Z}$.
\end{itemize}
Since $\widetilde{E}_i^{\ast}$ corresponds to $\widehat{S}_i$ in
$H_2(\Xi^5-\widehat{\nu_{Z_r}(E)};{\Bbb Z})$, with the notations
introduced above, one has thus proved

\bigskip

\noindent
{\bf Lemma 3.2.4.} {\it
 $$
  H_2(\Xi^5-\widehat{E};{\Bbb Z})\;
  =\; \mbox{\it Span}_{\subscriptsizeBbb Z}
       \{\,\widehat{e}_1, \cdots, \widehat{e}_r,
           \widehat{S}_1, \cdots, \widehat{S}_r\,\}/\sim\;
  =\;
   {\Bbb Z}^r\,\oplus\,
   {\Bbb Z}^r\hspace{-.4ex}/\hspace{-.2ex}
    \mbox{\raisebox{-.4ex}{$({\Bbb Z}^r\cdot A^t)$}}\,,
 $$
 where $\sim$ is generated by
 $\sum_{j=1}^r\,a_{ji}\,\widehat{S}_j$, $i=1,\ldots, r$,
 ${\Bbb Z}^r$ consists of the $r$-dimensional integral row
 vectors, and $A^t$ is the transpose of the coefficient matrix
 $A=(a_{ij})_{ij}$.
} 

\bigskip

\noindent
{\it Remark 3.2.5.} The above $\sim$ contains exactly all the
combinations of $\widehat{S}_i$ that bound a $3$-cycle in
$\Xi^5-\widehat{E}$. 

\bigskip

Now the exact sequence
$${\small
\begin{array}{ccccccccccc}
 \cdots & \rightarrow
  & H^2(\widehat{E};{\smallBbb Z})
  & \stackrel{\tilde{\delta}^{\ast}}{\rightarrow}
  & H^3(M^5(Z,X),\widehat{E};{\smallBbb Z}) & \rightarrow
  & H^3(M^5(Z,X);{\smallBbb Z}) & \rightarrow
  & H^3(\widehat{E};{\smallBbb Z})
  & \stackrel{\tilde{\delta}^{\ast}}{\rightarrow} & \cdots \\[.2ex]
 & & & & \| &
  & \hspace{3ex}\|\,\mbox{\sc p.d.} & & \| & &  \\[.2ex]
 & & &
  & H_2(M^5(Z,X)-\widehat{E};{\smallBbb Z})  &
  & H_2(M^5(Z,X);{\smallBbb Z})  & & 0  & &  \\[.2ex]
 & & & & \| & & & & & &  \\[.2ex]
 & & & & H_2(\Xi^5-\widehat{E};{\smallBbb Z})  & & & & & & 
\end{array}  }  
$$
implies that
$$
 H_2(M^5(Z,X);{\Bbb Z})\;
 =\; H_2(\Xi^5-\widehat{E};{\Bbb Z})/
       \tilde{\delta}^{\ast}(H^2(\widehat{E};{\Bbb Z}))
$$
and hence it is generated by
$\widehat{e}_1,\cdots,\widehat{e}_r,
 \widehat{S}_1,\cdots,\widehat{S}_r$ as well though 
the coboundary homomorphism $\tilde{\delta}^{\ast}$ is
expected to be different from $\delta^{\ast}$ in the exact
sequence for $(\Xi^5,\widehat{E})$ due to the twist by $\tau$.
We shall now work out the relations of these generators in the
quotient.
 
Since $E$ is a deformation retract of $\nu_{Z_r}(E)$, 
$$
 H^2(\widehat{E};{\Bbb Z})\;
 =\; H^2(\widehat{\nu_{Z_r}(E)};{\Bbb Z})\;
  \stackrel{\mbox{\sc l.p.d.}}{=}\;
   H_2(\nu_{Z_r}(E),\partial\nu_{Z_r}(E);{\Bbb Z}),
$$
where {\sc l.p.d.} stands for the Lefschetz-Poincar\'{e} duality.
Consider the relative $2$-cycle in
$(\nu_{Z_r}(E),\partial\nu_{Z_r}(E))$ represented by $2$-disk
$D_i$ that intersects $E_i$ at one point and is disjoint from all
other $E_j$ (cf.\ the shaded disks in {\sc Figure 3-3} with the
ambient space $N$ now being $Z_r$). They form a basis for
$H_2(\nu_{Z_r}(E),\partial\nu_{Z_r}(E);{\Bbb Z})$. Their lifting
to $\widehat{E}$ shall be denoted by $\widehat{D}_i$. Let
$\widehat{S}_i^{\sim}$ be a meridian $2$-sphere of
$\widehat{E}_i$ in $M^5(Z,X)$. Then 
$$
 \delta^{\ast}(\widehat{D}_i)\;=\;[\widehat{S}_i]\,,
  \hspace{1em}
 \widetilde{\delta}^{\ast}(\widehat{D}_i)
 =[\widehat{S}_i^{\sim}]
   \hspace{1em}\mbox{in}\hspace{1em}
  H_2(\Xi^5-\widehat{E};{\Bbb Z})
$$
after the various dualities. Furthermore, let $T^{\ast}$ be the
dual monodromy operator of $T$ on $H^2(\widehat{E};{\Bbb Z})$;
then by construction the difference
$$
 \widehat{S}_i^{\sim}-\widehat{S}_i\;
 =\; T^{\ast}\widehat{D}_i-\widehat{D}_i\;
 =\; (T-\mbox{\it Id})^{\ast}(\widehat{D}_i)
$$
lies in
$H_2(\widehat{\nu_{Z_r}(E)};{\Bbb Z})=H_2(\widehat{E};{\Bbb Z})$
since the restriction of the monodromy homeomorphism $\tau$ to
$\partial\widehat{\nu_{Z_r}(E)}$ is the identity map.

\bigskip

\noindent  
{\bf Lemma 3.2.6.} {\it Let $Q_{\nu_{Z_r}(E)}$ be the intersection
 matrix of $H_2(\nu_{Z_r}(E);{\Bbb Z})$ with respect to the basis
 $(E_1,\cdots,E_r)$. Then 
 $$
  \widehat{S}_i^{\sim}\;
  =\; \widehat{S}_i\,
         +\,\sum_{j=1}^r\, \left(\,
             (T-\mbox{\it Id})\,Q_{\nu_{Z_r}(E)}^{-1}\,
                               \right)_{ij}\,\widehat{E}_j\,,
 \hspace{1em} i=1,\ldots,r.
 $$
 and they generate
 $\widetilde{\delta}^{\ast}(H^2(\widehat{E};{\Bbb Z}))$
 in $H_2(\Xi^5-\widehat{E};{\Bbb Z})$.
} 

\bigskip

\noindent {\it Proof.}
The second statement is clear. For the first one, let
$(T-\mbox{\it Id})^{\ast}(\widehat{D}_i)\;
    =\; \sum_{j=1}^r\, b_{ij}\,\widehat{E}_j$;
then, by intersecting both sides with
$\widehat{E}_1,\cdots,\widehat{E}_r$ respectively, one has
$T-\mbox{\it Id}=BQ_{\nu_{Z_r}(E)}$, where $B$ is the
coefficient matrix $(b_{ij})_{ij}$. This concludes the lemma.

\noindent
\hspace{14cm} $\Box$  

\bigskip

Putting Lemma 3.2.4 and Lemma 3.3.6 together, one can now
conclude that
$$
 H_2(M^5(Z,X);{\Bbb Z})\;
 =\; \mbox{\it Span}_{\subscriptsizeBbb Z}
  \{\, \widehat{e}_1, \cdots, \widehat{e}_r,
        \widehat{S}_1, \cdots, \widehat{S}_r \,\}/\sim\,,
$$
where $\sim$ is generated by
$$
 \sum_{j=1}^r\,a_{ji}\widehat{S}_j
  \hspace{1em} \mbox{and} \hspace{1em}
 \widehat{S}_i\,+\,\sum_{j=1}^r\,\left(\,
              (T-\mbox{\it Id})\,Q_{\nu_{Z_r}(E)}^{-1}\,A
               \,\right)_{ij}\,\widehat{e}_j\,,
 \hspace{1em} \mbox{for $i=1,\ldots, r$}\,.
$$
These relators can be combined together to eliminate the
generators $\widehat{S}_i$ and give
$$
 H_2(M^5(Z,X);{\Bbb Z})\;
 =\; \mbox{\it Span}_{\subscriptsizeBbb Z}
   \{\, \widehat{e}_1, \cdots, \widehat{e}_r \,\}/\sim\,,
$$
where $\sim$ is generated by 
$$
 \eta_i\; =\; \left(\,
           A^t\,(T-\mbox{\it Id})\,Q_{\nu_{Z_r}(E)}^{-1}\,A
                                \right)_{ik}\,\widehat{e}_k\,.
$$

To give the last step of simplification, one has to recall the
following facts from [A-GZ-V] and [Di]:

\bigskip

\noindent
{\bf Fact 3.2.7 [Seifert matrix]} ([A-GZ-V] and [Di]) {\it
For an isolated hypersurface singularity in ${\Bbb C}^{n+1}$,
the intersection matrix $Q$ for the vanishing cycles in the
fiber of the Milnor fibration can be put into the form after
choosing appropriate basis:
$Q=-(U+(-1)^nU^t)$, where $U$ is upper triangular, unimodular,
and is called the {\rm Seifert matrix} with respect to this basis.
The monodromy operator $T$ is then related to $U$ 
by $T=(-1)^{n+1}\,U^{-1}U^t$.
} 

\bigskip

\noindent
Consequently, $(T-\mbox{\it Id})Q_E^{-1}=U^{-1}$ in our case
and we can conclude and summarize the whole discussion by the
following proposition.

\bigskip

\noindent
{\bf Proposition 3.2.8 [homology of boundary].} {\it
 Let $Z$ be a Gorenstein rational singular del Pezzo surface
 with $\mbox{\rm Pic}\,(Z)={\Bbb Z}$ that is embedded in a
 smooth Calabi-Yau threefold $X$. Denote the minimal resolution
 of $Z$ by $Z_r$, which is a smooth del Pezzo surface obtained
 by ${\Bbb C}{\rm P}^2$ blown up at $r$ many points, and the
 vanishing cycle of the resolution by $E$ with irreducible
 components $E_1,\cdots,E_r$. Then $\partial\nu_X(Z)$ is a
 simply-connected spin $5$-manifold with the second integral
 homology $H_2$ isomorphic to
 ${\Bbb Z}^r/\hspace{-.2ex}
  \mbox{\raisebox{-.4ex}{$\{{\Bbb Z}^r\,\cdot\,
    (A^t\,U^{-1}\,A)\}$}}$,
 where $U$ is the Seifert matrix associated to the singularity
 with respect to $(E_1,\cdots, E_r)$ and $A$ is the matrix whose
 $i$-th row is the coefficients of $E_i$ with respect to any basis
 for $K_{Z_r}^{\perp}$ in $H_2(Z_r;{\Bbb Z})$.
 (In the above expression, elements in ${\Bbb Z}^r$ are integral
  row vectors and the $i$-th column of $U$ is the coefficients of
  $U(E_i)$ with respect to $(E_1,\cdots,E_r)$.)
} 

\bigskip

Let $Q_E$ (resp.\ $Q_{K_{Z_r}^{\perp}}$) be the intersection
matrix of $E_1,\cdots,E_r$ (resp.\ $K_{Z_r}^{\perp}$) in
$H_2(Z_r;{\Bbb Z})$. Their determinant, whose value will be given
in the explicit computations below, are all non-zero. Since $U$ is
unimodular and hence 
$\det(A^tU^{-1}A)=(\det A)^2=\det Q_E/\det Q_{K_{Z_r}^{\perp}}$,
one has

\bigskip

\noindent
{\bf Corollary 3.2.9.} {\it
 $H_2(\nu_X(Z);{\Bbb Z})$ is a finite abelian group of
 order $|\det Q_E/\det Q_{K_{Z_r}^{\perp}}|$.
} 

\bigskip

\begin{flushleft}
{\bf $H_2(\partial\nu_X(Z);{\Bbb Z})$ and $\partial\nu_X(Z)$
     for our examples.}
\end{flushleft}
With Proposition 3.2.8, we can now work out
$H_2(\partial\nu_X(Z);{\Bbb Z})$ explicitly for the above kind of
singular del Pezzo surface $Z$ when it is embedded in a smooth
Calabi-Yau threefold $X$. This then determines the topology of
$\partial\nu_{X}(Z)$. All the notations follow from previous part 

\bigskip

\noindent
{\bf (a) Basic ingredients.}
The intersection matrix $Q$, monodromy operator $T$, and the
inverse of the Seifert matrix $U$ with respect to the set of
vanishing cycles for an $A$-$D$-$E$ singularity in a complex
surface are listed below:
(Note: $\cdots$ indicates the pattern and the blank indicates $0$.)
\begin{itemize}
 \item
 The $A_k$-singularity:
 $$
 {\scriptsize
  Q_{A_k}\;=\;\left[\,\begin{array}{rrrrrr}
                     -2 &  1 &    &    &  \\
                      1 & -2 &  1 &    &  \\
                      &  \ddots & \ddots &  \ddots & \\
               &    & 1 & -2 &  1 \\
               &    &   &  1 & -2
                    \end{array}\,\right]_{k\times k}\,,
       \hspace{1cm}
   T_{A_k}\;=\;\left[\,\begin{array}{rrrrrr}
                    0 &        &        &   & -1 \\
                    1 & 0      &        &   & -1 \\
                      & \ddots & \ddots &   & \vdots \\
                      &        & \ddots & 0 & -1 \\
                      &        &        & 1 & -1
                    \end{array}\,\right]_{k\times k}\,,
 } 
 $$
 $$
  \mbox{and} \hspace{2cm} {\scriptsize
   U_{A_k}^{-1}\;=\;\left[\,\begin{array}{rrrrrr}
                      1 & 1 &         & \cdots & 1 \\
                        & 1 &         & \cdots & 1 \\
                        &   & \ddots  & \cdots  & \vdots \\
                                  & & & 1 & 1 \\
                                    & & & & 1
                      \end{array}\,\right]_{k\times k}\,.
 } 
 $$

 \item
 The $D_k$-singularity, $k\ge 4$: 
 $$
  \hspace{-1cm} {\scriptsize
  Q_{D_k}\;=\;\left[\,\begin{array}{rrrrrrr}
                     -2 &  1 &    &    &  & & \\
                      1 & -2 &  1 &    &  & & \\
            &  \ddots & \ddots &  \ddots & & & \\
               &    & 1 & -2 &  1 &    & \\
               &    &   &  1 & -2 & 1  &  1 \\
               & & &         &  1 & -2 &    \\
               & & &         &  1 &    & -2
                    \end{array}\,\right]_{k\times k}\,,
       \hspace{1cm}
  T_{D_k}\;=\;\left[\,\begin{array}{rrrrrrr}
                       0 &   &    &    & 1 & -1 & -1 \\
                       1 & 0 &    &    & 1 & -1 & -1 \\
             &  \ddots & \ddots &  & \vdots & \vdots & \vdots \\
                       & & \ddots &  0 & 1 & -1 & -1 \\
                              & & &  1 & 1 & -1 & -1 \\
                              & & &  0 & 1 & -1 &  0 \\
                              & & &  0 & 1 &  0 & -1
                     \end{array}\,\right]_{k\times k}\,,
 } 
 $$
 $$
  \mbox{and} \hspace{2cm} {\scriptsize
  U_{D_k}^{-1}\;=\;\left[\,\begin{array}{rrrrrrr}
                       1 & 1 & & & \cdots & & 1 \\
                         & 1 & & & \cdots & & 1 \\
                           & & \ddots & & \cdots & & \vdots \\
                            & & & 1 & 1 & 1 & 1 \\
                              & & & & 1 & 1 & 1 \\
                              & & & &   & 1 & 0 \\
                              & & & &   &   & 1
                     \end{array}\,\right]_{k\times k}\,.
 } 
 $$

 \item
 The $E_k$-singularity, $k=6,7,8$:
 (Only the matrices for $k=8$ are shown; for $k=6$ or $7$, choose
  the corresponding last $6\times 6$ or $7\times 7$ submatrix.)
 $$\hspace{-1cm}{\scriptsize
  Q_{E_8}\;=\;\left[\,\begin{array}{rrrrrrrr}
                   -2 &  1 &    & & & & & \\
                    1 & -2 &  1 &    & & & & \\
                      &  1 & -2 &  1 & & & & \\
                         & &  1 & -2 &  1 & & & \\
                            & & &  1 & -2 &  1 &  1 & \\
                               & & & &  1 & -2 & & \\
                               & & & &  1 &    & -2 & 1 \\
                                & & & & & &  1 & -2
                    \end{array}\,\right]\,,
       \hspace{1cm}
  T_{E_8}\;=\;\left[\,\begin{array}{rrrrrrrr}
                    0 &   &   &   & 1 & -1 & 0 & -1 \\
                    1 & 0 &   &   & 1 & -1 & 0 & -1 \\
                      & 1 & 0 &   & 1 & -1 & 0 & -1 \\
                        & & 1 & 0 & 1 & -1 & 0 & -1 \\
                          & & & 1 & 1 & -1 & 0 & -1 \\
                            & & & & 1 & -1 & 0 &  0 \\
                           & & &  & 1 &  0 & 0 & -1 \\
                           & & & &  0 &  0 & 1 & -1
                    \end{array}\,\right]\,,
 } 
 $$
 $$
  \mbox{and} \hspace{2cm} {\scriptsize
  U_{E_8}^{-1}\;=\;\left[\,\begin{array}{rrrrrrrr}
                    1 & 1 & 1 & 1 & 1 & 1 & 1 & 1 \\
                      & 1 & 1 & 1 & 1 & 1 & 1 & 1 \\
                        & & 1 & 1 & 1 & 1 & 1 & 1 \\
                          & & & 1 & 1 & 1 & 1 & 1 \\
                            & & & & 1 & 1 & 1 & 1 \\
                              & & & & & 1 & 0 & 0 \\
                                & & & & & & 1 & 1 \\
                                  & & & & & & & 1
                    \end{array}\,\right]\,.
 } 
 $$
\end{itemize}

\bigskip

\noindent 
{\bf (b) The list of all possibilities.}
With respect to the canonical basis
$\varepsilon_0,\cdots,\varepsilon_r$ for $H_2(Z_r;{\Bbb Z})$,
one can choose the basis $e_1,\cdots,e_r$ for $K_{Z_r}^{\perp}$
to be
$$
 e_1\,=\,\varepsilon_1\,-\,\varepsilon_r\,,\hspace{1em}
  \cdots\,,\hspace{1em}
 e_{r-1}\,=\,\varepsilon_{r-1}\,-\,\varepsilon_r\,, \hspace{1em}
 e_r\,=\,\varepsilon_0\,-\,3\,\varepsilon_r\,.
$$
With respect to this,
$${\scriptsize
 Q_{K_{Z_r}^{\perp}}\;=\;
   \left[\,\begin{array}{rrrrrr}
     -2 & -1 & \cdots & -1 & -3 \\
     -1 & -2 &        & -1 & -3 \\
    \vdots & & \ddots & & \vdots \\
     -1 & -1 &        & -2 & -3 \\
     -3 & -3 & \cdots & -3 & -8 
   \end{array}\,\right]_{r\times r} } 
   \,,\hspace{1em}\mbox{where all the missing entries are $-1$.} 
$$
Its determinant is 
$$ \mbox{
\begin{tabular}{|c|rrrrrrrrrrr|}  \hline
 $r$            &  $3$ && $4$ && $5$ && $6$ && $7$ && $8$ \\ \hline
 {\footnotesize $\det Q_{K_{Z_r}^{\perp}}$}
   & $-6$ && $5$ && $-4$ && $3$ && $-2$ && $1$ \\ \hline
\end{tabular} }\,.
$$
The determinant of the intersection matrix $Q$ for the connected
components of $E$ involved in our problem is
$$ \mbox{
\begin{tabular}{|ccccccccc|}    \hline
 $A_k$           && $D_k$      && $E_6$ && $E_7$ && $E_8$  \\ \hline
 $(-1)^k\,(k+1)$ && $(-1)^k\,4$ && $3$  && $-2$  && $1$  \\ \hline
\end{tabular} }\,.
$$

When $E$ has several connected components $E_{(1)},\cdots$,
the corresponding $Q_E$ is the block diagonal matrix
$Q_{E_{(1)}}\oplus\cdots$. From the given data, we now give a
list of the order of $H_2(\partial\nu_X(Z);{\Bbb Z})$ below:

\noindent
\parbox[t]{2.7cm}{
 $$
  \begin{array}{l}
   (r=3): \\[1ex]
   \mbox{
    \begin{tabular}{|c|}  \hline
     $A_1+A_2$          \\ \hline
     1                  \\ \hline
    \end{tabular} }
  \end{array}
 $$
} \  
\parbox[t]{3cm}{
 $$
  \begin{array}{l}
   (r=4): \\[1ex]
   \mbox{
    \begin{tabular}{|c|}   \hline
     $A_4$          \\ \hline
     1              \\ \hline
    \end{tabular} }
  \end{array}
 $$
} \
\parbox[t]{3cm}{
 $$
  \begin{array}{l}
   (r=5): \\[1ex]
   \mbox{
    \begin{tabular}{|cc|}    \hline
     $2A_1+A_3$  & $D_5$     \\ \hline
      $4$  & $1$             \\ \hline
    \end{tabular} }
  \end{array}
 $$
} \ \hspace{1cm}
\parbox[t]{4cm}{
 $$
  \begin{array}{l}
   (r=6): \\[1ex]
   \mbox{
    \begin{tabular}{|ccc|}            \hline
     $A_1+A_5$  & $3A_2$ & $E_6$   \\ \hline
      $4$  & $9$  & $1$            \\ \hline
    \end{tabular} }               
  \end{array}
 $$
} \

\noindent\hspace{1.2ex} $(r=7):$
$$
 \mbox{
 \begin{tabular}{|ccccccccccc|}      \hline
  $A_1+2A_3$ && $A_2+A_5$ && $A_7$ &&  $3A_1+D_4$
       && $A_1+D_6$ && $E_7$ \\ \hline
  $16$       && $9$       && $4$   && $16$
       && $4$       && $1$   \\ \hline
 \end{tabular} }
$$

\noindent\hspace{1.2ex} $(r=8):$
$$
 \mbox{
 \begin{tabular}{|ccccccccccc|}    \hline
  $2A_1+2A_3$ && $A_1+A_2+A_5$ && $A_1+A_7$ && $4A_2$ && $2A_4$
       && $A_8$       \\ \hline
  $64$        && $36$          && $16$      && $81$   && $25$
       && $9$         \\ \hline
 \end{tabular} }
$$
$$
 \mbox{
 \begin{tabular}{|ccccccccccccc|}    \hline
  $2A_1+D_6$ && $A_3+D_5$ && $A_1+E_7$ && $A_2+E_6$ && $2D_4$
        && $D_8$    && $E_8$    \\ \hline
  $16$       && $16$      && $4$       && $9$       && $16$
        && $4$      && 1        \\ \hline
 \end{tabular} }
$$
By the classification theorem for abelian groups [Ja] and
the fact that the torsion part of the second integral
homology of a simply-connected spin closed $5$-manifold is
the direct sum of two identical subgroups (Fact 1.1.1 in Sec.\ 1.1),
the list of all abelian groups whose order appear above and have
such splitting property are:
$$
 \hspace{-2em}{\scriptsize
 \mbox{
 \begin{tabular}{|c|cccccccc|}      \hline
  order & $1$ & $4$ & $9$ & $16$ & $25$ & $36$  & $64$ & $81$ \\ \hline 
  group & $0$ \rule{0ex}{2.6ex}
        & ${\scriptsizeBbb Z}_2\oplus{\scriptsizeBbb Z}_2$
        & ${\scriptsizeBbb Z}_3\oplus{\scriptsizeBbb Z}_3$
        & $(\oplus_2{\scriptsizeBbb Z}_2)
             \oplus(\oplus_2{\scriptsizeBbb Z}_2)$
        & ${\scriptsizeBbb Z}_5\oplus{\scriptsizeBbb Z}_5$
        & ${\scriptsizeBbb Z}_6\oplus{\scriptsizeBbb Z}_6$
        & $(\oplus_3{\scriptsizeBbb Z}_2)
           \oplus(\oplus_3{\scriptsizeBbb Z}_2)$     
        & $(\oplus_2{\scriptsizeBbb Z}_3)
             \oplus(\oplus_2{\scriptsizeBbb Z}_3)$        \\[1ex]
     & & & & ${\scriptsizeBbb Z}_4\oplus{\scriptsizeBbb Z}_4$
     & & & $({\scriptsizeBbb Z}_2\oplus{\scriptsizeBbb Z}_4)
            \oplus({\scriptsizeBbb Z}_2\oplus{\scriptsizeBbb Z}_4)$
          & ${\scriptsizeBbb Z}_9
                  \oplus{\scriptsizeBbb Z}_9$ \\[1ex]
     & & & & & &
     & ${\scriptsizeBbb Z}_8\oplus{\scriptsizeBbb Z}_8$ & \\ \hline
 \end{tabular} }  } 
$$
These are the only groups, to which $H_2(\nu_X(Z);{\Bbb Z})$
can be isomorphic.

\bigskip

\noindent
{\bf (c) Conclusion.} From the tables given in (b) and comparing
with Smale's theorem, we can now conclude the possible topology of
$\partial\nu_X(Z)$ in each case as follows:
(See Fact 1.1.1 in Sec.\ 1.1 for the notation of manifolds.)

\noindent
\parbox[t]{2.7cm}{
 $$
  \begin{array}{l}
   (r=3): \\[1ex]
   \mbox{
    \begin{tabular}{|c|}  \hline
     $A_1+A_2$          \\ \hline
     $S^5$  \rule{0ex}{2.6ex} \\ \hline
    \end{tabular} }
  \end{array}
 $$
} \
\parbox[t]{3cm}{
 $$
  \begin{array}{l}
   (r=4): \\[1ex]
   \mbox{
    \begin{tabular}{|c|}   \hline
     $A_4$          \\ \hline
     $S^5$      \rule{0ex}{2.6ex} \\ \hline
    \end{tabular} }
  \end{array}
 $$
} \
\parbox[t]{3cm}{
 $$
  \begin{array}{l}
   (r=5): \\[1ex]
   \mbox{
    \begin{tabular}{|cc|}    \hline
     $2A_1+A_3$  & $D_5$     \\ \hline
      $M^5_2$  & $S^5$
      \raisebox{-1ex}{\rule{0ex}{3.6ex}} \\ \hline
    \end{tabular} }
  \end{array}
 $$
} \ \hspace{1cm}
\parbox[t]{4cm}{
 $$
  \begin{array}{l}
   (r=6): \\[1ex]
   \mbox{
    \begin{tabular}{|ccc|}            \hline
     $A_1+A_5$  & $3A_2$ & $E_6$   \\ \hline
      $M^5_2$  & $M^5_3$  & $S^5$
      \raisebox{-1ex}{\rule{0ex}{3.6ex}} \\ \hline
    \end{tabular} }
  \end{array}
 $$
} \

\noindent\hspace{1.2ex} $(r=7):$
$$
 \mbox{
 \begin{tabular}{|ccccccccccc|}      \hline
  $A_1+2A_3$   && $A_2+A_5$ && $A_7$  && $3A_1+D_4$ && $A_1+D_6$
                                             && $E_7$ \\ \hline
  ({\scriptsize $M^5_2\sharp M^5_2$}), $M^5_4$   && $M^5_3$
   && $M^5_2$ && $M^5_2\sharp M^5_2$, ({\scriptsize $M^5_4$})
   && $M^5_2$ && $S^5$
   \raisebox{-1ex}{\rule{0ex}{3.6ex}} \\ \hline
 \end{tabular} }
$$

\noindent\hspace{1.2ex} $(r=8):$
$$
 \hspace{-2em}\mbox{
 \begin{tabular}{|ccccccccccc|}    \hline
  $2A_1+2A_3$ && $A_1+A_2+A_5$ && $A_1+A_7$
   && $4A_2$                         && $2A_4$  && $A_8$  \\ \hline
  ({\scriptsize $\sharp_3 M^5_2$}), $M^5_2\sharp M^5_4$,
    ({\scriptsize $M^5_8$})    && $M^5_6$
    && ({\scriptsize $M^5_2\sharp M^5_2$}), $M^5_4$
   && $M^5_3\sharp M^5_3$, ({\scriptsize $M^5_9$})
   && $M^5_5$ && $M^5_3$
   \raisebox{-1ex}{\rule{0ex}{3.6ex}} \\ \hline
 \end{tabular} }
$$
$$
 \hspace{-2em}\mbox{
 \begin{tabular}{|ccccccccccccc|}    \hline
  $2A_1+D_6$                     && $A_3+D_5$
   && $A_1+E_7$ && $A_2+E_6$ && $2D_4$
    && $D_8$    && $E_8$    \\ \hline
  $M^5_2\sharp M^5_2$, ({\scriptsize $M^5_4$})
  && ({\scriptsize $M^5_2\sharp M^5_2$}), $M^5_4$
   && $M^5_2$   && $M^5_3$
   && $M^5_2\sharp M^5_2$, ({\scriptsize $M^5_4$})
    && $M^5_2$      && $S^5$
    \raisebox{-1ex}{\rule{0ex}{3.6ex}} \\ \hline
 \end{tabular} }
$$

\bigskip

Altogether there are eight cases in the above lists, in which
$\partial\nu_X(Z)$ is not uniquely determined by the order of 
$H_2(\partial\nu_X(Z);{\Bbb Z})$. By a computer check as explained
in Remark 3.2.10 below and Appendix, it turns out that those
$5$-manifolds that are put in parentheses are ruled out. This
determines $\partial\nu_X(Z)$ associated to each combination of
A-D-E singularities uniquely; and the discussion is thus complete.

\bigskip

\noindent
{\it Remark 3.2.10 [computer check].}
Let $\xi=a_0\varepsilon_0
       +a_1\varepsilon_1+\cdots+a_r\varepsilon_r$
in $K_{Z_r}^{\perp}$ such that $\xi\cdot\xi=-2$. Using the
inequality
$\frac{a_1^2+\cdots+a_r^2}{r}\ge(\frac{a_1+\cdots+a_r}{r})^2$,
one can show that $a_0^2\le\frac{2r}{9-r}$. Thus, due to the
symmetry of the equations for the coefficients $a_1,\cdots,a_r$,
one concludes after some elementary manipulation that the
following elements generate all $(-2)$-norm elements in
$K_{Z_r}^{\perp}$ by overall sign change and permutations of the
last $r$ coordinates:
\begin{quote}
 \parbox[t]{2cm}{$r=3$:}\
 \parbox[t]{11cm}{\scriptsize
                  $(0;1,-1,0)$,\hspace{1ex}
                  $(1;-1,-1,-1)$. }

 \parbox[t]{2cm}{$r=4$:}\
 \parbox[t]{11cm}{\scriptsize
                  $(0;1,-1,0,0)$,\hspace{1ex}
                  $(1;-1,-1,-1,0)$. }

 \parbox[t]{2cm}{$r=5$:}\
 \parbox[t]{11cm}{\scriptsize
                  $(0;1,-1,0,0,0)$,\hspace{1ex}
                  $(1;-1,-1,-1,0,0)$. }

 \parbox[t]{2cm}{$r=6$:}\
 \parbox[t]{12cm}{\scriptsize
                  $(0;1,-1,0,0,0,0)$,\hspace{1ex}
                  $(1;-1,-1,-1,0,0,0)$,\hspace{1ex}
                  $(2;-1,-1,-1,-1,-1,-1)$. }

 \parbox[t]{2cm}{$r=7$:}\
 \parbox[t]{12cm}{\scriptsize
                  $(0;1,-1,0,0,0,0,0)$,\hspace{1ex}
                  $(1;-1,-1,-1,0,0,0,0)$,\hspace{1ex}
                  $(2;-1,-1,-1,-1,-1,-1,0)$.}

 \parbox[t]{2cm}{$r=8$:}\
 \parbox[t]{11cm}{{\scriptsize
                  $(0;1,-1,0,0,0,0,0,0)$,\hspace{1ex}
                  $(1;-1,-1,-1,0,0,0,0,0)$,}\newline
                  {\scriptsize
                  $(2;-1,-1,-1,-1,-1,-1,0,0)$,\hspace{1ex}
                  $(3;-2,-1,-1,-1,-1,-1,-1,-1)$.} }
\end{quote}
Using this, one can write a computer code to first sort out all
the chains of $(-2)$-norm elements in $K_{Z_r}^{\perp}$ whose
intersection matrix is as indicated in the allowed combinations
of singularities, and then computing $A^tU^{-1}A$ and simplifying
it by integral row and column operations to render $A^tU^{-1}A$
an integral diagonal form
$$
 \mbox{\it Diag}\,(1,\cdots,1, d_1,d_1,\cdots,d_s,d_s)\,,$$
where
$d_i\ne 0,1$ and $d_i|d_j$ for $i\le j$. This will conclude that
$$
 H_2(\partial\nu_X(Z);{\Bbb Z})
 =({\Bbb Z}_{d_1}\oplus\cdots\oplus{\Bbb Z}_{d_s})
     \oplus({\Bbb Z}_{d_1}\oplus\cdots\oplus{\Bbb Z}_{d_s})
     \hspace{1em}\mbox{and}\hspace{1em}
 \partial\nu_X(Z)=M^5_{d_1}\sharp\cdots\sharp M^5_{d_s}
$$
and hence give the definite answer to the question of which 
of the candidate $5$-manifolds for $\partial\nu_X(Z)$ would
truely occur. The detail of the code is in the Appendix.

\bigskip

\section{Topologically admissible isolated singularities.}

Having studied the topology of $\partial\nu_X(Z)$ for a class of
del Pezzo surface $Z$ embedded in a Calabi-Yau threefold $X$,
we can now lay down many locally topologically admissible
pairs $(X,Z)$ and $(X^{\prime}, Z^{\prime})$ for some Calabi-Yau
threefolds $X$, $X^{\prime}$. Though we do not know at the moment
what $X$ and $X^{\prime}$ could be, we do know
from the discussion in Sec.\ 3  that, if such
extremal transition exists, then the topology
$\nu_X(Z)\cup_{\partial}\nu_{X^{\prime}}(Z^{\prime})$
of the link of the resulting isolated singularity $p_0$ in the
$7$-space $Y^7$ is independent of what $X$ and $X^{\prime}$
exactly are.
(See Remark 4.4 at the end for some notes on these
$6$-manifolds.)

\bigskip

\begin{flushleft}
{\bf The transition: 
$(X,Z)\;
 \stackrel{\mbox{\scriptsize K\"{a}hler deformation}}{\longrightarrow}\;
 (\overline{X},p_0)\;
 \stackrel{\mbox{\scriptsize K\"{a}hler deformation}}{\longleftarrow}\;
 (X^{\prime},Z^{\prime})$.
 }  
\end{flushleft}
Del Pezzo surfaces of the same group in the following list of
examples when embedded in some Calabi-Yau threefolds have the
chance to be pinched down to a point and puffed up to another
while this is impossible for those in different groups:

\bigskip

\noindent
{\bf Example 4.1.}

\begin{quote}
 \hspace{-1em}
 \parbox[t]{8em}{Group-$(S^5)$:}\
  \parbox[t]{11cm}{$(r=3:A_1+A_2)$,\hspace{1ex}
                   $(r=4:A_4)$,\hspace{1ex}
                   $(r=5:D_5)$,\hspace{1ex}
                   $(r=6:E_6)$,\hspace{1ex}
                   $(r=7:E_7)$,\hspace{1ex}
                   $(r=8:E_8)$.  }

 \hspace{-1em}
 \parbox[t]{8em}{Group-$(M^5_2)$:}\
  \parbox[t]{11cm}{$(r=5:2A_1+A_3)$,\hspace{1ex}
                   $(r=6:A_1+A_5)$,\hspace{1ex}
                   $(r=7:A_7)$,\hspace{1ex} \newline
                   $(r=7:A_1+D_6)$,\hspace{1ex}
                   $(r=8:A_1+E_7)$,\hspace{1ex}
                   $(r=8:D_8)$.  }

 \hspace{-1em}
 \parbox[t]{8em}{Group-$(M^5_2\sharp M^5_2)$:}\
  \parbox[t]{11cm}{$(r=7:3A_1+D_4)$,\hspace{1ex}
                   $(r=8:2A_1+D_6)$,\hspace{1ex}
                   $(r=8:2D_4)$.  }

 \hspace{-1em}
 \parbox[t]{8em}{Group-$(M^5_3)$:}\
  \parbox[t]{11cm}{$(r=6:3A_2)$,\hspace{1ex}
                   $(r=7:A_2+A_5)$,\hspace{1ex}
                   $(r=8:A_8)$,\hspace{1ex}
                   $(r=8:A_2+E_6)$.  }

 \hspace{-1em}
 \parbox[t]{8em}{Group-$(M^5_4)$:}\
  \parbox[t]{11cm}{$(r=7:A_1+2A_3)$,\hspace{1ex}
                   $(r=8:A_1+A_7)$,\hspace{1ex}
                   $(r=8:A_3+D_5)$  }
\end{quote}

\bigskip

\begin{flushleft}
{\bf The transition:
$(X,Z)\;
 \stackrel{\mbox{\scriptsize K\"{a}hler deformation}}{\longrightarrow}\;
 (\overline{X},p_0)\;
 \stackrel{\mbox{\scriptsize Complex deformation}}{\longleftarrow}\;
 (X^{\prime},Z^{\prime})$.
 }  
\end{flushleft}
Recall (e.g.\ [A-L], [B-G], [B-C-dlO], and [Gr].)
that, if one lets $b_3=\dim H_3(X^{\prime};{\Bbb Z})$ be the
third Betti number of $X^{\prime}$ and
$\{\alpha_i,\, \beta_j\}_{i,j=1}^{\frac{1}{2}b_3}$ be a canonical
basis for $H_3(X^{\prime};{\Bbb Z})$ such that
$$
 \alpha_i\cdot\beta_j\,=\,-\beta_j\cdot\alpha_i\,
  =\,\delta_{ij} \hspace{1em}\mbox{\rm and}\hspace{1em}
 \alpha_i\cdot\alpha_j\,
  =\,\beta_i\cdot\beta_j\,=\,0\,.
$$
Then the complex structure on $X^{\prime}$ is parametrized
projectively by the periods
$$
 \int_{\alpha_i}\,\Omega
  \hspace{2ex}\mbox{(the $\alpha$-periods)}
 \hspace{1em}\mbox{and}\hspace{1em}
 \int_{\beta_j}\,\Omega
  \hspace{2ex}\mbox{(the $\beta$-periods)}\,,
$$
where $\Omega$ is the unique holomorphic $3$-form (determined
by the complex structure) on $X^{\prime}$ up to a constant
multiple. It is known ([B-G]) that indeed the $\alpha$-periods
and the $\beta$-periods determine each other. For $X^{\prime}$ to
degenerate to a conifold via deformation of complex structures,
it is known that some disjoint collection of $3$-cycles, realized
as embedded $S^3$'s in $X^{\prime}$, get pinched into a collection
of isolated points, which is quite similar to the situation for
Riemann surfaces. If one chooses the canonical basis appropriately,
these conifolds may correspond to points in the moduli space of
complex structures on $X^{\prime}$, for which some of the periods
vanish. This suggests that there may exist severer
deformations of complex structures, for which a more complicated
chain of $S^3$'s get shrunk to points. Unfortunately, though
there is some work in this direction (e.g.\ [C-G-G-K]), there
seems not much that is known.

On the other hand, since any vector bundle over $S^3$ is trivial,
if more complicated union $Z^{\prime}$ of $S^3$ can get shrunk
to points, assuming that all the intersections are transverse,
then the tubular neighborhood $\nu_{X^{\prime}}(Z^{\prime})$
of $Z^{\prime}$ in $X^{\prime}$ is a plumbing of a collection
of trivial $D^3$-bundle $S^3\times D^3$, which depends only
on the intersection pattern and the sign of intersections.            
For this reason, let us boldly press on a little bit and see
locally how some isolated singularity of a singular Calabi-Yau
threefold from previous K\"{a}hler deformations may be resolved
by $3$-cycles.

In view that most of $\partial\nu_X(Z)$ we have seen in this
article is simply-connected, one may notice the following lemma:

\bigskip

\noindent
{\bf Lemma 4.2.} {\it
 Let $W^6$ be the $6$-manifold-with-boundary obtained by plumbing
 a collection of trivial $D^3$-bundle $S^3\times D^3$ over $S^3$
 (including possibly self-plumbing). Let $\Gamma$ be the plumbing
 diagram. If $\Gamma$ is not simply-connected, then $\partial W^6$
 cannot be simply-connected either.
}  

\bigskip

\noindent
{\it Proof.}
By construction, the closed $5$-manifold $\partial W^6$ is
obtained from a collection of $5$-manifold-with-boundary of
the form $(S^3-\nu_{S^3}\{\mbox{intersection points}\})\times S^2$
by pasting their boundary $S^2\times S^2$ with a twist that
interchanges the two $S^2$ factors. Recall from Sec.\ 1.1
the natural continuous pinching map $\phi$ from the plumbing $W^6$
to $\Gamma$. By construction, when restricted to $\partial W^6$,
$\phi^{-1}(p)$ is homeomorphic to
$(S^3-\nu_{S^3}\{\mbox{intersection points}\})\times S^2$ for $p$
a vertex of $\Gamma$ and to $S^2\times S^2$ for $p$ an interior
point of an edge of $\Gamma$. In particular, $\phi^{-1}(p)$ is
path-connected for all $p$. Consequently, one can lift any
non-null-homotopic loop in $\Gamma$ to a loop in $\partial W^6$,
which must be non-null-homotopic either. This concludes the proof.

\noindent
\hspace{14cm} $\Box$  

\bigskip

Consequently, for most of the generic K\"{a}hler deformations 
we have seen that pinch either an $S^2$ or a del Pezzo surface
$Z$ to a point $p_0$, if it is to be resolved by a smooth complex
deformation that puffs $p_0$ to a chain of immersed $S^3$'s that
may intersect among themselves transversely, then indeed all these
$S^3$ have to be embedded and the associated intersection diagram
has to be a tree. Let us give some examples below:

\bigskip

\noindent
{\bf Example 4.3.}
(Each $C^3_i$ below is an embedded $S^3$ and all the intersections
are transverse.)
\begin{quote}
 \parbox[t]{6em}{Group-$(S^5)$:}\
  \parbox[t]{11cm}{
   $Z^{\prime}=C^3_1\cup\cdots\cup C^3_{2k}$ with the intersection
   diagram the Dynkin diagram $A_{2k}$, where $k$ is any positive
   integer.}

 \parbox[t]{6em}{$(S^2\times S^3)$:}\
  \parbox[t]{11cm}{
   $Z$ is the del Pezzo surface $Z_1$ and
   $Z^{\prime}=C^3_1\cup\cdots\cup C^3_{2k-1}$ with the
   intersection diagram the Dynkin diagram $A_{2k-1}$, where $k$ is
   any positive integer.}
\end{quote}
These follow from the decomposition
$S^5=(D^3\times S^2)\cup_{\partial} S^2\times D^3$ induced by
the realization of $S^5$ as the join $S^2\ast S^2$.

\bigskip

Let us conclude this paper with the following remark:

\bigskip

\noindent
{\it Remark 4.4 [link of isolated singularity].}
The nature of the singularity $p_0$ in $Y^7$ certainly depends
on its link, the $6$-manifold
$\nu_X(Z)\cup_{\partial}\nu_{X^{\prime}}(Z^{\prime})$.
When $Z$ and $Z^{\prime}$ are both simply-connected as most of
the examples discussed in this article turn out to be, so are
$\nu_X(Z)$ and $\nu_{X^{\prime}}(Z^{\prime})$. Since
the common boundary of the latter manifolds is connected,
$\nu_X(Z)\cup_{\partial}\nu_{X^{\prime}}(Z^{\prime})$ is also
simply-connected by the van Kampen theorem.
It has been known ([Wa2], [Wil2]) that the structure of a
simply-connected closed $6$-manifold $M^6$ is closely coded in
the following triple of data:
(1) a {\it cubic form} on $H^2(M^6;{\Bbb Z})$ given by the
cup-product, (2) a {\it linear form} on $H^2(M^6;{\Bbb Z})$ given
by the cup product with the first Pontrjagin class $p_1$ or the
second Chern class $c_2$, and (3) the {\it middle cohomology}
$H^3(M^6;{\Bbb Z})$.
In view of this, first one likes to know which locally admissible
examples constructed in this paper are globally admissible, and
then, for those globally admissible ones, whether the above
triple of data provides us with some simple description of the
link of the corresponding isolated singularity in $Y^7$ and its
relations to the phenomenon of enhanced gauge symmetry in string
theory. These will require further much more demanding work in
the future.

\newpage

\begin{flushleft}
{\bf Appendix. Computer code.}
\end{flushleft}
The problem is to find all $r$-tuples $(E_1,\,\cdots\,, E_r)$ of
$(-2)$-norm elements $E_i\in K_{Z_r}^\perp$ that have the right
intersection matrix and then compute the smith normal form of
$A^t U^{-1} A$. Such a work can be achieved by a computer program.
Even so, checking all $r$-tuples straightforwardly is too
time-consuming for $r=8$, where we have (including permutations
and sign change) $240$ elements in $K_{Z_8}^\perp$ with norm $(-2)$,
which leads to ${{240!}\over{(240-8)!}}\approx 10^{19}$ possibilities.
Instead, we use backtracking with some simple optimizations.
Let us explain in less technical terms.

The {\it recursive} way to assemble all $r$-tuples of a set
$\{\xi_1,\,\cdots,\,\xi_n\}$ is to think of it as a tree where each
vertex is a $k$-tuple, $k\leq r$ and each edge corresponds to
appending a certain $\xi_i$ to the $k$-tuple:

\begin{figure}[htbp]
\centerline{\psfig{figure=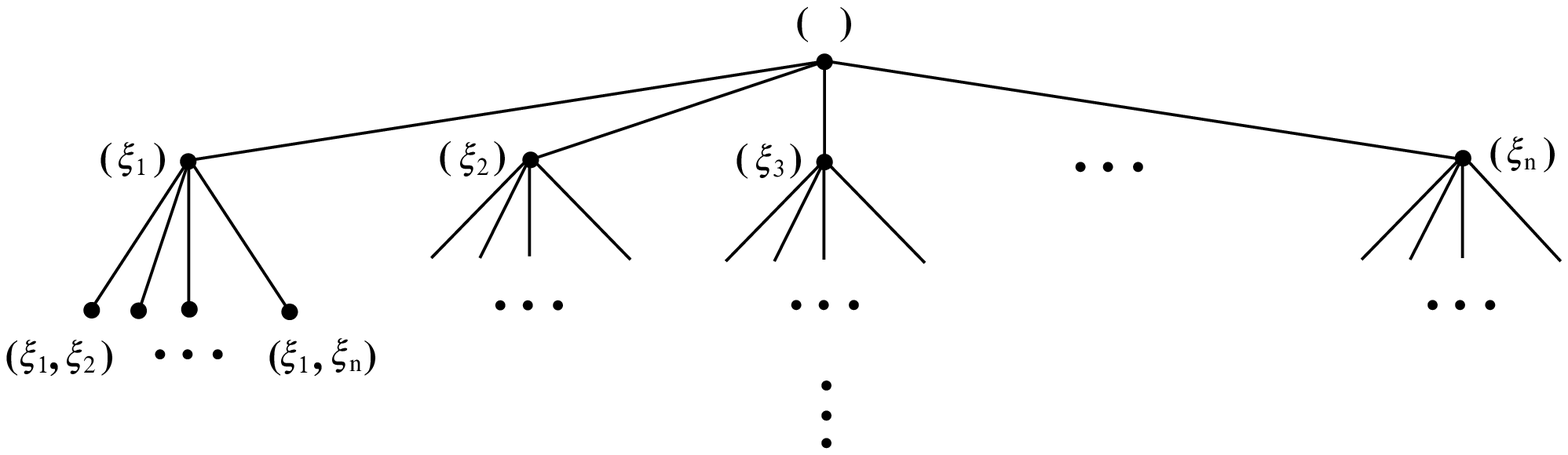,width=13cm,caption=}}
\end{figure}

\noindent
This can be implemented by a procedure that takes the current
$k$-tuple and the next element, appends the element to the list and
then calls itself with the new list and all possible choices for the
new next element. And of course the procedure has to save the tuple
and return for $k=r$. Backtracking means that we only allow edges
that lead to the ``right'' $(k+1)$-tuple, here those with the right
intersection numbers. In our code this is performed by the procedure
{\tt try\_E}.
For $r=7,8$ we optimize this algorithm further by starting not with
the empty list but with selected triplets, where we exclude triplets
that follow from previous ones by permutation or overall sign change.
Then the calculation finishes in each case within a few days on a
modern (year 1998) computer.

For completeness, the code in Maple is given below.

\medskip

\baselineskip 9pt

{\scriptsize
\begin{verbatim}
# cat ADE.m | nice maple -f -q > /tmp/run.log
# you can run the program like this under unix

# this is the path where the result will be saved
savepath:=`/u4/vrbraun/ade/`;

with(linalg): with(combinat):
with(share): readshare(normform, linalg):

# constructs all possible permutations / sign change
find_permutations:=
proc(a_1,a_list) local i,a,result;
  result:=permute(a_list);
  result:=map(  (i,a)->[a,op(i)], result, a_1);
  result:=[op(result), op(map(i->map(j->-j,i),result)) ];
  convert(convert(result,set),list);
end:

# find all
try_E:=
proc(E_list,E_next) local i,k,good,E_list_new; global A_searchfor,Q;

  k:=nops(E_list)+1;
  good:=true;
  # compare the k-th row of A_searchfor with the corresponding elements in Q
  # This checks that the new divisor has the right intersections.
  for i from 1 to nops(E_list) while good do
    good:=good and A_searchfor[i,k]=Q[E_list[i],E_next];
  od;
  if good then
    # the new element leads to the desired intersection pattern
    E_list_new:=[op(E_list),E_next];
    if k=rowdim(A_searchfor) then # we found a solution!
      add_solution(E_list_new);
      RETURN();
    fi;
    for i from 1 to rowdim(Q) do  # try adding E[i] to the list
      if not member(i,E_list_new) then # simple necessary condition
        try_E(E_list_new,i);  # do the next recursion step
      fi;
    od;
  fi;
end:

find_d:=
proc(E_index_list) local n,this_E,A,At,M,M_diag,i,d,M1,M2;
    global U,all_E_newbasis,e,evalm,examples;
  n:=nops(E_index_list);
  At:=blockmatrix(1,n,all_E_newbasis[E_index_list[i]]$i=1..n);
  A:=transpose(At);
  M:=multiply(At,U,A);
  M_diag:=ismithex(M,M1,M2);
  examples:=[evalm(A),evalm(M1),evalm(M2)];
  d:=[ M_diag[i,i]$i=1..rowdim(M_diag) ];
  map(abs,d);
end:

# this is called if we have found a configuration that has
# the right intersections.
add_solution:=
proc(E_list) local d; global searchfor_d,result,examples;
  d:=find_d(E_list); # The diagonal in the integer smith normal form
  if not member(d,result) then
    result:=[op(result),d];
    print(`New solution: `,d);
    examples:=[op(examples),E_list];
  fi;
end:

# simple optimisation: Find triplets to start with, remove permutations
find_good_triplets:=
proc()
  local i1,i2,i3,firstdigits,old_triplets,N,i,t,encoded_t,sort_under;
  global good_triplets,all_E,Q,A_searchfor;

  good_triplets:=[];
  firstdigits:=map(abs,convert(map(i->i[1],all_E),set));
  old_triplets:=table();
  N:=max(op(firstdigits));
  for i1 from -N to N do
    for i2 from -N to N do
      for i3 from -N to N do
        old_triplets[i1,i2,i3]:=[];
  od:od:od:
  for i1 from 1 to nops(all_E) do
    for i2 from 1 to nops(all_E) do
      for i3 from 1 to nops(all_E) do
        # check for right intersections at this level
        if Q[i1,i2]=A_searchfor[1,2] and
           Q[i1,i3]=A_searchfor[1,3] and
           Q[i2,i3]=A_searchfor[2,3] then
          t:=[all_E[i1],all_E[i2],all_E[i3]];
          encoded_t:=encode_tuple(t);
          sort_under:=op(encoded_t[1]);
          if member(encoded_t, old_triplets[sort_under] ) then next; fi;
          encoded_t:=encode_tuple( map(i->evalm(-i),t) );
          if member(encoded_t,old_triplets[sort_under]) then next; fi;
          # new triplet, permutations dont occur
          old_triplets[sort_under]:=[op(old_triplets[sort_under]),encoded_t];
           good_triplets:=[op(good_triplets),[i1,i2,i3]];
        fi;
  od:od:od:
end:

# returns a unique representative (same for all permutations)
encode_tuple:=
proc(t) local result,i,j;
  result:=[seq( [ 't[i][j]'$'i'=1..nops(t) ], j=1..vectdim(t[1]) )];
  result:=[result[1],{result[2..nops(result)]} ];
end:

# This is the main loop
find_all_d:=
proc() local i,b,r;  global good_triplets,Q,result,examples,A_searchfor;
  result:=[]; examples:=[];
  r:=rowdim(A_searchfor);
  if r>=7 then
  # we need the optimisation of starting with triplets instead of empty list
    find_good_triplets();
    for b in good_triplets do
      print(`testing `,b,` found so far `,result);
      for i from 1 to rowdim(Q) do
        try_E(b,i);
      od;
    od;
  else # r<7, computer is fast enough as it is
    for i from 1 to rowdim(Q) do
      print(`testing `,b,` found so far `,result);
      try_E([],i);
    od;
  fi;
end:

# to speed things up fill some tables:
calc_intersection:=
proc() local i; global n,e,Q,all_E,all_E_newbasis;
  n:=vectdim(all_E[1]);
  e:=array(sparse,1..n-1,1..n):
  for i from 1 to n-2 do e[i,i+1]:=1; e[i,n]:=-1; od:
  e[n-1,1]:=1: e[n-1,n]:=-3:
  Q:=matrix(nops(all_E),nops(all_E),
    (i,j)->2*all_E[i][1]*all_E[j][1]-innerprod(all_E[i],all_E[j]) ):
  all_E_newbasis:=map( i->linsolve(transpose(e), i), all_E):
  RETURN(); # we changed global variables, dont need to return data
end:

# set up A, D, E, U (U is $U^{-1}$) matrices.
for i from 1 to 8 do  #set up matrices for A series
  QA.i:=array(symmetric,sparse,1..i,1..i):
  for j from 1 to i do QA.i[j,j]:=-2; od:
  for j from 1 to i-1 do QA.i[j,j+1]:=1; od:
  UA.i:=array(sparse,1..i,1..i):
  for j from 1 to i do for k from j to i do UA.i[j,k]:=1; od; od;
od:
for i from 4 to 8 do  #set up matrices for D series
  QD.i:=array(symmetric,sparse,1..i,1..i):
  for j from 1 to i do QD.i[j,j]:=-2; od:
  for j from 1 to i-2 do QD.i[j,j+1]:=1; od;
  QD.i[i-2,i]:=1;  UD.i:=array(sparse,1..i,1..i):
  for j from 1 to i do for k from j to i do UD.i[j,k]:=1; od; od;
  UD.i[i-1,i]:=0;
od:
for i from 6 to 8 do  #set up matrices for E series
  QE.i:=array(symmetric,sparse,1..i,1..i):
  for j from 1 to i do QE.i[j,j]:=-2; od:
  for j from 1 to i-3 do QE.i[j,j+1]:=1; od:
  QE.i[i-3,i-1]:=1; QE.i[i-1,i]:=1;  UE.i:=array(sparse,1..i,1..i):
  for j from 1 to i do for k from j to i do UE.i[j,k]:=1; od; od;
  UE.i[i-2,i]:=0;  UE.i[i-2,i-1]:=0;
od:

# now calculate what we are interested in
# r=3
all_E:=map(vector,map(op,[
find_permutations( 0,[1,-1,0]), find_permutations( 1,[-1,-1,-1]) ])):
calc_intersection();

A_searchfor:=BlockDiagonal(QA1,QA2);  U:=BlockDiagonal(UA1,UA2);
find_all_d();  save(result,examples,all_E,cat(savepath,`ade_A1A2.dat`));

# r=4
all_E:=map(vector,map(op,[
find_permutations( 0,[1,-1,0,0]), find_permutations( 1,[-1,-1,-1,0]) ])):
calc_intersection();

A_searchfor:=BlockDiagonal(QA4);  U:=BlockDiagonal(UA4);
find_all_d();  save(result,examples,all_E,cat(savepath,`ade_A4.dat`));

# r=5
all_E:=map(vector,map(op,[
find_permutations( 0,[1,-1,0,0,0]), find_permutations( 1,[-1,-1,-1,0,0]) ])):
calc_intersection();

A_searchfor:=BlockDiagonal(QA1,QA1,QA3);  U:=BlockDiagonal(UA1,UA1,UA3);
find_all_d();  save(result,examples,all_E,cat(savepath,`ade_A1A1A3.dat`));

A_searchfor:=BlockDiagonal(QD5);  U:=BlockDiagonal(UD5);
find_all_d();  save(result,examples,all_E,cat(savepath,`ade_D5.dat`));

# r=6
all_E:=map(vector,map(op,[
find_permutations( 0,[1,-1,0,0,0,0]),
find_permutations( 1,[-1,-1,-1,0,0,0]),
find_permutations( 2,[-1,-1,-1,-1,-1,-1]) ])):
calc_intersection();

A_searchfor:=BlockDiagonal(QA1,QA5);  U:=BlockDiagonal(UA1,UA5);
find_all_d();  save(result,examples,all_E,cat(savepath,`ade_A1A5.dat`));

A_searchfor:=BlockDiagonal(QA2,QA2,QA2);  U:=BlockDiagonal(UA2,UA2,UA2);
find_all_d();  save(result,examples,all_E,cat(savepath,`ade_A2A2A2.dat`));

A_searchfor:=BlockDiagonal(QE6);  U:=BlockDiagonal(UE6);
find_all_d();  save(result,examples,all_E,cat(savepath,`ade_E6.dat`));

# r=7
all_E:=map(vector,map(op,[
find_permutations( 0,[1,-1,0,0,0,0,0]),
find_permutations( 1,[-1,-1,-1,0,0,0,0]),
find_permutations( 2,[-1,-1,-1,-1,-1,-1,0]) ])):
calc_intersection();

A_searchfor:=BlockDiagonal(QA1,QA3,QA3);  U:=BlockDiagonal(UA1,UA3,UA3);
find_all_d();  save(result,examples,all_E,cat(savepath,`ade_A1A3A3.dat`));

A_searchfor:=BlockDiagonal(QA2,QA5);  U:=BlockDiagonal(UA2,UA5);
find_all_d();  save(result,examples,all_E,cat(savepath,`ade_A2A5.dat`));

A_searchfor:=BlockDiagonal(QA7);  U:=BlockDiagonal(UA7);
find_all_d();  save(result,examples,all_E,cat(savepath,`ade_A7.dat`));

A_searchfor:=BlockDiagonal(QA1,QA1,QA1,QD4);
U:=BlockDiagonal(UA1,UA1,UA1,UD4);
find_all_d();  save(result,examples,all_E,cat(savepath,`ade_A1A1A1D4.dat`));

A_searchfor:=BlockDiagonal(QA1,QD6);  U:=BlockDiagonal(UA1,UD6);
find_all_d();  save(result,examples,all_E,cat(savepath,`ade_A1D6.dat`));

A_searchfor:=BlockDiagonal(QE7);  U:=BlockDiagonal(UE7);
find_all_d();  save(result,examples,all_E,cat(savepath,`ade_E7.dat`));

# r=8
all_E:=map(vector,map(op,[
find_permutations( 0,[1,-1,0,0,0,0,0,0]),
find_permutations( 1,[-1,-1,-1,0,0,0,0,0]),
find_permutations( 2,[-1,-1,-1,-1,-1,-1,0,0]),
find_permutations( 3,[-2,-1,-1,-1,-1,-1,-1,-1]) ])):
calc_intersection();

A_searchfor:=BlockDiagonal(QA1,QA1,QA3,QA3);
U:=BlockDiagonal(UA1,UA1,UA3,UA3);
find_all_d();  save(result,examples,all_E,cat(savepath,`ade_A1A1A3A3.dat`));

A_searchfor:=BlockDiagonal(QA1,QA2,QA5);  U:=BlockDiagonal(UA1,UA2,UA5);
find_all_d();  save(result,examples,all_E,cat(savepath,`ade_A1A2A5.dat`));

A_searchfor:=BlockDiagonal(QA1,QA7);  U:=BlockDiagonal(UA1,UA7);
find_all_d();  save(result,examples,all_E,cat(savepath,`ade_A1A7.dat`));

A_searchfor:=BlockDiagonal(QA2,QA2,QA2,QA2);
U:=BlockDiagonal(UA2,UA2,UA2,UA2);
find_all_d();  save(result,examples,all_E,cat(savepath,`ade_A2A2A2A2.dat`));

A_searchfor:=BlockDiagonal(QA4,QA4);  U:=BlockDiagonal(UA4,UA4);
find_all_d();  save(result,examples,all_E,cat(savepath,`ade_A4A4.dat`));

A_searchfor:=BlockDiagonal(QA8);  U:=BlockDiagonal(UA8);
find_all_d();  save(result,examples,all_E,cat(savepath,`ade_A8.dat`));

A_searchfor:=BlockDiagonal(QA1,QA1,QD6);  U:=BlockDiagonal(UA1,UA1,UD6);
find_all_d();  save(result,examples,all_E,cat(savepath,`ade_A1A1D6.dat`));

A_searchfor:=BlockDiagonal(QA3,QD5);  U:=BlockDiagonal(UA3,UD5);
find_all_d();  save(result,examples,all_E,cat(savepath,`ade_A3D5.dat`));

A_searchfor:=BlockDiagonal(QA1,QE7);  U:=BlockDiagonal(UA1,UE7);
find_all_d();  save(result,examples,all_E,cat(savepath,`ade_A1E7.dat`));

A_searchfor:=BlockDiagonal(QA2,QE6);  U:=BlockDiagonal(UA2,UE6);
find_all_d();  save(result,examples,all_E,cat(savepath,`ade_A2E6.dat`));

A_searchfor:=BlockDiagonal(QD4,QD4);  U:=BlockDiagonal(UD4,UD4);
find_all_d();  save(result,examples,all_E,cat(savepath,`ade_D4D4.dat`));

A_searchfor:=BlockDiagonal(QD8);  U:=BlockDiagonal(UD8);
find_all_d();  save(result,examples,all_E,cat(savepath,`ade_D8.dat`));

A_searchfor:=BlockDiagonal(QE8);  U:=BlockDiagonal(UE8);
find_all_d();  save(result,examples,all_E,cat(savepath,`ade_E8.dat`));
\end{verbatim}
}



\newpage
\baselineskip 14pt
{\footnotesize
\begin{thebibliography}{AAAAaa}
%
\bibitem[Al]{} O.\ Alvarez,
 {\it Strings, branes, and the structure of space-time},
 talk given at the AMS meeting at Chattanooga, Tennessee,
 October 1996.

\bibitem[As]{} P.S.\ Aspinwall,
 {\it K3 surfaces and string duality}, {\tt hep-th/9611137}.

\bibitem[At]{} M.F.\ Atiyah,
 {\it On analytic surfaces with double points},
 {\sl Proc.\ Royal Soc.}\ {\bf A247} (1958), pp.\ 237 - 244.

\bibitem[A-G-M]{} P.S.\ Aspinwall, B.R.\ Greene, and D.R.\ Morrison,
 {\it Calabi-Yau moduli space, mirror manifolds and spacetime
      topology change in string theory},
 {\sl Nucl.\ Phys.}\ {\bf B416} (1994), pp.\ 414 - 480.

\bibitem[A-GZ-V]{} V.I.\ Arnold, S.M.\ Gusein-Zade,
                   and A.N.\ Varchenko,
 {\sl Singularities of differentiable maps,
      Vol.\ II: Monodromy and asymptotics of integrals},
 Monographs Math.\ 83, Birkh\"{a}user, 1988.

\bibitem[A-L]{} P.S.\ Aspinwall and C.A.\ L\"{u}tken,
 {\it Quantum algebraic geometry of superstring compactifications},
 {\sl Nucl.\ Phys.}\ {\bf B355} (1991), pp.\ 482 - 510.

\bibitem[Ba]{} D.\ Barden,
 {\it Simply connected five-manifolds},
 {\sl Ann.\ Math.}\ {\bf 82} (1965), pp.\ 365 - 385.

\bibitem[Bou]{} N.\ Bourbaki,
 {\sl El\'{e}m\'{e}nts de Math\'{e}matique, Livre II, Alg\`{e}bre},
 Hermann, 1959.

\bibitem[B-B-S]{} K.\ Becker, M.\ Becker, and A.\ Strominger,
 {\it Fivebranes, membranes and non-perturbative string theory},
 {\sl Nucl.\ Phys.}\ {\bf B456} (1995), pp.\ 130 - 152.

\bibitem[B-C-D]{} D.\ Berenstein, R.\ Corrado, and J.\ Distler,
 {\it On the moduli spaces of M(atrix)-theory compactifications},
 {\tt hep-th/9704087}.

\bibitem[B-C-dlO]{} P.\ Berglund, P.\ Candelas,
   and X.C.\ de la Ossa,
 {\it Periods for Calabi-Yau and Landau-Ginzburg vacua},
 {\sl Nucl.\ Phys.}\ {\bf B419} (1994), pp.\ 352 - 403.

\bibitem[B-G]{} R.L.\ Bryant and P.A.\ Griffiths,
 {\it Some observations on the infinitesimal period relations
      for regular threefolds with trivial canonical bundle},
 in {\sl Arithmetic and geometry - vol II: Geometry},
 M.\ Artin and J.\ Tate eds.
 Progr.\ Math.\ 36, Birkh\"{a}user, 1983.

\bibitem[B-I-K-M-S-V]{} M.\ Bershadsky, K.\ Intriligator,
   S.\ Kachru, D.R.\ Morrison, V.\ Sadov, and C.\ Vafa,
 {\it Geometric singularities and enhanced gauge symmetries},
 {\sl Nucl.\ Phys.}\ {\bf B481} (1996), pp.\ 215 - 252.

\bibitem[B-P-VV]{} W.\ Barth, C.\ Peters, and A.\ Van de Ven,
 {\sl Compact complex surfaces},
 Ser.\ Modern Surveys Math.\ 4, Springer-Verlag, 1984.

\bibitem[B-T]{} R.\ Bott and L.W.\ Tu,
 {\sl Differential forms in algebraic topology},
 GTM 82, Springer-Verlag, 1982.

\bibitem[B-V-S1]{} M.\ Bershadsky, C.\ Vafa, and V.\ Sadov,
 {\it D-strings on D-manifolds},
 {\sl Nucl.\ Phys.}\ {\bf B463} (1996), pp.\ 398 - 414.

\bibitem[B-V-S2]{} --------,
 {\it D-branes and topological field theory},
 {\sl Nucl.\ Phys.}\ {\bf B463} (1996), pp.\ 420 - 434.

\bibitem[C-dlO]{} P.\ Candelas and X.C.\ de la Ossa,
 {\it Moduli space of Calabi-Yau manifolds},
 {\sl Nucl.\ Phys.}\ {\bf B355} (1991), pp.\ 455 - 481.

\bibitem[C-G-G-K]{} T.-M. Chiang, B.R.\ Greene, M.\ Gross, and
                    Y.\ Kanter,
 {\it Black hole condensation and the web of Calabi-Yau manifolds},
 {\tt hep-th/9511204}.

\bibitem[C-G-H]{} P.\ Candelas, P.S.\ Green, and T.\ H\"{u}bsch,
 {\it Rolling among Calabi-Yau vacua},
 {\sl Nucl.\ Phys.}\ {\bf B330} (1990), pp.\ 49 - 102.

\bibitem[C-H-S-W]{} P.\ Candelas, G.T.\ Horowitz, A.\ Strominger,
                      and E.\ Witten,
 {\it Vacuum configurations for superstrings},
 {\sl Nucl.\ Phys.}\ {\bf B258} (1985), pp.\ 46 - 74.

\bibitem[C-P-R]{} P.\ Candelas, E.\ Perevalov, and G.\ Rajesh,
 {\it Toric geometry and enhanced gauge symmetry of
      F-theory/heterotic vacua},
 {\tt hep-th/9704097}.

\bibitem[Da]{} V.I.\ Danilov,
 {\it The geometry of toric varieties},
 {\sl Russian Math.\ Surveys} {\bf 33} (1978), pp.\ 97 - 154. 

\bibitem[De]{} M.\ Demazure,
 {\it Surfaces de Del Pezzo},
 in {\sl S\'{e}mimaire sur les singularit\'{e}s des surfaces},
 M.\ Demazure, H.\ Pinkham, and B.\ Teissier eds.,
 Lect.\ Notes Math.\ 777, Springer-Verlag, 1980.

\bibitem[Di]{} A.\ Dimca,
 {\sl Singularities and topology of hypersurfaces},
 Springer-Verlag, 1992.

\bibitem[D-G-M]{} J.\ Distler, B.R.\ Greene, and D.R.\ Morrison,
 {\it Resolving singularities in $(0,2)$ models},
 {\sl Nucl.\ Phys.}\ {\bf B481} (1996), pp.\ 289 - 312.

\bibitem[D-G-W]{} R.\ Donagi, A.\ Grassi, and E.\ Witten,
 {\it A non-perturbative superpotential with $E_8$-symmetry},
 {\sl Mod.\ Phys.\ Lett.}\ {\bf A11} (1996), pp.\ 2199 - 2212.

\bibitem[D-K-L]{} M.J.\ Duff, R.\ Khuri, and J.X.\ Lu,
 {\it String solitons},
 {\sl Phys.\ Reports} {\bf 259} (1995), pp.\ 213 - 326.

\bibitem[D-M]{} M.R.\ Douglas and G.\ Moore,
 {\it D-branes, quivers, and ALE instantons},
 {\tt hep-th/9603167}.

\bibitem[D-T]{} S.K.\ Donaldson and R.P.\ Thomas,
 {\it Gauge theory in higher dimensions},
 Oxford preprint, 1997.

\bibitem[Fr]{} R.\ Friedman,
 {\it Simultaneous resolution of threefold double points},
 {\sl Math.\ Ann.}\ {\bf 274} (1986), pp.\ 671 - 689.

\bibitem[Ful]{} W.\ Fulton,
 {\sl Introduction to toric varieties},
 Ann.\ Math.\ Study 131, Princeton Univ.\ Press, 1993.

\bibitem[Fur]{} M.\ Furushima,
 {\it Singular del Pezzo surfaces and analytic compactifications
  of $3$-dimensional complex affine space
  ${\footnotesizeBbb C}^3$},
 {\sl Nagoya Math.\ J.}\ {\bf 104} (1986), pp.\ 1 - 28.

\bibitem[F-K-M]{} S.\ Ferrara, R.R.\ Khuri, and R.\ Minasian,
 {\it M-theory on a Calabi-Yau manifold},
 {\sl Phys.\ Lett.}\ {\bf B375} (1996), pp.\ 81 - 88.

\bibitem[Go]{} R.E.\ Gompf,
 {\sl $4$-dimensional manifolds},
 course given at the Department of Mathematics, U.T.\ Austin,
 fall 1997; also many private discussions after class.

\bibitem[Gr]{} B.R.\ Greene,
 {\it String theory on Calabi-Yau manifolds}, lectures given at
  TASI-96 summer school on {\sl Strings, Fields, and Duality},
 {\tt hep-th/9702155}.

\bibitem[G-H]{} P.\ Griffiths and J.\ Harris,
 {\sl Principles of algebraicc geometry},
 John Wiley \& Sons, Inc., 1978.

\bibitem[G-M-S]{} B.R.\ Greene, D.R.\ Morrison, and A.\ Strominger,
 {\it Black hole condensation and the unification of string vacua},
 {\sl Nucl.\ Phys.}\ {\bf B451} (1995), pp.\ 109 - 120.  

\bibitem[G-M-V]{} B.R.\ Greene, D.R.\ Morrison, and C.\ Vafa,
 {\it A geometric realization of confinement},
 {\sl Nucl.\ Phys.}\ {\bf B481} (1996), pp.\ 513 - 538.

\bibitem[G-Mu-V]{} M.\ Green, J.\ Murre, and C.\ Voisin,
 {\sl Algebraic cycles and Hodge theory},
 Lect.\ Notes Math.\ 1594, Springer-Verlag, 1994.

\bibitem[G-P]{} V.\ Guillemin and A.\ Pollack,
 {\sl Differential topology},
 Prentice-Hall, 1974.

\bibitem[G-S]{} R.E.\ Gompf and A.I.\ Stipsicz,
 {\sl An introduction to $4$-manifolds and Kirby calculas},
 preliminary version - July 1997, to be published in book form
 by the American Mathematical Society.

\bibitem[Ha]{} A.\ Haefliger,
 {\it Plongements diff\'{e}rentiables de vari\'{e}t\'{e}s dans
  vari\'{e}t\'{e}s},
 {\sl Comment.\ Math.\ Helv.}\ {\bf 36} (1961), pp.\ 47 - 82.

\bibitem[Hiro]{} H.\ Hironaka,
 {\it Resolution of singularities of an algebraic variety over
  a field of characteristic zero. I $\&$ II.}
 {\sl Ann.\ Math.}\ {\bf 79} (1964),
 pp.\ 109 -203 and pp.\ 205 - 326.

\bibitem[Hirs]{} M.W.\ Hirsch,
 {\sl Differential topology},
 GTM 33, corrected 4th printing, Springer-Verlag, 1991.

\bibitem[H\"{u}]{} T.\ H\"{u}bsch,
 {\sl Calabi-Yau manifolds -- a bestiary for physicists},
 World Scientific, 1992.

\bibitem[H-K-K]{} J.\ Harer, A.\ Kas, and R.\ Kirby,
 {\sl Handlebody decompositions of complex surfaces},
 Memoirs Amer.\ Math.\ Soc.\, no.\ 350, Amer.\ Math.\ Soc., 1986.

\bibitem[H-W1]{} P.\ Ho\v{r}ava and E.\ Witten,
 {\it Heterotic and type I string dynamics from eleven dimnsions},
 {\sl Nucl.\ Phys.}\ {\bf B460} (1996), pp.\ 506 - 524.

\bibitem[H-W2]{} --------,
 {\it Eleven-dimensional supergravity on a manifold with boundary},
 {\sl Nucl.\ Phys.}\ {\bf B475} (1996), pp.\ 94 - 114.

\bibitem[I-M-S]{} K.\ Intriligator, D.R.\ Morrison, N.\ Seiberg,
 {\it Five-dimensional supersymmetric gauge theories and degenerations
      of Calabi-Yau spaces},
 {\sl Nucl.\ Phys.}\ {\bf B497} (1997), pp.\ 56 - 100.

\bibitem[Ja]{} N.\ Jacobson,
 {\sl Basic algebra I},
 W.H.\ Freeman $\&$ Co., 1974.

\bibitem[Jo1]{} D.D.\ Joyce,
 {\it Compact Riemannian 7-manifolds with holonomy $G_2$. I},
 {\sl J.\ Diff.\ Geom.}\ {\bf 43} (1996), pp.\ 291 - 328.

\bibitem[Jo2]{} --------,
 {\it Compact Riemannian 7-manifolds with holonomy $G_2$. II},
 {\sl J.\ Diff.\ Geom.}\ {\bf 43} (1996), pp.\ 329 - 375.

\bibitem[Ki]{} R.C.\ Kirby,
 {\sl The topology of $4$-manifolds},
 Lect.\ Notes Math.\ 1374, Springer-Verlag, 1989.

\bibitem[Kl]{} A.\ Klemm,
 {\it On the geometry behind $N=2$ supersymmetric effective
  actions in four dimensions},
 lectures presented at the Trieste Summer School 1996 and the
 33rd Karpacz School on String Duality 1997, {\tt hep-th/9705131}.

\bibitem[Kod]{} K.\ Kodaira,
 {\sl Complex manifolds and deformations of complex structures},
 translated by K.\ Akao, Grund.\ Math.\ Wiss.\ 283,
 Springer-Verlag, 1986.

\bibitem[Kol]{} J.\ Koll\'{a}r,
 {\it Flops},
 {\sl Nagoya Math.\ J.}\ {\bf 113} (1989), pp.\ 15 - 36.

\bibitem[K-L-R-Y]{} A.\ Klemm, B.\ Lian, S.-S.\ Roan,
  and S.T.\ Yau,
 {\it Calabi-Yau fourfolds for M- and F-theory compactifications},
 {\tt hep-th/9701023}.

\bibitem[K-MK]{} S.\ Keel and J.\ McKernan,
 {\it Rational curves on quasi-projective surfaces},
 {\tt alg-geom/9707016}.

\bibitem[Le]{} W.\ Lerche,
 {\it Introduction to Seiberg-Witten theory and its stringy
  origin},
  contribution to the {\sl Proceedings of the "Spring School and
  Workshop on String Theory, Gauge Theory and Quantum Gravity",
  I.C.T.P.}, Trieste, Italy, March 18 - 29, 1996,
 {\tt hep-th/9611190.}

\bibitem[Ma1]{} E.\ Martinec,
 {\it Geometric structures of M-theory},
 {\tt hep-th/9608017}.

\bibitem[Ma2]{} --------,
 {\it M-theory and $N=2$ strings},
 talk given at the NATO Advanced Study Institute on Strings,
  Branes, and Duality, Cargese France, summer 1997,
 {\tt hep-th/9710122}.

\bibitem[Mi1]{} J.W.\ Milnor,
 {\sl Morse theory},
 Ann.\ Math.\ Study 51, Princeton Univ.\ Press, 1963.

\bibitem[Mi2]{} --------,
 {\sl Singular points of complex hypersurfaces},
 Ann.\ Math.\ Study 61, Princeton Univ.\ Press, 1968.

\bibitem[Mu]{} J.R.\ Munkres,
 {\sl Elements of algebraic topology},
 Addison-Wesley Publ.\ Co., 1984.

\bibitem[M-S]{} J.W.\ Milnor and J.D.\ Stasheff,
 {\sl Characteristic classes},
 Ann.\ Math.\ Study 76, Princeton Univ.\ Press, 1974.

\bibitem[M-V1]{} D.R.\ Morrison and C.\ Vafa,
 {\it Compactifications of F-theory on Calabi-Yau threefolds (I),}
 {\sl Nucl.\ Phys.}\ {\bf B473} (1996), pp.\ 74 - 92.              1

\bibitem[M-V2]{} --------,
 {\it Compactifications of F-theory on Calabi-Yau threefolds (II),}
 {\sl Nucl.\ Phys.}\ {\bf B476} (1996), pp.\ 437 - 469.

\bibitem[M-Z]{} M.\ Miyanishi and D.Q.\ Zhang,
 {\it Gorenstein log del Pezzo surfaces of rank one},
 {\sl J.\ Algebra}, {\bf 118} (1988), pp.\ 63 - 84.

\bibitem[Od]{} T.\ Oda,
 {\sl Convex bodies and algebraic geometry
       - an introduction to the theory of toric varieties},
 Springer-Verlag 1988.

\bibitem[Po]{} V.\ Po\'{e}naru,
 {\it On the geometry of differentiable manifolds},
 in {\sl Studies in modern topology}, P.J.\ Hilton ed.,
 MAA Studies in Math., vol.\ 5,
 Math.\ Asso.\ Amer.\ \& Prentice-Hall Inc., 1968.

\bibitem[Re1]{} M.\ Reid,
 {\it Canonical threefolds}, in
 {\sl G\'{e}om\'{e}trie alg\'{e}brique}, Angers 1979,
 A.\ Beauville ed., pp.\ 273 - 310, Sijthoff and Noordhoff,
 Alphen aan den Rijn, 1980.

\bibitem[Re2]{} --------,
 {\it Young person's guide to canonical singularities},
 in {\sl Algebraic geometry}, Bowdoin 1985,
 S.J.\ Bloch with H.\ Clemens, D.\ Eisenbud, W.\ Fulton,
 D.\ Gieseker, J.\ Harris, R.\ Hartshorne, and S.\ Mori ed.,
 Proc.\ Symp.\ Pure Math.\ vol.\ 46 - part 1, pp.\ 345 - 414,
 Amer.\ Math.\ Soc.\ 1987.

\bibitem[Ro]{} D.\ Rolfsen,
 {\sl Knots and links},
 2nd printing, Publish or Perish, 1990.

\bibitem[Sa]{} S.\ Salamon,
 {\sl Riemannian geometry and holomony group},
 Pitman Research Notes Math.\ Ser.\ 201,
 Longman Scientific \& Technical, 1989.

\bibitem[Sh1]{} I.R.\ Shafarevich,
 {\sl Basic algebraic geometry},
 {\sl vol I: varieties in projective space},
 {\sl vol II: schemes and complex manifolds},
 Springer-Verlag, 1994.

\bibitem[Sh2]{} -------- ed.,
 {\sl Algebraic geometry II:}
 {\it Cohomology of algebraic varieties} by V.I.\ Danilov;
 {\it Algebraic surfaces} by V.A.\ Iskovskikh and I.R.S.,
 Springer-Verlag, 1996.

\bibitem[Sl1]{} P.\ Slodowy,
 {\sl Simple singularities and simple algebraic groups},
 Lect.\ Notes Math.\ 815, Springer-Verlag, 1980.

\bibitem[Sl2]{} --------,
 {\it Platonic solids, Kleinian singularities, and Lie groups},
 in {\sl Algebraic geometry}, I.\ Dolgachev ed., pp.\ 102 - 138,
 Lect.\ Notes Math.\ 1008, Springer-Verlag, 1983.

\bibitem[Sm1]{} S.\ Smale,
 {\it Generalized Poincar\'{e} conjecture in dimensions greater
   than four},
 {\sl Ann.\ Math.}\ {\bf 74} (1961), pp.\ 391 - 406.

\bibitem[Sm2]{} --------,
 {\it On the structure of $5$-manifolds},
 {\sl Ann.\ Math.}\ {\bf 75} (1962), pp.\ 38 - 46.

\bibitem[Sp]{} E.H.\ Spanier,
 {\sl Algebraic topology},
 Springer-Verlag, 1966.

\bibitem[Sto]{} R.E.\ Stong,
 {\sl Notes on cobordism theory},
 Princeton Math.\ Notes, Princeton Univ.\ Press, 1958.

\bibitem[Str]{} A.\ Strominger,
 {\it Massless black holes and conifolds in string theory},
 {\sl Nucl.\ Phys.}\ {\bf B451} (1995), pp.\ 96 - 108.

\bibitem[Sw]{} R.M.\ Switzer,
 {\sl Algebraic topology - homotopy and homology},
 Grund.\ Math.\ Wiss.\ 212, Springer-Verlag, 1975.

\bibitem[Ti]{} G.\ Tian,
 {\it Smoothness of the universal deformation space of compact
  Calabi-Yau manifolds and its Peterson-Weil metric}, in
 {\sl Mathematical aspects of string theory}, S.T.\ Yau ed.,
 pp.\ 629 - 646, World Scientific, 1987. 

\bibitem[To]{} P.K.\ Townsend,
 {\it Four lectures on M-theory},
 To appear in the {\sl Proceedings of the 1996 ICTP Summer School in
  High Energy Physics and Cosmology} (Trieste, June 10 - 26),  
 {\tt hep-th/9612121}.

\bibitem[Vo]{} C.\ Voisin,
 {\sl Sym\'{e}trie miroir},
 Panoramas et Synthe\`{e}ses 2, Soc.\ Math.\ de France 1996,
 distributed by Amer.\ Math.\ Soc..

\bibitem[Wa1]{} C.T.C.\ Wall,
 {\it Determination of the cobordism ring},
 {\sl Ann.\ Math.}\ {\bf 72} (1960), pp.\ 292 - 311.

\bibitem[Wa2]{} --------,
 {\it Classification problems in differential topology. V:
      On certain $6$-manifolds},
 {\sl Invent.\ Math.}\ {\bf 1} (1966), pp.\ 355 - 374.

\bibitem[Wa3]{} --------,
 {\it Real forms of smooth del Pezzo surfaces},
 {\sl J.\ Reine Angew.\ Math.}\ {\bf 375/376} (1987),
 pp.\ 47 - 66.

\bibitem[Wil1]{} P.M.H.\ Wilson,
 {\it Calabi-Yau manifolds with large Picard number},
 {\sl Invent.\ Math.}\ {\bf 98} (1989), pp.\ 139 - 155.

\bibitem[Wil2]{} --------,
 {\it The K\"{a}hler cone on Calabi-Yau threefolds},
 {\sl Invent.\ Math.}\ {\bf 107} (1992), pp.\ 561 - 583.

\bibitem[Wit1]{} E.\ Witten,
 {\it String theory dynamics in various dimensions},
 {\sl Nucl.\ Phys.}\ {\bf B443} (1995), pp.\ 85 - 126.

\bibitem[Wit2]{} --------,
 {\it Phase transitions in M-theory and F-theory},
 {\sl Nucl.\ Phys.}\ {\bf B471} (1996), pp.\ 195 - 216.

\bibitem[Wit3]{} --------,
 {\it Non-perturbative superpotentials in string theory},
 {\sl Nucl.\ Phys.} {\bf B474} (1996), pp.\ 343 - 360.

\bibitem[W-W]{} R.S.\ Ward and R.O.\ Wells Jr.\
 {\it Twistor geometry and field theory},
 Cambridge Univ.\ Press, 1990.

\end{thebibliography}
}
\enddocument